\documentclass{jfm}
\usepackage{epsfig,graphics,graphicx,amssymb,amsmath,color,natbib}

\usepackage[sfdefault=cmbr]{isomath}

\def\e{\epsilon}
\def\v{\vectorsym}
\def\t{\tensorsym}

\include{help}
\title[Hydrodynamics of self-propulsion near a boundary]{Hydrodynamics of self-propulsion near a boundary: predictions and accuracy of far-field approximations}

\author[S. E. Spagnolie and E. Lauga]%
{S\ls A\ls V\ls E\ls R\ls I\ls O \ls E. \ls S\ls P\ls A\ls G\ls N\ls O\ls L\ls I\ls E$^{1,2}$
\and E\ls R\ls I\ls C  \ls  L\ls A\ls U\ls G\ls A$^2$}

\affiliation{$^1$School of Engineering,
Brown University,
182 Hope Street, Providence, RI 02912, USA\\
$^2$Department of Mechanical and Aerospace Engineering,
University of California, San Diego,
9500 Gilman Drive, La Jolla, CA 92093, USA}

\pubyear{2011}
\volume{??}
\pagerange{??--??}
\date{\today}
\usepackage{graphicx,color,bm}
\graphicspath{{converted_graphics/}}
\usepackage{graphicx}
\begin{document}

\maketitle

\begin{abstract}
The swimming trajectories of self-propelled organisms or synthetic devices in a viscous fluid can be altered by hydrodynamic interactions with nearby boundaries. We explore a multipole description of swimming bodies and provide a general framework for studying the fluid-mediated modifications to swimming trajectories. The validity of the far-field description is probed for a selection of model swimmers of varying geometry and propulsive activity by comparison with full numerical simulations. The reduced model is then used to deliver simple but accurate predictions of hydrodynamically generated wall attraction and pitching dynamics, and may help to explain a number of experimental results.
\end{abstract}

\section{Introduction}

The swimming kinematics and trajectories of many microorganisms are altered by the presence of nearby boundaries, be they solid or deformable, and often in perplexing fashion. This activity takes place at extremely low Reynolds numbers, a regime in which fluid motion is dominated by viscous dissipation. An important factor for swimming at such scales is the long range nature of hydrodynamic interactions, either between immersed bodies, or between an immersed body and a surface \citep{lp09}. When the swimming dynamics of an organism vary near such boundaries a question arises naturally: is the change in behavior biological, fluid mechanical, or perhaps mediated by other physical laws? {\it E. coli} cells, for instance, have been observed to swim in large circles when in the presence of a solid boundary, which has been accounted for in a purely fluid mechanical consideration by \cite{ldlws06}. Other organisms have been shown to reverse direction at boundaries by inverting the orientation of flagellar rotation, resulting in a departure from the boundary which is clearly not a passive hydrodynamic effect \citep{cgdk07}. In an attempt to help differentiate such observations, we seek a general framework for determining the extent to which fluid mechanics can passively alter the swimming trajectories of microorganisms near surfaces.

Surface effects on motility lead to varied and important consequences in a number of engineering and biological systems. \cite{vllnz90} note that surfaces are the major site of microbial activity in natural environments, and refer to \cite{hy80} who showed that almost all detectable bacteria in a marsh estuary were associated with particles. Correspondingly, the attraction of certain microorganisms to surfaces has a major impact on the development of biofilms, which can begin with the adhesion of individual cells to a surface \citep{vllnz90,otkk00}. Biofilms are responsible for numerous microbial infections, and can play an important role in such phenomena as biological fouling \citep{llsc03,Harshey03,kg06}. A recent review on the mathematical modeling of microbial biofilms has been presented by \cite{kd10}. Meanwhile, in a lab setting it is common that microorganisms are in near contact with microscope slides or are directed through microfluidic channels in which boundaries can play  significant roles. The migration of bacteria through small-diameter capillary tubes was studied by \cite{bt90}, and that of infectious bacteria along medically implanted surfaces was considered by \cite{hdf92}. More recently, \cite{el10} have shown that the presence of a wall can lead to a change in the waveform expressed by actuated flagella, which in turn results in an increase or decrease in its propulsive force depending on the type of actuation.

The study of microbial attraction to surfaces reaches back to the observations of \cite{Rothschild63}, who measured the distribution of bull spermatozoa swimming between two glass plates and found the cell distribution to be nonuniform with the cell density strongly increasing near the walls. By modeling swimmers as dipolar pushers, it has been argued by \cite{btbl08} that this hydrodynamic consideration alone can account for the attraction. Immersed boundary simulations of swimming bodies with undulating flagella have also shown a hydrodynamic attraction towards a wall \citep{fm95}. More recently, \cite{sgbk09} have explored numerically the wall effects on geometrically accurate swimming human spermatozoa. They have demonstrated that hydrodynamic interactions can trap the body in a stable orbit near a boundary, in some cases with counter-intuitive orientation and at finite separation distance from the wall. The numerical results of \cite{sgs10} and \cite{giy10} show the existence of a stable swimming distance from the boundary in swimming {\it E. coli} that depends upon the shape of the cell body and the flagellum. \cite{gnbnm05} have also detected an equilibrium pitching angle for a given wall separation distance. Recent experiments showing the upstream swimming of bacteria in a shear flow by \cite{hkmk07} suggest that the geometry and orientation of hydrodynamically bound swimming organisms can be important. Meanwhile, \cite{ddcgg11} have measured experimentally the flow field generated by the swimming of an individual {\it E. coli} bacterium near a solid surface and have shown that steric collisions and near-field lubrication forces dominate any long-range fluid dynamical effects on these length scales.

Other recent studies on the dynamics of swimming bodies near walls of a more theoretical nature includes work by \cite{znm09}, who studied the dynamical motion of a three-sphere swimmer near a wall, and \cite{co10}, who studied a  simple two-dimensional model of a swimmer using methods of complex analysis \citep[see also][]{Crowdy11}. A different avenue of inquiry has also seen much recent activity, the effect of boundaries on swimming suspensions of microorganisms. For instance, \cite{houg09} have studied model swimmers composed of dipolar pushing beads, and have shown that the additional length scales introduced by confinement can suppress the onset of large scale structures in the suspension. 

Frequently it is the case that the surface of interest does not impose a no-slip condition on the fluid velocity, for instance at a free boundary between water and air. \cite{tcdwkg05} have considered the development of large scale fluid structures driven by a competition between oxygen-taxis near the surface of a sessile drop and gravitational effects. \cite{dldaai11} considered the hydrodynamic interactions of a swimming bacterium with a stress-free surface, which can be analyzed by placing a mirror image of the swimming organism opposite the free-surface. The circular trajectories studied by \cite{ldlws06} were found to be reversed in this setting. 

Outside of the fundamental benefits of scientific inquiry, a more complete understanding of the hydrodynamic interactions between self-propelled bodies will continue to drive the development of engineering applications as well. Synthetic swimming particles have been designed to perform tasks at an exceptionally small length scales, including chemically driven bimetallic nano-rods \citep{pkossacmlc04, fbamo05, rk07}, magnetic nanopropellers \citep{gf09,pgwl11}, and undulatory chains of magnetic colloidal particles \citep{dbrfsb05} \citep[see also][]{Wang09}. Sorting and rectification devices which lean upon asymmetries in microbial interactions with walls have been explored by \cite{gkca07}, while \cite{dladariscmdadf10} have considered the driven motion of gear-like ratchets in bacterial suspensions. Another application of more recent interest is in the production of biofuels, where suspensions of algae are shuttled through long channels \citep[see][]{bc10}. Exploring the hydrodynamic interactions between self-propelled bodies and surfaces not only allows us to understand the biological realm with greater sophistication, but may also allow for the development of manmade devices of increasing complexity and creativity.

In the present study, we utilize a multipole representation of self-propelled organisms in order to improve our understanding of swimming behaviors near a surface from a generalized perspective. The modeling of swimming organisms by Stokeslet dipole singularities has become commonplace, but here we take one systematic step further in the far-field expansion of the flow generated by self-propelling bodies. The inclusion of higher order singularities will be shown to have important consequences on swimming trajectories. Full scale simulations of the Stokes equations are used as a benchmark to explore the regions of validity and limitations of the reduced model for two types of model swimmers, namely ellipsoidal Janus particles with prescribed tangential surface actuation and bacteria-like spheroid-rod swimmers. The far-field approximation leads to very good quantitative agreement with the full simulation results in some cases down to a tenth of a body length away from the wall. Exploiting the quantitative predictions from our singularity approach, the reduced model is further shown to provide good predictive power for the initial attraction/repulsion to the wall and the rotation induced by the presence of the wall, and even surface scattering in the particular case of spheroidal {\it squirmers}.

This paper is organized as follows. In \S II, the Stokeslet and higher order singularity solutions of the Stokes equations are introduced, and a general axisymmetric swimmer is described in terms of a Stokeslet dipole, a source dipole, a Stokeslet quadrupole, and a rotlet dipole. The wall effects on the  trajectory of a swimming body are described through the contribution of each singularity in \S III. In \S IV we address the question of how accurately this multipole representation of swimming trajectories captures the real wall effects observed in full numerical solutions of the Stokes equations. The reduced (singularity) model is used to provide a simple description of the wall-induced rotation for model Janus particles, as well as to describe the complete swimming dynamics of a squirming spheroid in \S V. In \S VI we consider model polar swimmers that are  bacteria-like in their geometry, and develop an approximate Fax\'en Law for their study. In \S VII we then employ the reduced model to study a transition in the wall-induced rotations experienced by bacteria-like swimmers for a critical flagellum length. We finally discuss the accuracy and limitations of the reduced model in describing the geometry and dynamics of trapped self-propelled bodies near surfaces.

\section{Singularity representation of motion in a viscous fluid}

\subsection{The Stokes equations and singularity solutions}
The length and velocity scales which describe the locomotion of microorganisms are extremely small. The fluid flow generated by their activity is dictated almost entirely by viscous dissipation, as summarized by \cite{Purcell77}. The Reynolds number describing the flow, $Re=\rho\, U L/\mu$ is likewise very small, where $\rho$ is the fluid density, $\mu$ is the dynamic viscosity, and $L$ and $U$ are length and velocity scales characteristic of the organism. The swimming of {\it E. coli}, for example, is characterized by a Reynolds number $Re\approx 10^{-4}$ \citep[see][]{Childress81}. The fluid behavior is therefore described well by the Stokes equations,
\begin{gather}
\nabla \cdot \t{\sigma}=-\nabla p + \mu\,\Delta \v{u}=0,\label{E:Stokes}\\
\nabla\cdot \v{u}=0,\label{E:Incompressibility}
\end{gather}
where $\t{\sigma}=-p\,\t{I}+2\mu \t{E}$ is the Newtonian fluid stress tensor, $p$ is the pressure, $\v{u}$ is the fluid velocity, $\t{I}$ is the identity operator, and $\t{E}=(\nabla \v{u}+\nabla \v{u}^T)/2$ is the symmetric rate-of-strain tensor. The fluid velocity is assumed to decay in the far-field, and the boundary condition assumed on the swimming body depends on the specific organism, as will be described below. In situations where we include the presence of a plane wall of infinite extent at $z=0$, we shall also assume a no-slip condition there $(\v{u}(z=0)=0)$. Now classical treatises on zero Reynolds number fluid dynamics have been written by \cite{hb65} and \cite{kk91}.

The linearity of the Stokes equations allows for the introduction and exploitation of Green's functions. The description of the fluid behavior far from an actively motile organism, for instance, can be described accurately using only the first few terms in a multipole expansion of fundamental singularities, which will be our approach here. The utilization of fundamental singularities allowed for a series of exact solutions to fundamental problems in Stokesian fluid dynamics to be derived by \cite{cw75}. 

A free-space Green's function for the Stokes equations is derived by placing a point force in an otherwise quiescent infinite fluid, $\v{f}\hat{\delta}(\v{x_0})$ (where $\hat{\delta}(\v{x_0})$ is the Dirac delta function centered at $\v{x_0}$). With the point force directed along the unit-vector $\v{e}$ (and defining $\v{f}=f\v{e}$), the solution to the forced system produces the so-called Stokeslet singularity,
\begin{gather}
\v{u}(\v{x})=\frac{f}{8\pi\mu}\v{G}(\v{x-x_0};\v{e}),
\end{gather}
where 
\begin{gather}
\v{G}(\v{x-x_0};\v{e})=\frac{1}{R}\left(\v{e}+\frac{[\v{e\cdot (x-x_0)}](\v{x-x_0})}{R^2}\right)\label{Stokeslet1}
\end{gather}
is the $\v{e}$-directed Stokeslet, and $R=|\v{x-x_0}|$. Derivatives of the Stokeslet singularity produce other higher-order singularity solutions of the Stokes equations. The first three such singularities are the Stokeslet dipole, quadrupole, and octupole, described by
\begin{align}
\v{G_D}(\v{x-x_0};\v{d},\v{e})&=\v{d}\cdot \nabla_0 \v{G}(\v{x-x_0};\v{e}),\\
\v{G_Q}(\v{x-x_0};\v{c},\v{d},\v{e})&=\v{c}\cdot \nabla_0 \v{G_D}(\v{x-x_0};\v{d},\v{e}),\\
\v{G_O}(\v{x-x_0};\v{b},\v{c},\v{d},\v{e})&=\v{b}\cdot \nabla_0 \v{G_Q}(\v{x-x_0};\v{c},\v{d},\v{e}),
\end{align}
respectively, where the gradient ($\nabla_0$) acts on the singularity placement $\v{x_0}$. The vectors $\v{b,c}$ and $\v{d}$ indicate the directions along which each derivative is taken. As the solutions above are regular outside of the singular point there are many possible identities that may be observed by rearranging the order in which these derivatives are taken. Tensorial expressions of the singularities above are provided by \cite{poz} \citep[see also][]{cw75}, and we have included the full vector expressions of these singularities in Appendix A. In addition to the solutions above, there are singular potential flow solutions to the Stokes equations which are associated with Laplace's equation ($p=0$ in \ref{E:Stokes}). The source, source dipole, source quadrupole, and source octupole singularity solutions are, respectively,
\begin{align}
\v{U}(\v{x-x_0})&=\frac{\v{x-x_0}}{R^3},\\
\v{D}(\v{x-x_0};\v{e})&=\v{e}\cdot \nabla_0 \v{U}(\v{x-x_0}),\\
\v{Q}(\v{x-x_0};\v{d},\v{e})&=\v{d}\cdot \nabla_0 \v{D}(\v{x-x_0};\v{e}),\\
\v{O}(\v{x-x_0};\v{c},\v{d},\v{e})&=\v{c}\cdot \nabla_0 \v{Q}(\v{x-x_0};\v{d},\v{e}). 
\end{align}
The source singularities are related to the force singularities through the relation 
\begin{gather}
\v{D}(\v{x-x_0;e})=-\frac{1}{2}\nabla_0^2 \v{G}(\v{x-x_0;e})\label{DQrelation},
\end{gather}
and its derivatives. A notable combination of the above singularities has been named alternately the couplet or rotlet,
\begin{align}
\v{R}(\v{x-x_0};\v{e})&=\frac{1}{2}\left(\v{G_D}(\v{x-x_0};\v{e^{\perp}},\v{e^{\perp\perp}})-\v{G_D}(\v{x-x_0};\v{e^{\perp\perp}},\v{e^{\perp}})\right)\nonumber \\
&=\frac{\v{e}\times \v{(x-x_0)}}{R^3},
\end{align}
where the vectors $\v{e,e^\perp,e^{\perp\perp}}$ form an orthonormal basis with $\v{e\times e^\perp=e^{\perp\perp}}$. A rotlet dipole may then be written simply as
\begin{gather}
\v{R_D}(\v{x-x_0};\v{d},\v{e})=\v{d}\cdot \nabla_0\v{R}(\v{x-x_0};\v{e}).
\end{gather}
Finally, a combination of the above singularities that we will require later is termed the Stresslet, which may be written as
\begin{gather}
\v{T(x-x_0;d,e)}=2(\v{d\cdot e})\v{U}(\v{x-x_0})+\left(\v{G_D(x-x_0;d,e)}+\v{G_D(x-x_0;e,d)}\right)\label{Stresslet}\\
=-\frac{6(\v{x-x_0})}{R^5}\left[\v{d}\cdot(\v{x-x_0})\right]\left[\v{e}\cdot(\v{x-x_0})\right],
\end{gather}
representing the fluid force on a plane with normal $\v{d}$ corresponding to the $\v{e}$-directed Stokeslet velocity field.

\subsection{Far-field description of a swimming body}

In the present study, we will consider only microorganisms which are to a good approximation axisymmetric along a direction indicated by the unit-vector $\v{e}$. Such an organism, with its centroid at a point $\v{x_0}$, generates fluid motion in the far-field of the form
\begin{gather}
\v{u}(\v{x})=\alpha \,\v{G_D}(\v{e,e})+\beta\,\v{D}(\v{e})+\gamma\, \v{G_Q}(\v{e,e,e})+\tau\,  \v{R_D}(\v{e,e})+O\left(|\v{x-x_0}|^{-4}\right).\label{Farfieldu}
\end{gather}
Here we have introduced the shorthand notation $\v{G_D}(\v{x-x_0;e,e})=\v{G_D}(\v{e,e})$. The coefficient $\alpha$ has units of $[Velocity]\times[Length]^2$, while $\beta,\gamma$ and $\tau$ have units of $[Velocity]\times[Length]^3$. The values of the coefficients $\alpha,\beta,\gamma$, and $\tau$ must be determined for each microorganism, and depend on the specific body geometry and propulsive mechanism. 

An illustration of the singularity decomposition above is provided in Fig. 1. At leading order, a swimming {\it E. coli} organism can be modeled as a force dipole (decaying as $1/R^2$). This leading order representation has been used by many authors to consider the effects of nearby walls \citep{btbl08}, and many-swimmer interaction dynamics \citep[see for instance][]{hosg05,ss08,hs10}. A swimmer such as the one illustrated in Fig. 1, in which a flagellar propeller pushes a load through the fluid, is generally referred to as a {\it pusher}, in contrast to such organisms as {\it Chlamydomonas} which {\it pulls} a cell body through the fluid with a pair of flagella. At the next order (decaying as $1/R^3$), the flow in the far-field varies due to the length asymmetry between the backward-pushing propeller and the forward-pushing cell body (producing a Stokeslet quadrupole), due to the finite size of the cell body (producing a source dipole), and due to the rotation of the flagellum and counter-rotation of the cell body (producing a rotlet dipole). Vector field cross-sections of the Stokeslet dipole, source dipole, and Stokeslet quadrupole singularities are shown in Fig.~\ref{Figure2}. The strengths of these singularities have been measured experimentally for the organisms {\it Volvox carteri} and {\it Chlamydomonas reinhardtii} by \cite{dgmpt10}, and for {\it E. coli} by \cite{ddcgg11}. The effects of the Stokeslet quadrupole component of spermatozoan swimming has been suggested by \cite{sb09}, and force-quadrupole hydrodynamic interactions of {\it E. coli} have been studied by \cite{lsdkw07}.

\begin{figure}
\vspace{.15in}
\begin{center}
\includegraphics[width=3in]{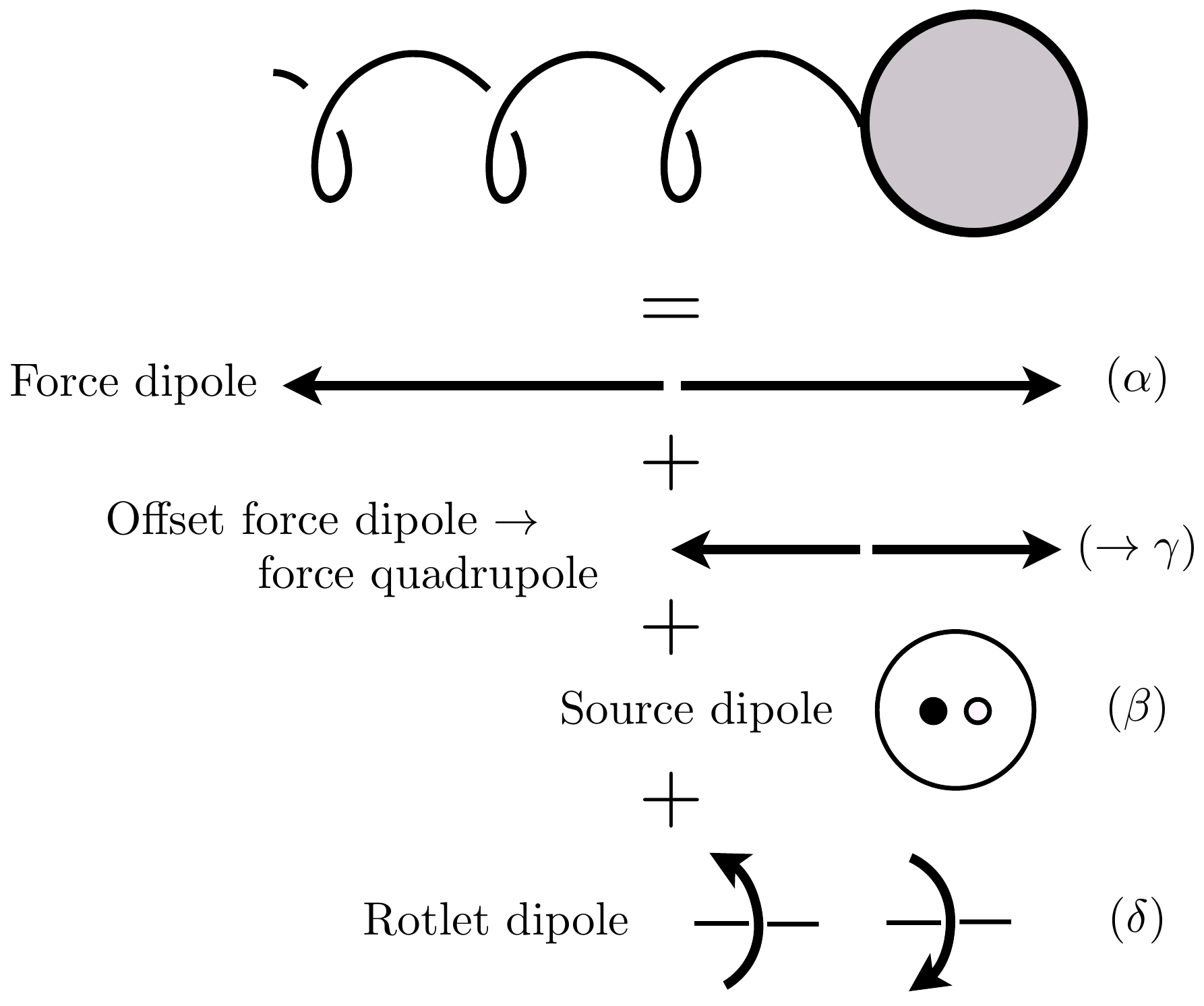}
\caption{The fluid velocity far from a swimming {\it E. coli} is modeled at leading order as that of a Stokeslet dipole. At the next order, the flow in the far-field varies due to the length asymmetry between the backward-pushing propeller and the forward-pushing cell body (producing a Stokeslet quadrupole), the finite size of the cell body (producing a source dipole), and the rotation of the flagellum and counter-rotation of the cell body (producing a rotlet dipole). }
\label{Figure1}
\end{center}
\end{figure}

While the flow field is set up instantaneously in Stokes flow upon the variation of an organism's geometry, the means of propulsion of a particular organism might be unsteady. In general the singularity strengths can be time-dependent, varying for example with the different phases of a swimming stroke pattern. As an example, the highly time dependent flow field generated by the oscillating motions of \textit{C. reinhardtii} has been examined by \cite{gjg10}. Nevertheless, for a first broad look at the far-field representation above we will restrict our attention to constant values of the singularity strengths for the remainder of our study. Also, we have assumed in the description given by \eqref{Farfieldu} that there are no net body forces or torques on the organism, which would require the inclusion of Stokeslet and rotlet singularity terms as well \citep[as explored for the organism {\it Volvox} by][]{dltipg09}. While some organisms are not neutrally buoyant and do experience a body force or torque due to gravity, many others (including most bacteria) live on such a scale that such effects are negligible. In addition, we assume that there is no mass flux through such mechanisms as fluid extrusion, as studied by \cite{sl10}, which can present a source singularity in addition to those included in the expression above.

\begin{figure}
\vspace{.15in}
\begin{center}
\includegraphics[width=5.3in]{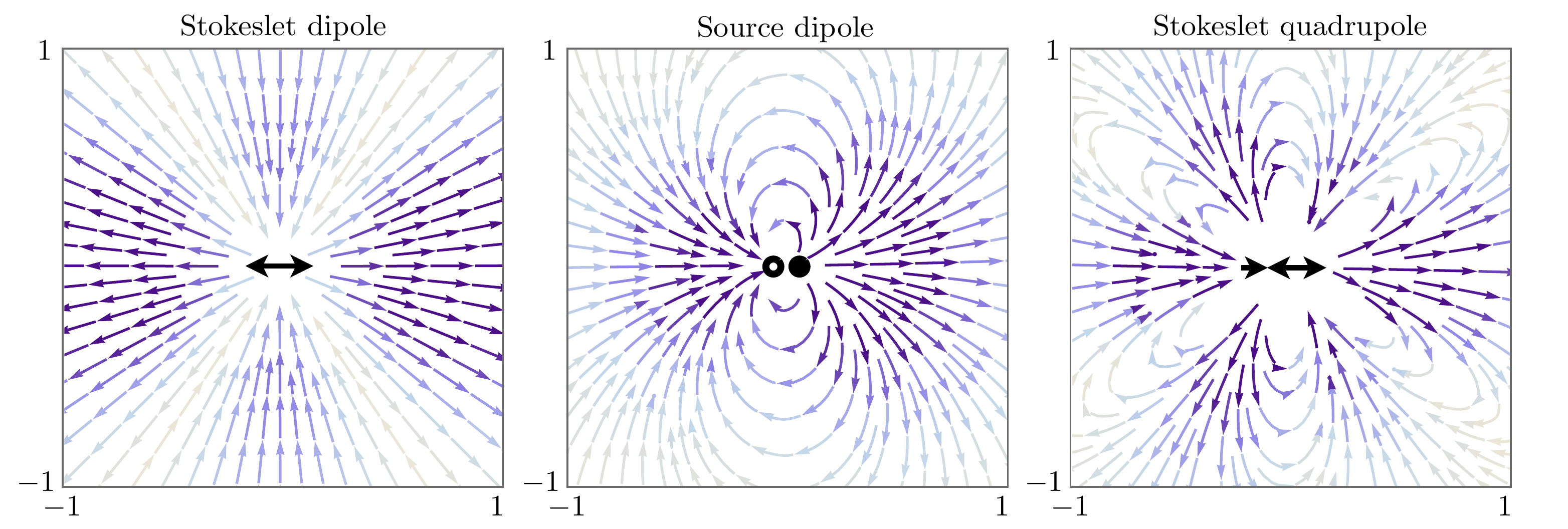}
\caption{Velocity field cross-sections of (a) a Stokeslet (force) dipole, which decays as $1/R^2$; (b) a source dipole, which decays as $1/R^3$; and (c) a force quadrupole, which decays as $1/R^3$, all in free-space. Arrow intensity correlates with the magnitude of the velocity. The effects of a nearby boundary may be intuited by imagining the wall to follow the streamlines.}
\label{Figure2}
\end{center}
\end{figure}

\subsection{The surface effect: Fax\'en's Law}

In a fluid of infinite extent, the fluid velocity in the far-field generated by an active body behaves as described in \eqref{Farfieldu}. When a boundary such as a plane wall is present, however, the velocity everywhere is altered due to the additional boundary condition. Borrowing an approach which has seen a long history in electrodynamics, the boundary condition on the surface can be satisfied by the placement of additional singularities at the image point $\v{x_0^*}=\v{x_0}-2(\v{x_0\cdot \hat{z}})\v{\hat{z}}$ inside the wall {(where $\v{\hat{z}}$ is the unit vector normal to the surface)}. 

The image singularities required to cancel the effects of Stokeslet singularities placed parallel or perpendicular to a no-slip wall have been presented by \cite{bc74}, each requiring a Stokeslet, Stokes doublet, and source dipole, as described in Appendix B. The image system for a ``tilted'' Stokeslet (a Stokeslet directed at an angle relative to the wall) is simply a linear combination of the wall-parallel and wall-perpendicular image systems. The image systems for higher order singularities, however, are not simply linear combinations of wall-parallel and wall-perpendicular image systems. The images for each of the axisymmetric singularities in \eqref{Farfieldu} will be denoted by an asterisk (see Appendix B). For instance, the effect of a Stokeslet dipole along with its image, evaluated on the wall surface $z=0$, returns $\v{G_D}(z=0;\v{e,e})+\v{G_D^*}(z=0;\v{e,e})=0$. Likewise, we denote by $\v{u^*}(\v{x})$ the fluid velocity generated by the entire collection of image singularities needed to cancel, on the no-slip wall, the swimmer-generated velocity description in \eqref{Farfieldu}. \cite{Magnaudet03} have recently taken a similar approach to studying the deformation and migration of a drop moving near a surface, and have provided a valuable review of Fax\'en's technique.

The flow generated by the image singularities indicates the alteration to the fluid motion everywhere due to the presence of the wall. The effects of this induced fluid motion on the swimming trajectory are provided by Fax\'en's Law, which can be written exactly for a prolate ellipsoidal body geometry \citep{kk91}. For an ellipsoid of major and minor axis lengths $2a$ and $2b$, with the major axis aligned with the unit vector $\v{e}$, the translational velocity $\v{\tilde{u}}$ and rotational velocity $\v{\tilde{\Omega}}$ induced on the swimmer due to its experience of the flow $\v{u^*}(\v{x})$ may be written as
\begin{gather}
\v{\tilde{u}}=\v{u}^*(\v{x_0})+O\left(a^2\nabla^2\v{u}^*|_{\v{x_0}}\right),\label{FaxenU}\\
\v{\tilde{\Omega}}=\frac{1}{2}\nabla \times \v{u}^*(\v{x_0})+\Gamma \v{e}\times \left(\t{E^*}(\v{x_0})\cdot \v{e}\right)+O\left(a^2\nabla^2\left(\nabla \times \v{u}^* \right)|_{\v{x_0}}\right),\label{FaxenO}
\end{gather}
where $\v{x_0}$ is the position of the body centroid, $\t{E^*(x)}=(\nabla \v{u^*}+[\nabla \v{u^*}]^T)/2$, and $\Gamma=(1-e^2)/(1+e^2)$, with $e=b/a$ the body aspect-ratio. Denoting the swimming speed attained by the organism in free-space by $U$, we have therefore that the body swims with velocity $U\v{e}+\v{\tilde{u}}$ and changes swimming direction via $\v{\dot{e}}=\v{\tilde{\Omega}}\times \v{e}$. 

At the order of our consideration in this paper (via \ref{Farfieldu}) the strength of the singularities representing the body motion are not changed by the presence of the wall. If the singularities required to represent the motion differed in rate of decay by more than one degree of separation (we currently include only terms decaying at order $R^{-2}$ and $R^{-3}$), then Fax\'en's Law above would indicate a problematic interaction of the wall effect with the measurement of the singularity strengths. The approach above, then, must be handled with more care in the event that a Stokeslet singularity is required, or if higher order terms than those considered here are to be included in \eqref{Farfieldu}. So long as the distance of the body to the wall is sufficiently large relative to the body size, the higher order terms in \eqref{FaxenU}-\eqref{FaxenO} may be neglected. 

The expressions above do not extend easily to geometries that are not ellipsoidal. In order to study a body geometry more like that illustrated in Fig. 1 we will need to develop an approximate ``Fax\'en Law.'' First, however, {let us} consider the consequences of singularity images on a prolate ellipsoidal body to develop some intuition.

\section{The surface effect, singularity by singularity}

\begin{figure}
\vspace{.25in}
\begin{center}
\includegraphics[width=2.3in]{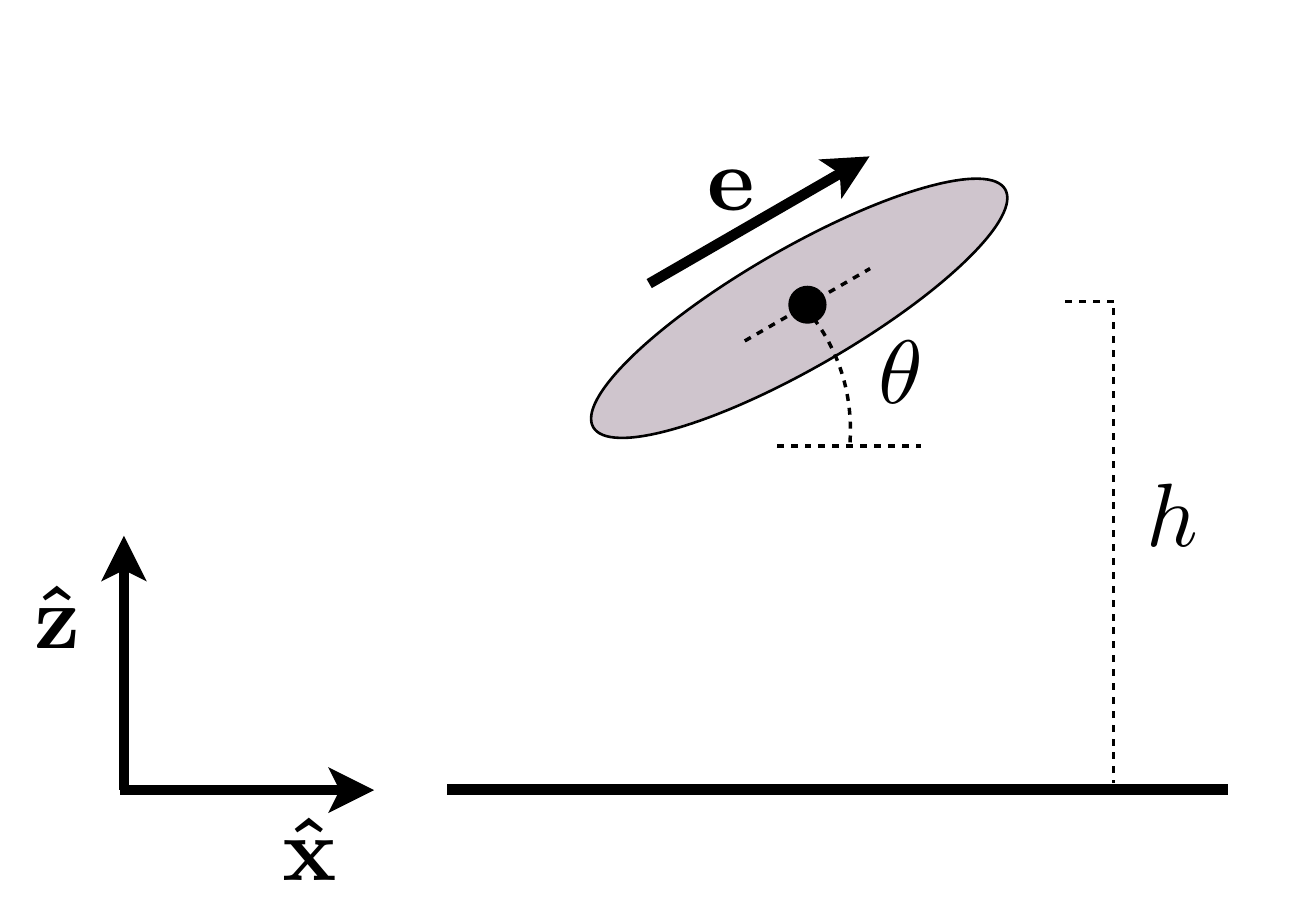}
\caption{Schematic representation of the generic problem studied in this paper: an inclined swimmer near a solid surface. The distance of the body centroid from the wall is denoted by $h$, measured along the direction normal to the wall $\v{\hat{z}}$. The pitch angle {of the body's director $\v{e}$} with respect to the wall is denoted by $\theta$; the body is swimming directly away from the wall when $\theta=\pi/2$, directly towards the wall when $\theta=-\pi/2$, and parallel to the wall when $\theta=0$.}
\label{Figure3}
\end{center}
\end{figure}

A swimming body first begins to experience the hydrodynamic effects of a wall through the singularities which decay least rapidly in space. We now list each singularity in the multipole representation and describe the corresponding effect on the swimming speed and orientation of the body. First, however, the system is made dimensionless by scaling velocities upon the free-space swimming speed, $U$, lengths upon the body semi-major axis length $a$, and forces upon $\mu a U$. Henceforth all variables are understood to be dimensionless. The unit vector $\v{\hat{z}}$ is normal to the wall and points into the fluid, and the dimensionless distance between the wall and the body centroid is denoted by $h$ (see Fig.~\ref{Figure3}). The pitch angle with respect to the wall is denoted by $\theta$; the body is swimming directly away from the wall when $\theta=\pi/2$, directly towards the wall when $\theta=-\pi/2$, and parallel to the wall when $\theta=0$.

\subsection{Force dipole}

Since we will not consider external body forces or torques, the least rapidly decaying singularity (in space) generated by the activity of an organism is a Stokeslet dipole directed along the swimming direction, $\v{G_D}(\v{e,e})$, which induces a dimensionless attraction to the wall (or repulsion from the wall) by \eqref{FaxenU} of the form
\begin{gather}
\v{\hat{z}}\cdot \v{\tilde{u}}=\tilde{u}_z=-\frac{3\alpha}{8h^2}\left(1-3\sin^2\theta \right)\label{dipoleuz}.
\end{gather}
Further details are provided in Appendix B, along with the effects induced not by a wall but instead by a stress-free surface such as a fluid-air interface. From \eqref{dipoleuz}, the surface-induced velocity of a pusher ($\alpha>0$) is towards the wall when $|\sin(\theta)|<1/\sqrt{3}$. For small orientation angles $\theta\sim0$ (when the body swims almost parallel to the wall, and $\sin(\theta)\sim \theta$), combining the wall effect above with the vertical component of the free-space swimming speed, $\sin(\theta)$, we see that the body will move towards the wall when $\theta<3\alpha/(8 h^2)$. Hence, for a pusher that is swimming nearly parallel to the wall, the first effect of the hydrodynamic interaction with the wall is an attraction. It has been argued by \cite{btbl08} that this hydrodynamic consideration can account for observations of the entrapment of {\it E. coli} near surfaces, as well as the observations of \cite{Rothschild63}, who measured the distribution of bull spermatozoa swimming between two glass plates and found the cell density to increase near the walls. Experimental measurements of force dipole strengths generated by swimming {\it E. coli} were evaluated by \cite{ddcgg11}, who found rotational diffusion to dominate hydrodynamic effects in that particular regime.

Now, what of the body orientation dynamics? The pitch angle $\theta$ is assumed to be constant in time absent the presence of a wall, so the variation in $\theta$ is due only to wall-induced rotational effects. The leading order effect is again that of the slowly decaying Stokeslet dipole term, which generates the rotation rate
\begin{gather}
\dot{\theta}=-\frac{3\alpha}{8h^3}\left(1+\frac{\Gamma}{2}\right)\theta+O(\theta^2),
\end{gather}
where the approximation given is appropriate when the body is nearly parallel to the wall (see Appendix B for the full expression). Hence, for $\theta\sim 0$ the induced rotation acts to align the body with the wall for $\alpha>0$ (pushers), and perpendicular to the wall for $\alpha<0$ (pullers), with no qualitative dependence upon the aspect-ratio of the body. The rotation induced by the force dipole may be intuited based on a consideration of the velocity fields shown in Fig.~\ref{Figure2}a; imagining a wall to follow the streamlines, the body is seen to be drawn into the wall, and based on velocity gradients to rotate towards $\theta=0$. The nature of the wall effect is even more predictable for the source dipole and force quadrupole from a similar consideration of Figs.~\ref{Figure2}b-c, as we will show. As we now proceed to consider the next order of singularities, we shall find that the leading order wall effects described above can be rather deceptive if they are used to predict the full trajectory of a given swimmer.

\subsection{Source dipole}
As a swimming body comes into closer contact with a wall, or when the body is swimming parallel to the wall, higher order singularities will begin to affect the  trajectory of the swimming organism. The source dipole singularity, which enters due to the presence of a cell body of finite size (such as the spherical head in Fig. 1) contributes an induced attraction/repulsion relative to the wall of the form
\begin{gather}
\tilde{u}_z=-\beta\frac{\sin\theta}{h^3}\cdot
\end{gather}
For a swimmer such as that shown in Fig. 1 it is common to have $\beta<0$, since motion of an inert sphere through a viscous fluid can be represented by a Stokeslet singularity and a source dipole with $\beta<0$ placed at its center \citep{kk91}. In this case, the source dipole term contributes a wall repulsion when the body is pitched away from the wall (``nose up''), and contributes an attraction to the wall when the body is pitched down (``nose down''). 

When $\theta=0$, the Stokeslet dipole term no longer determines the rate of rotation; instead, it is set by the higher order singularity structure. The rotational velocity induced (for $\theta\sim0$) by the source dipole is
\begin{gather}
\dot{\theta}\sim \frac{3\beta}{8h^4}\left(1+\frac{3\Gamma}{2}\right)\label{tdotbeta},
\end{gather}
which acts to rotate the nose downward towards the wall if $\beta<0$. In general, these effects will compete with those generated by the Stokeslet dipole and quadrupole to set the  trajectory and equilibrium states of the self-propelled body.

If instead a body swims by activity on its surface (as is the case for ciliated organisms, often modeled as so-called {\it squirmers}), we can find $\beta>0$, as will be shown in the following section. The noted effects above are thus reversed for such an organism, which we will explore in greater detail in \S V. 

\subsection{Force quadrupole}

At the same order of decay as the source dipole, the Stokeslet quadrupole enters and induces a wall-perpendicular velocity of the form
\begin{gather}
\tilde{u}_z= \gamma\frac{\sin(\theta)}{4 h^3}\left(7-9\sin^2\theta\right),
\end{gather}
and contributes an induced rotation rate of 
\begin{gather}
\dot{\theta}=\frac{-3\gamma}{8 h^4}\left(1+\frac{11\Gamma}{4}\right)\label{quadrupolerot},
\end{gather}
for $\theta\sim0$. The attraction/repulsion and induced rotation rate depend on the sign of $\gamma$, which itself depends upon the fore-aft body asymmetry, indicated in Fig. 1. From studying swimmers with exact singularity expressions (to be described below), we expect $\gamma<0$ for such swimmers as shown in Fig. 1 with large cell bodies and short flagella, and $\gamma>0$ for those with small cell bodies and long flagella. Like the source dipole, this singularity also acts to rotate the swimmer when $\theta=0$.

\subsection{Rotlet dipole}

The rotlet dipole term can account for at least one surprising behavior of locomotion near surfaces, the circular swimming trajectories of {\it E. coli} as studied by \cite{ldlws06}. For a body swimming parallel to the wall, $\theta=0$, the rotation about the $\v{\hat{z}}$ axis is given by
\begin{gather}
\v{\hat{z}}\cdot \v{\tilde{\Omega}}=-\frac{3\tau}{32 h^4}\left(1-\Gamma\right) \label{RotDipTraj},
\end{gather}
the effect disappearing for infinitely slender swimmers (for fixed $\tau$). Fixing the distance to the wall $h$, the body is thus predicted to swim in circles with a (dimensionless) radius
\begin{gather}
R_\tau=\frac{32\,h^4}{3|\tau|(1-\Gamma)}\cdot 
\end{gather}
The cell bodies of {\it E. coli} bacteria rotate clockwise as seen from the distal end (behind the organisms) during their forward swimming runs, and the net torque is balanced by the counter-clockwise rotation of the propelling flagellar bundle. This situation corresponds to $\tau>0$, and hence \eqref{RotDipTraj} predicts a large clockwise circular trajectory (as seen from above) parallel to the plane of the wall, which is consistent with the experimental observations of \cite{ldlws06}.

Note that the same organism moving near a stress-free boundary (like an air-water interface) experiences a passive rotation in the opposite direction (see Appendix B), as studied by \cite{dldaai11}. While the rotlet dipole contributes to three dimensional swimming dynamics, this component of the propulsion has no bearing on the wall-attraction/repulsion or pitching dynamics of a swimmer: $\v{\tilde{u}}=0$ and $\dot{\theta}=0$. This is assured by the kinematic reversibility of Stokes flow. We will focus here on wall-attraction/repulsion and pitching dynamics, and thus for the remainder of our consideration we will set $\tau=0$ in \eqref{Farfieldu}.

\section{Where is the multipole singularity representation valid?}

The central question that we wish to answer in this paper is: how accurately are the wall effects predicted by the multipole singularity representation \eqref{Farfieldu} and accompanying Fax\'en Law \eqref{FaxenU}-\eqref{FaxenO}? In order to provide an answer, we must have at our disposal a means of computing the full fluid-body interaction. For this purpose we will utilize the method of images with regularized Stokeslets, as derived by \cite{adebc08}, in which the boundary integral formulation of the Stokes equations is accompanied by image singularity kernels which cancel the fluid velocity on the wall. In this approach the wall need not be discretized. We will make a necessary adjustment to this framework to allow for the inclusion of a slip velocity on the body surface. Equipped with a means of computing solutions to the full Stokes equations in the half-space bounded by a no-slip wall, we can study the accuracy of the multipole representation and wall effects for a selection of model swimmers.

\subsection{Model swimmers}

\begin{figure}
\vspace{.15in}
\begin{center}
\includegraphics[width=5in]{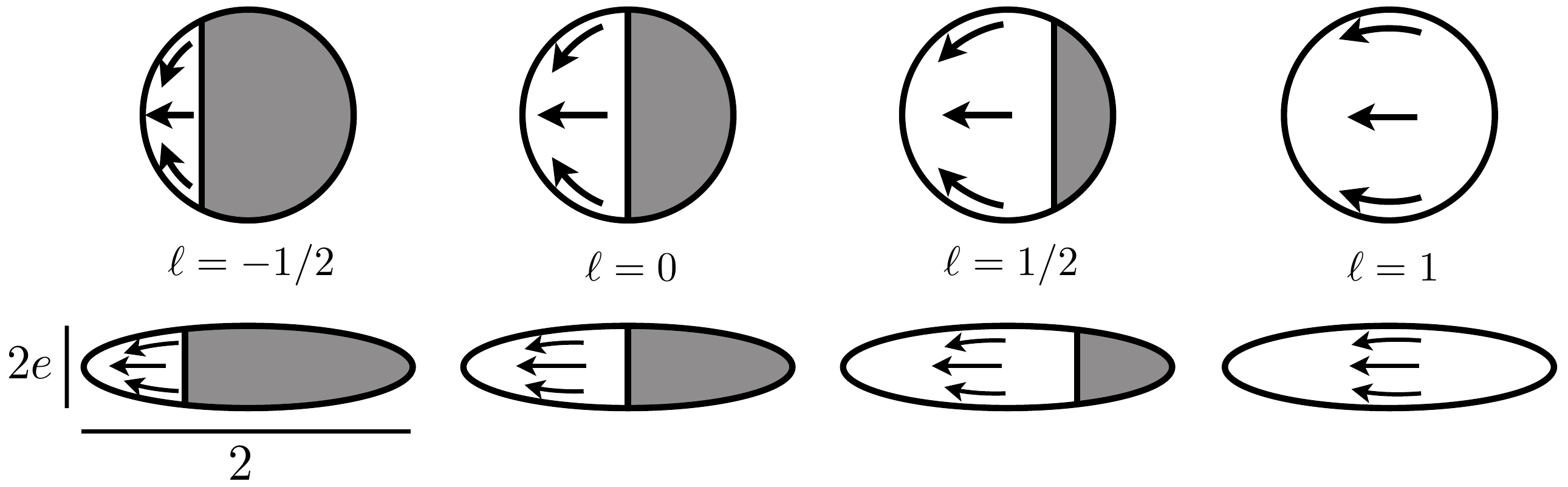}
\caption{A selection of model swimmers are illustrated. Spheres ($e=1$) and ellipsoids ($e=1/4$) are shown with activity lengths $\ell=-1/2,0,1/2$, and 1. Each body shown shuttles fluid along the active part of the surface to the left (the propulsive activity is indicated by arrows), and thus swims to the right. The gray regions are inert, where a no-slip condition is assumed.}
\label{Figure4}
\end{center}
\end{figure}

In order to probe the accuracy of the far-field representation we will enlist the help of model ``Janus swimmers,'' as illustrated in Fig.~\ref{Figure4}. The model swimmers are prolate ellipsoids, chosen so that the Fax\'en expressions \eqref{FaxenU}-\eqref{FaxenO} may be applied exactly. For a given pitch angle $\theta$, the surface is parameterized as $\v{x}=\v{x_0}+\cos(\xi)\v{e}+e\sin(\xi)[\cos(\eta)\v{e^\perp}+\sin(\eta)\v{e^{\perp\perp}}]$, with $\xi\in[0,\pi]$ and $\eta\in[0,2\pi]$, and where $\v{e,e^\perp,e^{\perp\perp}}$ form an orthonormal basis with $\v{e\times e^\perp=e^{\perp\perp}}$. Recall that $e$ is the body aspect-ratio. The unit tangent vector everywhere on the surface is denoted by $\v{s}(\xi,\eta)$, and the unit normal is denoted by $\v{n}(\xi,\eta)$.

The propulsive mechanism is a prescribed axisymmetric distribution of a slip velocity $\v{u_s}$ which acts tangentially to the body surface for $-1 \leq \cos(\xi) \leq \ell$, where $\ell \in[-1,1]$ is a dimensionless ``activity length,'' while the remainder of the body surface is inert (where a no-slip condition is applied). The entire body surface is active for $\ell=1$ (a squirmer). More specifically, the prescribed slip velocity distribution is chosen to be
\begin{gather}
\v{u_s}(\xi,\eta)=u_s^0\sqrt{\frac{2(1+\cos\xi)(\ell-\cos\xi)}{(1+e^2)-(1-e^2)\cos(2\xi)}}\,\v{s}(\xi,\eta).\label{slipus}
\end{gather}
The constant $u_s^0$ is selected so that the free-space (no wall) swimming velocity is unity, and is determined numerically. A no-penetration condition is applied on the entire body surface. 

The squirmer model of ciliated organisms was introduced by \cite{Lighthill52} and extended by \cite{Blake71}. Squirmer models (where either the slip velocity or the surface stress is specified) have been used to study multiple-organism interactions by \cite{isp06} and \cite{kst10}, hydrodynamically bound states by \cite{dltipg09}, efficiency optimization in ciliary beating by \cite{ml10}, fluid stirring effects by \cite{ltc11}, and motion in a polymeric fluid by \cite{zdqlb11}. {Swimmers with partially activated surfaces ($\ell<1$) have recently been designed and studied with great enthusiasm; see for instance the work of \cite{pkossacmlc04}, \cite{gla07} and \cite{jys10}, where the activity is generated by self-phoretic and thermophoretic surface effects.} Migration of similar `slip-stick' spheres in an ambient flow has been studied by \cite{sk08}.

\subsection{Full numerical simulation approach}

An application of Green's theorem to the Stokes equations \eqref{E:Stokes} reveals a representation of the fluid velocity everywhere based solely on integrations of the stress and velocity on the immersed boundaries (the swimming body, in this case) \citep{poz}. Accounting for the presence of the wall by including image singularities, the fluid velocity everywhere may be written as  
\begin{gather}
\v{u}(\v{x})=\mathcal{K}[\v{f}](\v{x})+\mathcal{K}^*[\v{f}](\v{x})-\left(\mathcal{T}[\v{u}](\v{x})+\mathcal{T}^*[\v{u}](\v{x})\right),\label{BoundaryIntegral}
\end{gather}
where
\begin{eqnarray}
\mathcal{K}[\v{f}](\v{x})&=&\frac{1}{8\pi}\int_{\partial D}\v{G}(\v{x-y};\v{f}(\v{y}))\,dS_y,\\\mathcal{K}^*[\v{f}](\v{x})&=&\frac{1}{8\pi}\int_{\partial D}\v{G}^*(\v{x-y};\v{f}(\v{y}))\,dS_y,\\
\mathcal{T}[\v{u}](\v{x})&=&\frac{1}{8\pi}\int_{\partial D} \v{T}(\v{x-y};\v{n(y)},\v{u(y)-u(x)})\,dS_y,\\
\mathcal{T^*}[\v{u}](\v{x})&=&\frac{1}{8\pi}\int_{\partial D} \v{T}^*(\v{x-y};\v{n(y)},\v{u(y)-u(x)})\,dS_y,
\end{eqnarray}
with $\v{f}$ the dimensionless fluid force per unit area, $\v{G}$ the Stokeslet singularity \eqref{Stokeslet1}, $\v{T}$ the Stresslet singularity \eqref{Stresslet}, and $dS_y$ the differential surface area element for the integration variable $\v{y}$. The image singularities $\v{G_D}^*$ and $\v{T}^*$ ensure that the no-slip condition is satisfied on the wall at $z=0$, and are provided in Appendix B. The single-layer integral $\mathcal{K}$ is weakly singular, which presents both theoretical and numerical difficulties. One approach to computing this integral is through the use of a regularized kernel, $\v{G}_\delta$, where a small regularization parameter $\delta$ is introduced. This is the approach of \cite{adebc08}, who derive the necessary adjustments which must be made to account for this regularization, and who also derive the image-singularities ($\mathcal{K}^*$) which must accompany such an approach when a wall is present. This is the approach taken in the present work, though we must include the double layer integral $\mathcal{T}$ and its image in order to accommodate the slip velocity $\v{u_s}$. 

We briefly recount the method of images with regularized Stokeslets as presented by \cite{adebc08} \citep[itself a modification to the method derived by][]{Cortez02}. The surface is discretized by $M$ points, located at $\v{x}_{k,0}$ for $k=1,2,...,M$. For a given point $\v{x}$ in the fluid or on the body or wall, we define $\v{x}_k^*=\v{x}-\v{x}_{k,0}$, and $\v{x}_k=\v{x}-(\v{x}_{k,0}-2(\v{\hat{z}}\cdot \v{x}_{k,0})\v{\hat{z}})$. Absorbing the surface integration into the force so that we may simply write $\int_{\partial D} \v{f}\,dS_y=\sum_k \v{f}_k$, and choosing a blob function of $\phi(\v{x})=15\e^4/8\pi [r^2+\e^2]^{7/2}$ (used to spread the singular effect of a point force to a small finite area), it may be shown that
\begin{gather}
\mathcal{K}[\v{f}](\v{x})=\sum_{k=1}^M\left[\v{f}_k H_1(|\v{x_k^*}|)+(\v{f_k\cdot x_k^*})\v{x_k^*}H_2(|\v{x_k^*}|)\right],\\
\mathcal{K}^*[\v{f}](\v{x})=-\left[\v{f}_k H_1(|\v{x_k}|)+(\v{f_k\cdot x_k})\v{x_k}H_2(|\v{x_k}|)\right]-h_k^2[\v{g}_kD_1(|\v{x_k}|)+(\v{g}_k\cdot \v{x}_k)\v{x}_kD_2(|\v{x}_k|)]\nonumber\\
-2 h_k\left[\frac{H_1'(|\v{x}_k|)}{|\v{x}_k|}+H_2(|\v{x}_k|) \right](\v{L}_k\times \v{x}_k)+2 h_k\Big[(\v{g}_k\cdot \v{\hat{z}})\v{x}_kH_2(|\v{x}_k|)+(\v{x}_k\cdot \v{\hat{z}})H_2(|\v{x}_k|)\nonumber\\
+(\v{g}_k\cdot \v{x}_k)\v{\hat{z}}\frac{H_1'(|\v{x}_k|)}{|\v{x}_k|}+(\v{x}_k\cdot \v{\hat{z}})(\v{g}_k\cdot \v{x}_k)\v{x}_k\frac{H_2'(|\v{x}_k|)}{|\v{x}_k|}\Big],\label{RegularizedStokeslets}
\end{gather}
where $\delta$ is a regularization parameter (discussed below), $\v{g}_k=2(\v{f}_k\cdot\v{\hat{z}})\v{\hat{z}}-\v{f}_k$, $\v{L}_k=\v{f}_k\times\v{\hat{z}}$, and
\begin{gather}
H_1(r)=\frac{1}{8\pi\mu(r^2+\delta^2)^{1/2}}+\frac{\delta^2}{8\pi\mu(r^2+\delta^2)^{3/2}},\,\,\,\,H_2(r)=\frac{1}{8\pi\mu(r^2+\delta^2)^{3/2}},\\
D_1(r)=\frac{1}{4\pi\mu(r^2+\delta^2)^{3/2}}-\frac{3\delta^2}{4\pi\mu(r^2+\delta^2)^{5/2}},\,\,\,\,D_2(r)=-\frac{3}{4\pi\mu(r^2+\delta^2)^{5/2}}\cdot
\end{gather}
Having subtracted off the velocity at the target point $\v{x}$, the integrand of the double layer integral $\mathcal{T}[\v{u}]$ is finite with a jump discontinuity at $\v{y=x}$ \citep{pm87,poz}. Insertion of rigid body motion velocities into the integrals returns zero, so only the tangential slip velocity $\v{u_s}$ need be considered. Since the integrands in $\mathcal{T}[\v{u}]$ and $\mathcal{T}^*[\v{u}]$ are known and finite, they are computed using adaptive quadrature to computer precision accuracy. 

Having absorbed the surface integration into the definition of $\v{f_k}$, the net (dimensionless) force on a boundary is computed simply as 
\begin{gather}
\v{F}=\int_{\partial D} \v{f}(\v{x})\,dS=\sum_{k=1}^M \v{f}_k,
\end{gather}
which must return zero in the case of self-propelled swimming (where body forces such as gravity have been neglected, as previously noted).

For a given body position and orientation, a linear system must be solved to determine the swimming velocity and rotation rate, $\v{U}$ and $\v{\Omega}$, along with the scaled force $\v{f}_k$, via the boundary integral relation in \eqref{BoundaryIntegral}. The linear system is closed by requiring that the boundary conditions hold as follows. Denoting the inert part of the body by $\partial D_I$ and the active part of the body by $\partial D_A$,
\begin{gather}
\v{u}(\v{x}\in\partial D_I)=\v{U}+\v{\Omega}\times(\v{x-x_0}),\\
\v{u}(\v{x}\in\partial D_A)=\v{U}+\v{\Omega}\times(\v{x-x_0})+\v{u_s}(\xi,\eta).
\end{gather}
The no-slip condition on the wall is satisfied automatically by the inclusion of the image kernels. Continuing to follow \cite{adebc08}, the ellipsoidal surface is discretized by dividing the azimuthal angle $\xi$ into $N_\xi+1$ points, $\xi_m=\pi m/N_\xi$ for $m=0,1,...,N_\xi$. At each station $\xi_m$, the polar angle is discretized into $N_m$ points, $\eta_n=2\pi n/N_m$ for $n=1,2,...,N_m$, where $N_m$ is the smallest integer larger than $2 N_\xi\sin(\xi_m)$. Taking as a representative discretization size $h_d=\sqrt{S/N'}$, with $N'$ the total number of gridpoints and $S$ the {(dimensionless)} ellipsoidal surface area, the regularization parameter chosen for all the problems considered herein is $\delta=0.8 h_d^{0.9}$, and we set $N_\xi=30$ to capture the swimming behavior with sufficient accuracy for our purpose.

\subsection{Computing the singularity strengths: $\alpha,\beta$, and $\gamma$}
In order to compare the full system with the singularity representation in \eqref{Farfieldu} we must determine the singularity strengths for the swimming bodies under consideration. Having recovered $\v{f(x)}$ numerically, these singularity strengths may be computed as follows. Assuming that the body is directed along $\v{e}=\v{\hat{x}}$ and centered about $\v{x_0}=\mathbf{0}$, for the sake of presentation, we have by an involved inspection (expanding \eqref{RegularizedStokeslets} for $|\v{x}|\gg |\mathbf{0}|$ in free-space, and matching the $x^3$ term for the dipole and the $x^4$ term and $(y^2+z^2)x^2$ terms for the higher order singularities):
\begin{align}
\alpha=\frac{1}{16\pi}\int_{\partial D}&(3x f_x-\v{x\cdot f})\,dS+\frac{3}{8\pi}\int_{\partial D}n_x u_x\,dS,\label{SingularityStrengths}\\
\beta=\frac{1}{16\pi}\int_{\partial D}&(x^2-|\v{x}|^2)f_x\,dS \nonumber\\
+&\frac{1}{8\pi}\int_{\partial D}2 x \,n_x u_x -(u_x\,\v{x\cdot n}+n_x\,\v{x\cdot u}-2x\,\v{n\cdot u})\,dS,\\
\gamma=\frac{1}{32\pi}\int_{\partial D}&(5x^2-|\v{x}|^2) f_x-2x\,(\v{x\cdot f})\,dS\nonumber\\
+&\frac{1}{8\pi}\int_{\partial D}5 x\, n_x u_x -(u_x\,\v{x\cdot n}+n_x\,\v{x\cdot u}+x\,\v{n\cdot u})\,dS.\label{SingularityStrengths3}
\end{align}
For purely rigid body motion, $\v{u}=\v{U+\Omega\times (x-x_0)}$, the integrals involving $\v{u}$ vanish, so we need only insert the slip velocity $\v{u_s}$ into these integral expressions. Note that the expressions above are geometry dependent; other expressions would need to be derived for non-ellipsoidal body shapes.

\begin{figure}
\vspace{.15in}
\begin{center}
\includegraphics[width=5.2in]{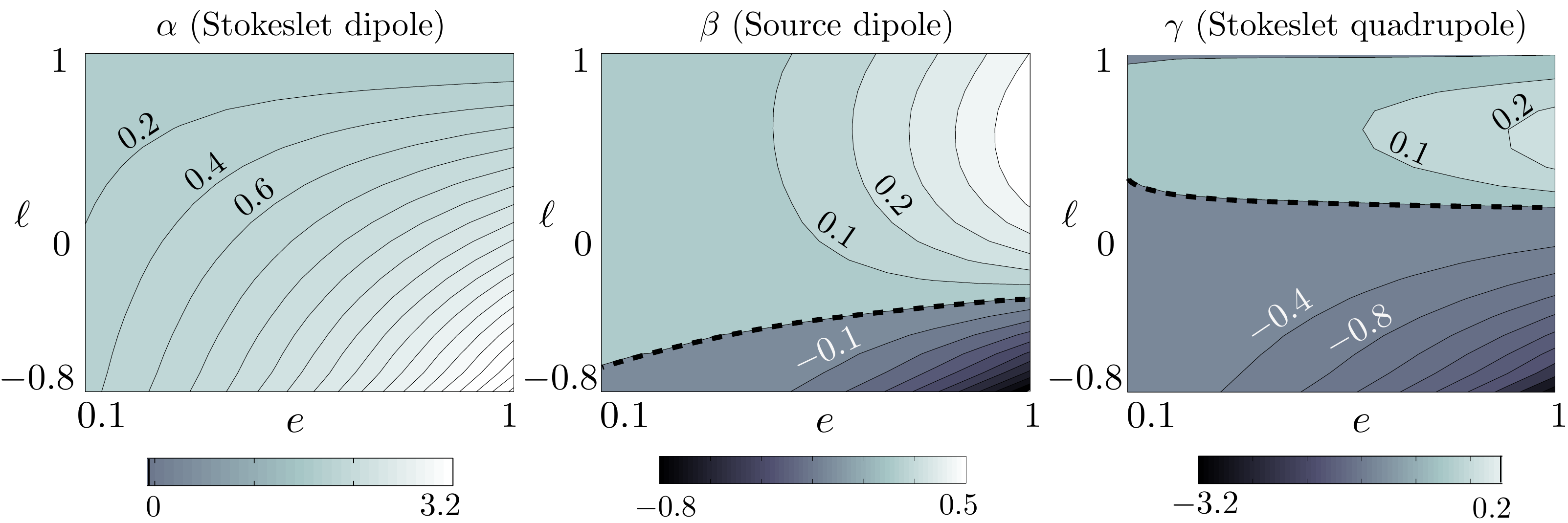}
\caption{The singularity strengths $\alpha,\beta$, and $\gamma$ are shown for the model swimmers illustrated in Fig.~\ref{Figure4}, but for a wide range of aspect-ratios $e$ and activity lengths $\ell$. The Stokeslet dipole strength, $\alpha$, is positive for all such swimmers (they are all {\it pushers}). The source dipole strength, $\beta$, and the Stokeslet quadrupole strength, $\gamma$, change sign where indicated by dashed lines.}
\label{Figure5}
\end{center}
\end{figure}

The singularity strengths $\alpha,\beta$, and $\gamma$ computed for a range of aspect-ratios $e$ and activity lengths $\ell$ are shown in Fig.~\ref{Figure5}. For all such model swimmers as shown in Fig.~\ref{Figure4}, we find that $\alpha>0$ (the swimmers are {\it pushers}). However, both $\beta$ and $\gamma$ change sign, as indicated by dashed lines. For a body that is primarily inactive ($\ell\approx -1$), the source dipole term is negative, $\beta<0$, which can be predicted by considering the exact singularity solution for a sedimenting solid ellipsoid as derived by \cite{cw75}. However, when the body is primarily active $(\ell\approx 1)$, we find $\beta>0$, which can be predicted from the exact solution for a squirming ellipsoid as derived by \cite{kw77}. The source dipole term decreases in magnitude with $e^2$, and also decreases in magnitude as the body becomes less geometrically active ($\ell\rightarrow -1$). Meanwhile, $\gamma$ is negative for bodies with small active surface areas $\ell \lesssim 0$ and positive for bodies with large active surfaces $\ell \gtrsim 0$.

As a simple example, consider a slender rod of dimensionless length $2$ which satisfies a no-slip condition for arc-lengths $s\in[\ell,1]$ and has specified active forcing $\v{e\cdot f}=-\mathcal{F}/(1+\ell)$ for arc-lengths $s\in[-1,\ell]$. A leading order approximation (in the small aspect-ratio of the body) of the force on the no-slip part of the body is simply $\mathcal{F}/(1-\ell)$. From \eqref{SingularityStrengths}-\eqref{SingularityStrengths3} we find that $\alpha=\mathcal{F}/(8\pi)$, $\beta=0$, and $\gamma= \ell\,\mathcal{F} /(24\pi)$. The swimmer is a pusher for $\mathcal{F}>0$, has no source dipole term due to its slenderness, and generates a Stokeslet quadrupole term which changes sign precisely when $\ell=0$, distinguishing bodies which are more active than inert or vice versa. The direction of rotation induced on such a swimmer by the presence of the wall thus depends here critically upon the asymmetry of the propulsive mechanism [through \eqref{quadrupolerot}], and not the body geometry, which is symmetric about its centroid. {Namely, when the swimmer is parallel to the wall ($\theta=0$) then $\dot{\theta}>0$ for $\mathcal{F}\ell<0$ and $\dot{\theta}<0$ for $\mathcal{F}\ell>0$. A long inert segment pushed by a short active segment $(\mathcal{F}>0$) leads to an upward pitching motion, while a short inert segment pushed by a long active segment leads to a downward pitching motion, with the situation reversed if the swimmer is a {\it puller} ($\mathcal{F}<0$).}

The singularity strengths for a squirming (potential flow, $\ell=1$) spheroid are even simpler, with $\alpha=\gamma=0$, and $\beta>0$. \cite{kw77} have shown that such a swimmer may be described in free-space exactly as an integration over source doublet singularities distributed along the body centerline. Specifically, writing
\begin{gather}
\v{u}(\v{x})=\frac{3\beta}{4c^3}\int_{-c}^c (c^2-s^2)\v{D}(\v{x_0}-s\,\v{e})\,ds=\beta\,\v{D}(\v{x_0})+O(h^{-5}),\\
\beta=\frac{2 e^2 c^3}{6 c-3 e^2 \log\left([1+c]/[1-c]\right)},\label{Squirmbeta}
\end{gather}
where $c=\sqrt{1-e^2}$ is the ellipsoidal focal length, then on the body boundary the velocity satisfies the rigid body motion (with speed unity) and the tangential slip velocity prescribed in \eqref{slipus} with $\ell=1$. For a spherical squirmer we have $\beta=1/2$, and $u_s^0=3/2$ in \eqref{slipus}, and the $O(h^{-5})$ term above is precisely zero. A squirming potential flow spheroid therefore always rotates away from the wall surface {[through \eqref{tdotbeta}]}, which we will explore in greater detail in \S V.

\subsection{How accurate is the far-field representation?}

\begin{figure}
\vspace{.15in}
\begin{center}
\includegraphics[width=5.15in]{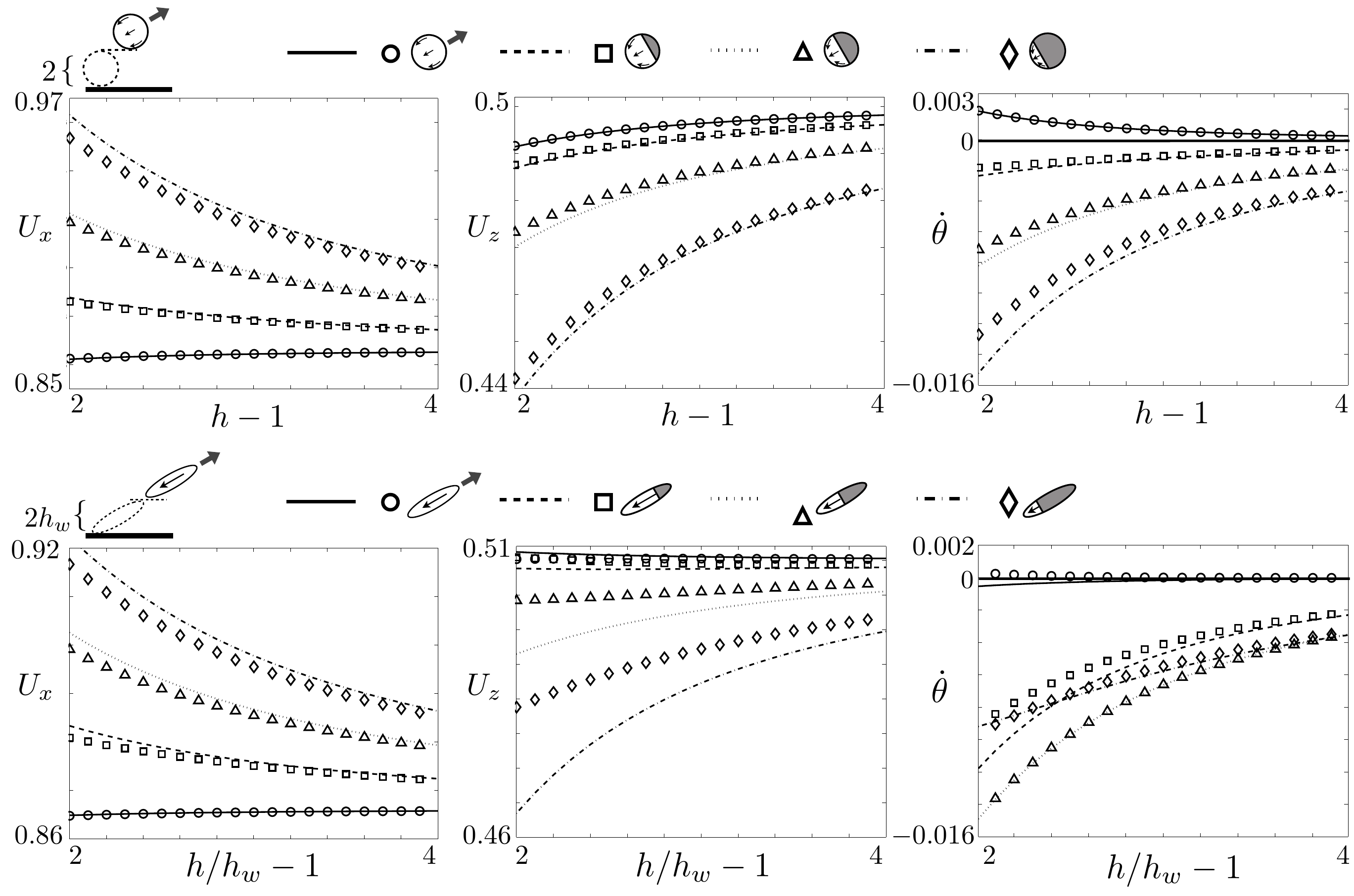}
\caption{The wall-parallel {(``horizontal,'' $U_x$)} and wall-perpendicular  {(``vertical,'' $U_z$)} velocities and rotational velocities ($\dot{\theta}$) of the eight model swimmers shown in Fig.~\ref{Figure4} are shown for a range of distances to the wall $h$, with computed values indicated by symbols and far-field predictions indicated by lines. The pitching angle is fixed to $\theta=\pi/6$.  {Top row: spherical swimmers. Bottom row: spheroidal swimmers.} The far-field approximation is seen to  reproduce the full simulation results, and generally remains valid for the range considered. The scaling $h_w$ is the smallest value of $h$ for which there is no wall contact (defined in the text).}
\label{Figure6}
\end{center}
\end{figure}

Given the computed singularity strengths for the model Janus swimmers we can compare the far-field representation of the swimming dynamics with the full simulation results. Figure~\ref{Figure6} shows the ``horizontal'' (wall-parallel, in the direction $\v{\hat{x}}$) and ``vertical'' (wall-perpendicular, in the direction $\v{\hat{z}}$) velocities and the rotational velocities of the eight model swimmers shown in Fig.~\ref{Figure4}, for a range of distances from the wall, and with pitching angle fixed at $\theta=\pi/6$. Here $h_w$ is the minimal distance of the centroid to the wall without penetration, $h_w=\sqrt{(1+e^2)-\left(1-e^2\right) \cos(2 \theta)}/\sqrt{2}$. The far-field approximation from \eqref{Farfieldu} generally matches the full simulation results with great accuracy in this range. Inspection of log-log plots (not shown here) affirm that the singularity strengths have been computed correctly, or alternatively verify the accuracy of the full numerical simulations. All the bodies with the exception of the squirmers (for which $\alpha=0$) swim horizontally with greater velocity due to the presence of the wall at this pitch angle. Also note that the spherical squirmer is the only swimmer of the eight that rotates away from the wall at this pitching angle; the rest all passively rotate towards the wall. 

Importantly, we find that the rotation rates are very small relative to the swimming speed, even when the swimmer is as close to the wall as one body length away. For swimming organisms directed towards the wall a collision may be inevitable. {Meanwhile,} when a body is swimming parallel to a wall, this small rotation and trajectory adjustment may be sufficient to entirely determine whether or not the swimmer collides with the wall and enters a hydrodynamically or otherwise bound state, or possibly swims away. 

\begin{figure}
\vspace{.15in}
\begin{center}
\includegraphics[width=5.15in]{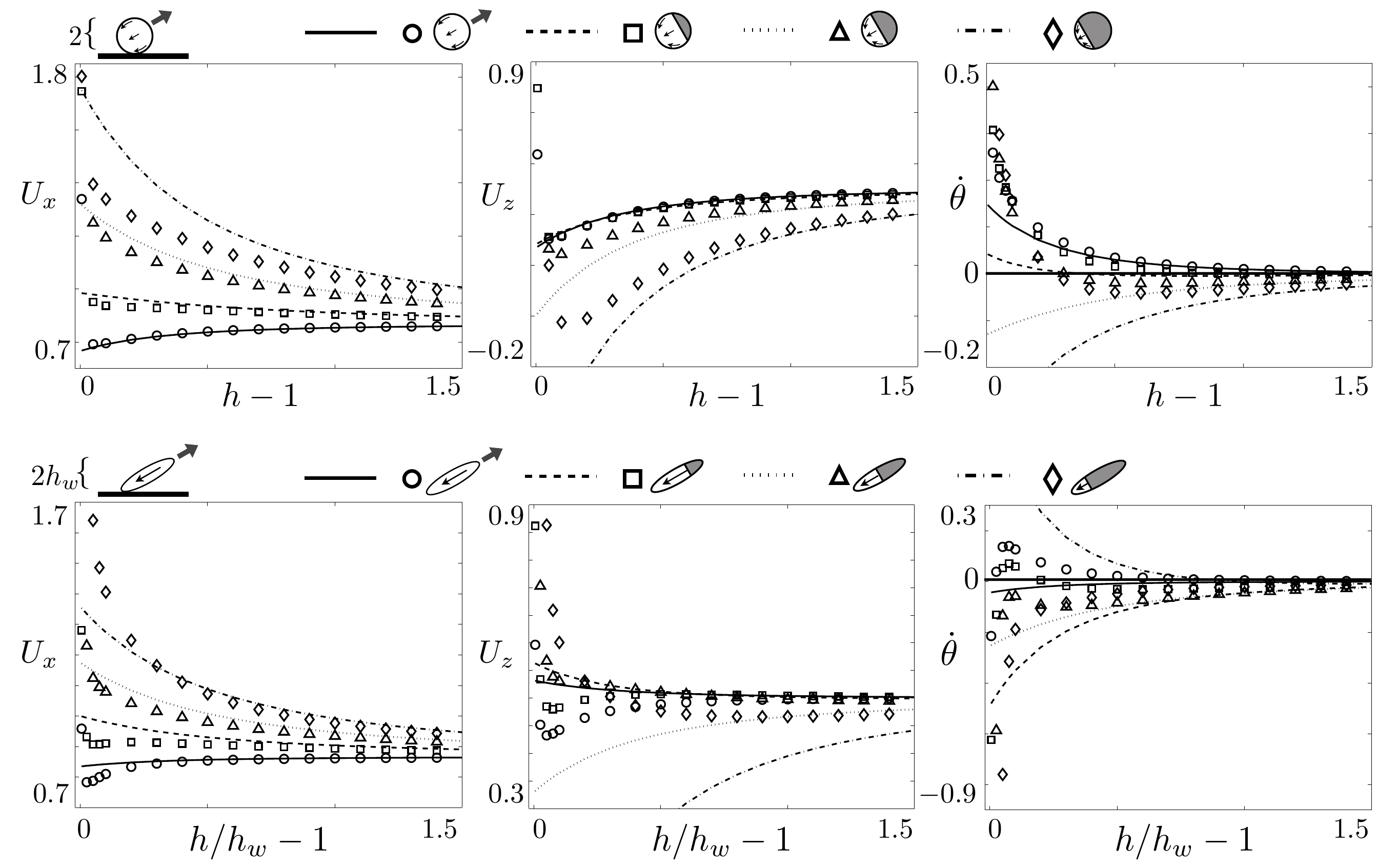}
\caption{The horizontal, vertical, and rotational velocities of the eight model swimmers shown in  Fig.~\ref{Figure4}, just as in Fig.~\ref{Figure6}, but for smaller distances from the wall. Computed values are indicated by symbols and far-field predictions by lines. The far-field approximation is surprisingly accurate, in some cases all the way down to one-tenth the body radius distance from the wall. Other velocities begin to vary not only quantitatively but qualitatively as the body approaches wall contact.}
\label{Figure7}
\end{center}
\end{figure}

How close to the wall can the far-field expressions be extended? Figure~\ref{Figure7} shows the same measurements but for a range of even smaller body distances from the wall. We find that the far-field approximation is surprisingly accurate, in some cases all the way down to a distance of one-tenth a body length from the wall. In particular, the horizontal swimming velocities of both the spherical and ellipsoidal swimmers is generally matched quite well, as are the vertical and rotational velocities of the spherical bodies. The behaviors of the ellipsoidal bodies are still captured qualitatively in this regime, but we begin to see a departure of the far-field approximation from the full simulation results. This is to be expected, as the far-field approximation does not capture effects such as lubrication, and higher order terms in the multipole expansion eventually become comparable to those kept in \eqref{Farfieldu}. For the swimmers considered here, however, the far-field approximations appear to be surprisingly accurate, and we expect that much can be predicted using the simple framework encapsulated by \eqref{Farfieldu}. This said, the predictions can become qualitatively inaccurate when the body is almost touching the wall, yielding non-physical results in some cases, so care clearly must be taken when attempting to apply the far-field theory to near-wall or near-swimmer interactions.

\section{Employment of the reduced model: Janus swimmers}

\subsection{Equilibrium pitching angles}

\begin{figure}
\vspace{.15in}
\begin{center}
\includegraphics[width=5.3in]{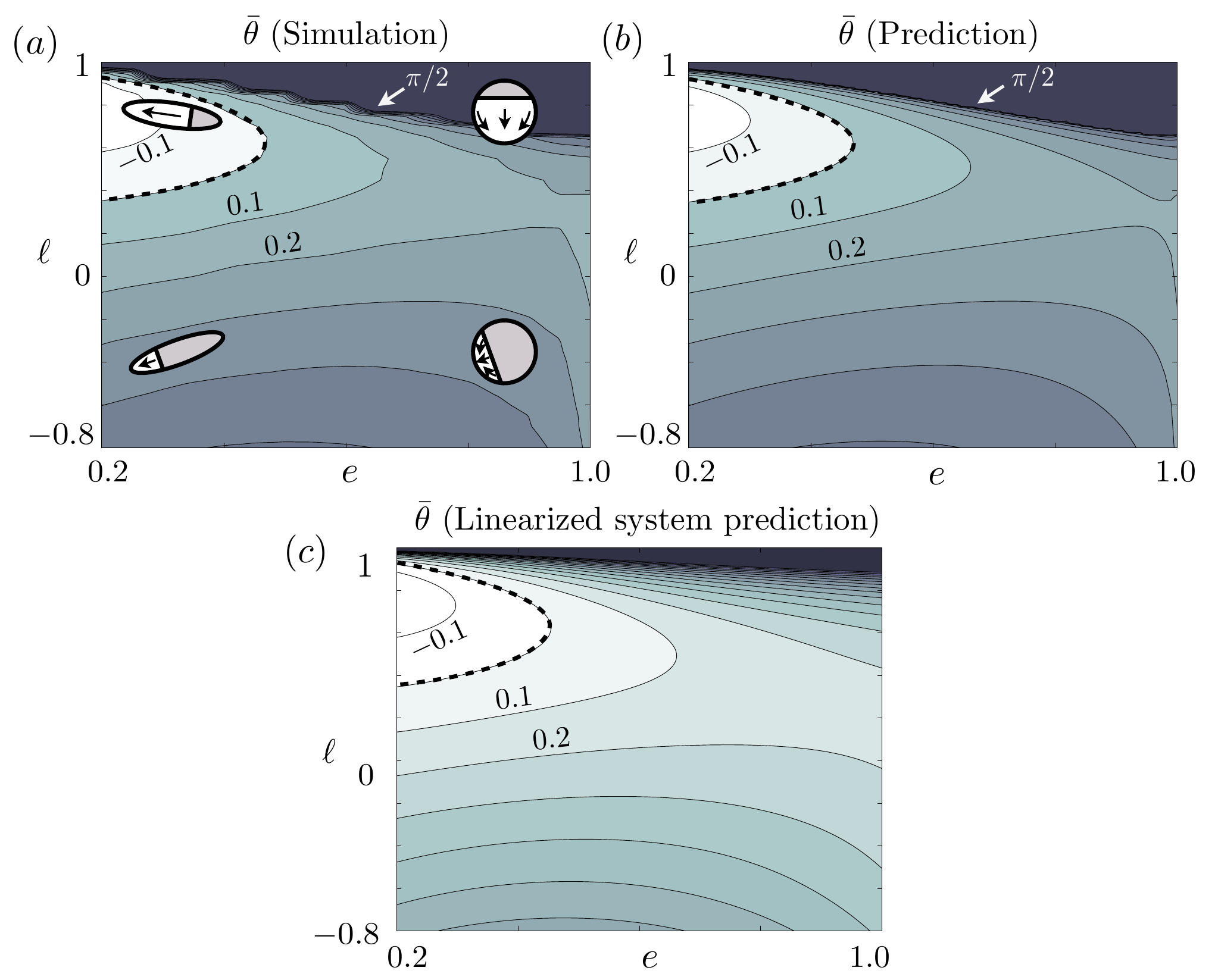}
\caption{Comparison of the full simulations to the far-field predictions for fixed distance $h=2$. (a) Computed contours of the stable equilibrium pitching angle at fixed distance $h=2$. (b) The same, as predicted with the far-field theory. (c) Analytically predicted equilibrium angle using the far-field approximation, linearized about $\theta=0$ (taking the minimum value between the prediction in \eqref{ffeq} and $\pi/2$). Bodies that are both slender and sufficiently active (but not completely active, $\ell=1$) exhibit pitching equilibria with their noses down towards the wall. Bodies that are not sufficiently slender or not sufficiently active exhibit pitching equilibria with their noses turned up away from the wall.}
\label{Figure8}
\end{center}
\end{figure}

Our interest now turns to the rotations induced on a self-propelled body by the presence of a wall. Absent any boundaries, the swimmer is assumed to move along a straight path, with $\dot{\theta}=0$. When a wall is present, the body will pitch up away from the wall or pitch down towards the wall due to hydrodynamic interactions. For a given fixed distance from the wall, $h$, this rotation will persist until an equilibrium angle $\bar{\theta}$ is reached. Stable equilibrium angles so found using the full numerical simulations are shown in Fig.~\ref{Figure8}a, where we have fixed the centroid distance to one body length away from the wall, $h=2$. Figure~\ref{Figure8}b shows the same plot but generated using the predictions of the far-field theory. Both the full simulations and reduced model predict that there exists a class of swimming bodies for which the hydrodynamic interaction with the wall results in a negative equilibrium pitching angle. Namely, those that are simultaneously slender and more geometrically active than inert, though only to a point. For potential flow squirmers, $\ell=1$, we find that for any aspect-ratio the pitching equilibrium angle is oriented directly away from the wall. Inclusive of the potential flow squirmers, we detect a barrier above which the only equilibrium angle is $\bar{\theta}=\pi/2$, and we note that the transition from a small orientation angle to no equilibrium other than $\bar{\theta}=\pi/2$ is extremely sensitive to the geometry of activation along this curve (a small change in $\ell$ can produce a dramatically different behavior).

In order to derive a simple analytical estimate of the equilibrium angle, we linearize the rotational contribution of each singularity about $\theta=0$, yielding
\begin{gather}
\dot{\theta}=-\frac{3\alpha}{8 h^3}\left(1+\frac{\Gamma}{2}\right)\theta-\frac{3}{8 h^4}\left(\gamma-\beta+\frac{\Gamma}{4}(11\gamma-6\beta)\right),\label{dotthetacompare}
\end{gather}
and hence an equilibrium swimming angle $\bar{\theta}$ (for a fixed distance $h$) is predicted to exist at
\begin{gather}
\bar{\theta}=\frac{1}{\alpha h\left(1+\Gamma/2\right)}\left(\gamma-\beta+\frac{\Gamma}{4}(11\gamma-6\beta)\right)\cdot\label{ffeq}
\end{gather}
The estimate from the linearized far-field theory is show in Fig.~\ref{Figure8}c. The results match the full far-field theory and simulations well where the equilibrium angle is small, as expected, showing an overestimate in the angle for bodies which are less geometrically active ($\ell\approx-1$). That the completely active bodies ($\ell=1$) have no stable equilibrium pitching angle outside of $\bar{\theta}=\pi/2$, in which the body is swimming directly away from the wall, is explored in greater detail {in the following section}.

The dashed curve marking the transition from pitched-up to pitched-down equilibria also separates bodies that will rotate away from the wall and those that will rotate towards the wall when $\theta=0$ (which may be inferred by continuity). Pausing to consider this rotation rate for wall-parallel swimming ($\theta=0$), we refer to Fig.~\ref{Figure5} and note that the rotation is towards the wall in a region where $0<\beta\ll1$ and $\gamma>0$, and also where $\Gamma=(1-e^2)/(1+e^2)\gtrsim 0.5$. For $\ell>0$, Fig.~\ref{Figure5} shows that $\beta$ grows more rapidly than $\gamma$ as the aspect-ratio is increased. For $\ell>0$, then, both $(\gamma-\beta)$ and $\Gamma(11\gamma-6\beta)/4$ will become negative for sufficiently large aspect-ratio, and $\dot{\theta}$ will become positive [see \eqref{dotthetacompare}]. In other words, the effect of the source dipole will overwhelm the effects of the Stokeslet quadrupole for sufficiently large aspect-ratio when $\ell>0$. When $\ell<0$ the situation is reversed: $\gamma$ decreases more rapidly than does $\beta$ as the aspect-ratio is increased. In this regime the Stokeslet quadrupole effects dominate those of the source dipole, both $(\gamma-\beta)$ and $\Gamma(11\gamma-6\beta)/4$ are negative in \eqref{dotthetacompare}, and thus $\dot{\theta}>0$.

To summarize, bodies that are both slender and sufficiently active exhibit pitching equilibria with their noses down towards the wall, though completely active bodies ($\ell=1$) always rotate away from the wall. Bodies that are not sufficiently slender or are not sufficiently active exhibit pitching equilibria with their noses turned up away from the surface. In addition, there is a boundary in parameter space beyond which the only stable equilibrium angle is $\theta=\pi/2$, with the body swimming directly away from the wall.

\subsection{Full swimming trajectories of squirmers}

Unsurprisingly, the far-field approximation cannot generally be counted on for quantitative (and in many cases qualitative) predictions of the entire swimming behavior when the body is in very near contact with the boundary. The exact form of the propulsive mechanism (in this case the form of the tangential slip velocity) will specify the nature of near-wall contact, be that a hydrodynamically trapped state or a trajectory that leads to a departure from the surface. However, we have found one class of swimmers for which the far-field approximation can be used to predict the entire interaction with the boundary, namely for squirmers ($\ell=1$), which we now describe in detail. That the far-field theory provides an accurate depiction of the full dynamics of a treadmilling swimmer was found in a two dimensional setting by \cite{co10} and \cite{Crowdy11}. The interaction of a squirmer with a wall has also been studied recently by \cite{lp10}.

We comment briefly on the numerical method. Time does not enter into the Stokes equations explicitly, and since the means of propulsion studied here is steady there are no variations in the dynamics with time outside of the trajectories described by the distance of the centroid from the wall, $h(t)$, and the pitching angle, $\theta(t)$. An adaptive time-stepping algorithm for stiff systems (ode15s in Matlab) is used to integrate the swimming trajectory to small enough error tolerance so that the sole error in the dynamics is due to discretization errors and the associated quadrature errors in evaluating $\mathcal{K}[\v{f}]$ (through the regularization parameter $\delta$). Hydrodynamic interactions with the wall are sufficient to prevent body-wall collisions in some but not in all cases. Following \cite{bb85} and \cite{ip07}, we include a screened electrostatic-type body repulsion force which acts only at very small distances from the boundary of the form $\v{F}_{rep}=A e^{-B d}/[1-e^{-B d}]\v{\hat{z}}$, where $d$ is the minimum distance between the body surface and the wall, and we take $A=100$, $B=10$. These values are selected so that the body does not come closer than approximately {$h/h_w=1.05$}, where the numerical method for computing the fluid velocity just begins to lose accuracy \citep[though not dramatically; see][]{adebc08}. For elongated squirmers ($e<1$) an associated torque is included as well. More physically realistic near-contact interaction effects have been discussed by \cite{pbnb02}. 

\begin{figure}
	\vspace{.15in}
	\begin{center}
	\includegraphics[width=5.3in]{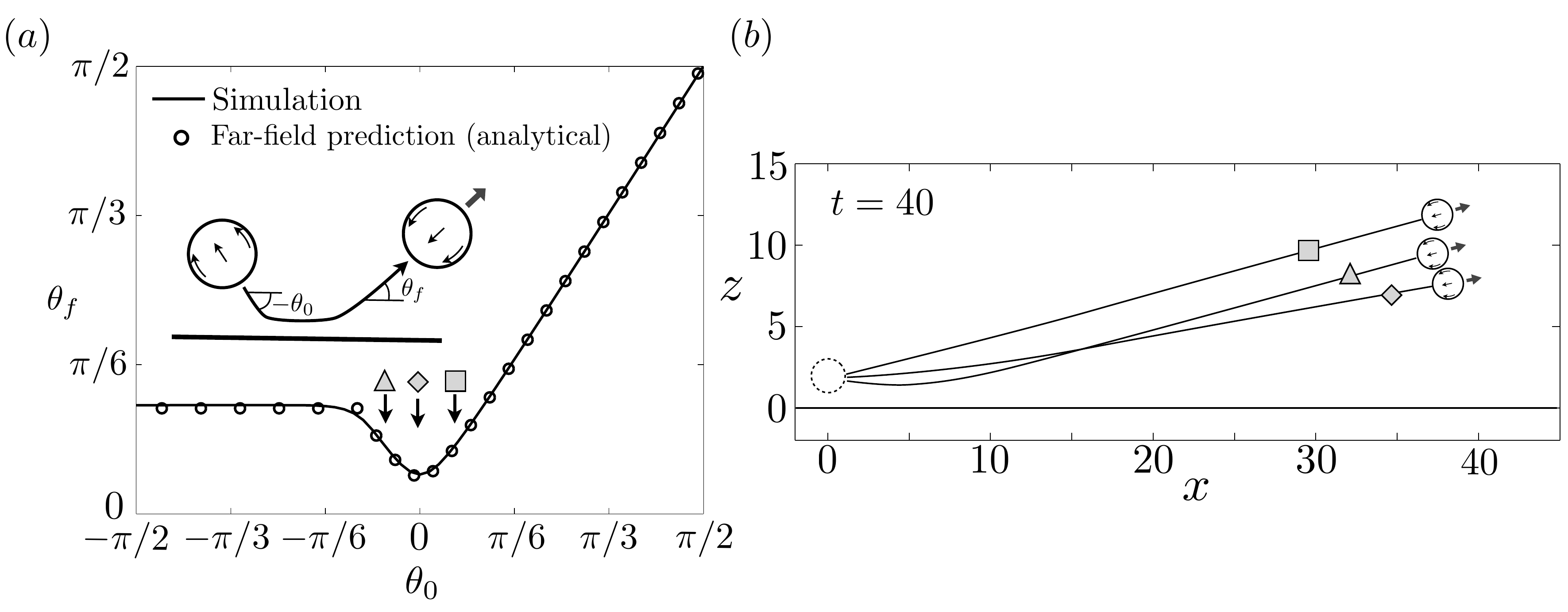}
	\caption{(a) The scattering angle exhibited by a squirming sphere ($\ell=1$ and $e=1$), for an initial  {centroid distance} $h=2$. Solid line: results from the full simulation. Circles: analytical far-field prediction, neglecting the surface attraction/repulsion, from \eqref{ScatteringSphere1} and \eqref{ScatteringSphere2}. (b) The trajectories of spherical squirmers with three different initial pitching angles, illustrating the non-monotonicity indicated in (a). A trajectory that takes the body nearer to the wall leads to a greater net rotation before the swimmer escapes.}
	\label{Figure9}
	\end{center}
	\end{figure}

We first describe the results of the full simulations by focusing on a spherical squirmer ($e=1$). By scanning the parameter space of distances $h$ and pitching angles $\theta$, we have observed that the body rotates away from the wall regardless of its distance from the boundary and orientation (save for the special case of swimming directly towards the wall at $\theta=-\pi/2$, though this orientation is found to be unstable). Setting the body initially at a distance one body radius from the wall ($h=2$), for initial pitching angles $\theta_0\gtrsim -0.4$ the body rotates as it moves through the fluid and does not come into close contact with the surface. It then settles into a final pitching angle $\theta_f>0$ as it swims away from the wall, never to return. This scattering angle, as determined from full simulations, is shown in Fig.~\ref{Figure9}a as a solid line. For initial angles $\theta_0<0$ the body swims towards the wall, which increases the rotational effect on the body and hence increases the final pitching angle once it departs, accounting for the non-monotonicity of the scattering curve. Three trajectories have been picked out with the intention of illustrating this non-monotonicity and are included as Fig.~\ref{Figure9}b.

Meanwhile, for initial pitching angles $\theta_0\lesssim -0.4$ the squirmer ``impacts'' the wall (realized here by blocks of time in which there is negligible wall-perpendicular velocity), but continues to rotate while in near-wall contact. Eventually the squirmer escapes the surface and swims away until settling to a final pitching angle $\theta_f\approx 0.4$. For non-impacting trajectories, the swimming trajectory must be symmetric about the point of parallel swimming, $\theta=0$, by the time-reversibility of the Stokes equations (reversing the direction of time is indistinguishable from reversing the direction of surface activation and swimming speed). In such a case we must have that the pitching angle reaches a value $-\theta_0$ when the swimmer has returned to the distance $h=2$ on its journey away from the surface. Having determined in this case that a critical initial pitching angle for wall-impact is given by $\theta_0\approx -0.4$, it might then have been predicted correctly that all wall-impacting trajectories in this case lead to final pitching angles of $\theta_f\approx 0.4$. The near-wall contact behavior simply acts to remove any information about the initial pitching angle until the trajectory matches that of the critical case (outside of a horizontal shift along the wall).

For the spherical squirmer that swims with speed unity in a quiescent fluid we have $\beta=1/2$ (see \ref{Squirmbeta}), and the dynamics predicted by the reduced model are set by
\begin{gather}
\dot{h}=\left(1+\frac{\beta}{h^3}\right)\sin(\theta),\,\,\,\,\dot{\theta}=\frac{3\beta}{8h^4}\left(1+\frac{3\Gamma}{2}\right)\cos(\theta).\label{htbeta}
\end{gather}	

Can we predict the scattering behavior described above analytically? In order to derive a simple estimate of the swimming behavior, let us linearize the motion about $\theta=0$ and assume $\beta\theta \ll h^3$. Then the translational swimming speed is not varied by the wall beyond its effect on the swimming angle $\theta$, and we have
\begin{gather}
\dot{h}=\theta,\,\,\,\,\dot{\theta}=\frac{3\tilde{\beta}}{8h^4},
\end{gather}	
with $h(0)=h_0$ and $\theta(0)=\dot{h}(0)=\theta_0$, where the geometrical dependence has been absorbed into the singularity strength, $\tilde{\beta}=\left(1+3\Gamma/2\right)\beta$. Integrating this system, we find (for trajectories that do not impact the wall) that
\begin{gather}
\frac{\dot{h}^2}{2}=\frac{\theta_0^2}{2}+\frac{\tilde{\beta}}{8}\left(\frac{1}{h_0^3}-\frac{1}{h^3}\right)\label{zdotEqn},
\end{gather}
or equivalently,
\begin{gather}
\frac{\theta^2}{2}-\frac{\theta_0^2}{2}=\frac{\tilde{\beta}}{8}\left(\frac{1}{h_0^3}-\frac{1}{h^3}\right)\cdot
\end{gather}
Taking $t\rightarrow \infty$, we know from the simulations that all squirmer trajectories have $\dot{h}\rightarrow \theta_f$ and $h\rightarrow \infty$; upon insertion, we find a final pitch angle for non-impacting squirmers of
\begin{gather}
\theta_f=\sqrt{\theta_0^2+\frac{\tilde{\beta}}{4 h_0^3}}\cdot\label{ScatteringSphere1}
\end{gather}

The distance to the wall $h=h_w(\theta)=\sqrt{(1+e^2)-(1-e^2)\cos(2\theta)}/\sqrt{2}$ here corresponds to wall impact. First we ask: which initial distances and orientations lead to wall impact? Writing the pitching angle at the time of impact as $\theta_w$, the centroid will be located at a distance $h_w(\theta_w)$ at that time. Inserting $\theta_w$ and $h_w$ into \eqref{zdotEqn} gives
\begin{gather}
\frac{\theta_w^2}{2}-\frac{\theta_0^2}{2}=\frac{\tilde{\beta}}{8}\left(\frac{1}{h_0^3}-\frac{1}{h_w(\theta_w)^3}\right)\cdot
\end{gather}
Given the simple expression for the translational velocity $\dot{h}=\theta$, and that $\dot{\theta}>0$ at all times, a wall-impacting squirmer must have $\theta_w<0$. A curve in $(h_0,\theta_0)$ space separating initial conditions for which a squirmer does or does not impact the wall may then be deduced by setting $\theta_w=0$, leaving
\begin{gather}
\frac{\theta_0^2}{2}+\frac{\tilde{\beta}}{8 h_0^3}=\frac{\tilde{\beta}}{8e^3},
\end{gather}
or
\begin{gather}
\theta_0=-\sqrt{\frac{\tilde{\beta}}{4}\left(\frac{1}{e^3}-\frac{1}{h_0^3}\right)}\cdot\label{ScatteringSphere2}
\end{gather}
In the case of spherical squirmers,  {for $h_0=2$ as in Fig.~\ref{Figure9},} this evaluates to approximately $\theta_0=-0.33$, which is slightly smaller in magnitude than the critical angle $\theta_0\approx -0.4$ found in the full simulations. Initial pitching angles smaller than $\theta_0=-0.33$ are predicted to lead to wall collisions. As the initial position becomes more distant from the wall, the critical angle increases in magnitude, and depending on the value of $\tilde{\beta}$ there may not exist an initial orientation such that the swimmer impacts the wall. (The special case of swimming directly towards the wall and guaranteeing impact, $\theta_0=-\pi/2$, is not accounted for in the linearized system.)

After the wall contact there is rotation for a time $T(\theta_w)$ until the pitching angle reaches $\theta=0$, at which point the body is predicted  {in this approximation} to separate from the wall and depart to a final pitching angle. The final pitching angle may be determined by setting $\theta_0=0$ and $h_0=e$ in \eqref{ScatteringSphere1}, giving
\begin{gather}
\theta_f=\sqrt{\frac{\tilde{\beta}}{4e^3}},\label{ScatteringSphere3}
\end{gather}
which in the spherical squirmer example returns the value $\theta_f\approx 0.35$. The predicted scattering angles from \eqref{ScatteringSphere1} (and \ref{ScatteringSphere3} for $\theta_0\leq -0.33$) are shown in Fig.~\ref{Figure9}a as circles. The simple estimates derived above provide a remarkably accurate depiction of the full interaction dynamics with the surface.

The role of the body geometry is noted here simply by inserting the exact source dipole strength as a function of aspect-ratio, which is a monotonically increasing function of $e$ from its minimum of $0$ for a slender squirmer to $1/2$ for a spherical squirmer (see \ref{Squirmbeta}). Scattering behavior depends not only upon the source dipole strength, however, but also upon the geometrical factor $(1+3\Gamma/2)$. In particular, we note that \eqref{ScatteringSphere3} may without difficulty be written solely as a function of the aspect-ratio, $e$, and we find that $\theta_f$ is a monotonically decreasing function on the domain $e\in[0,1]$. As $e\rightarrow 1$ (a slender rod), we have $\theta_f\rightarrow\sqrt{2}/4$. The wall interaction is predicted to be strongest for a spherical squirmer, and the final angle at which the body swims away from the wall is predicted to be greatest in this case. 

\begin{figure}
	\vspace{.15in}
	\begin{center}
	\includegraphics[width=5.6in]{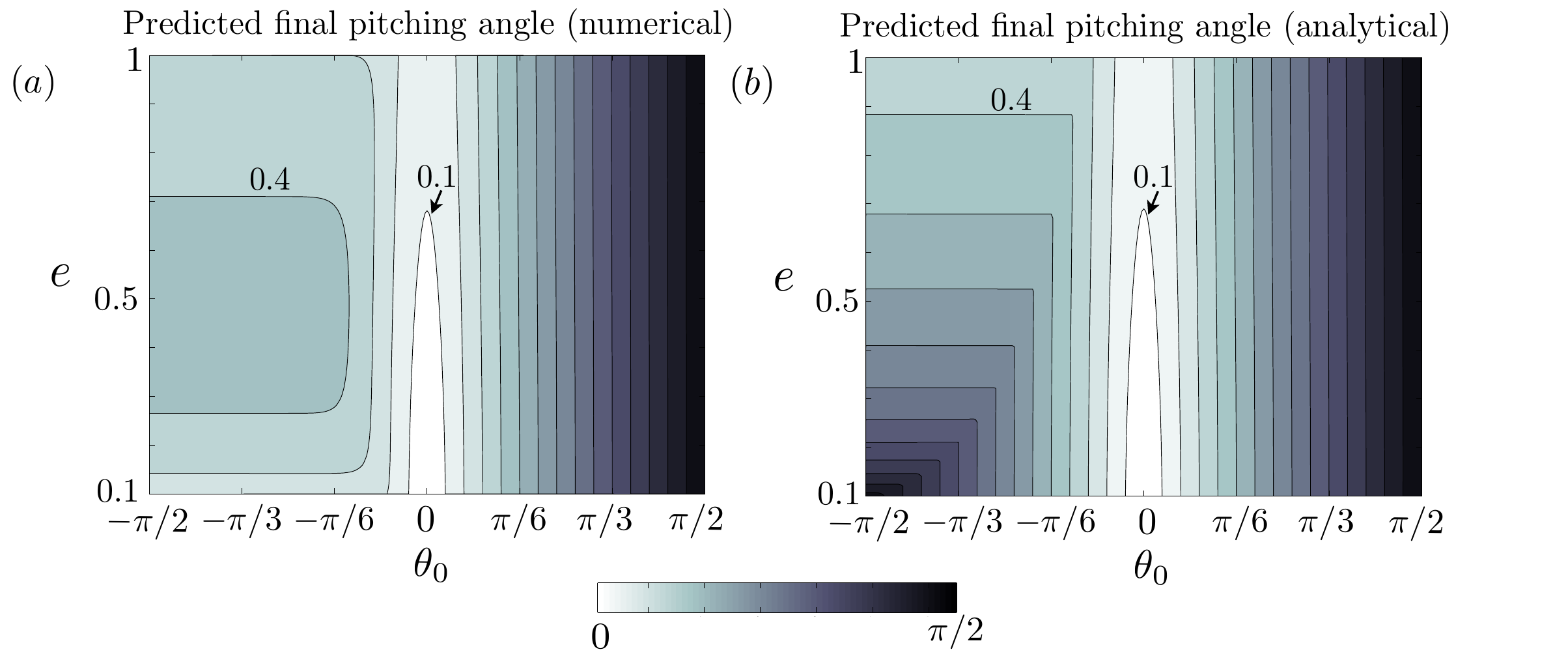}
	\caption{Contours of the final pitching angle $\theta_f$ of a squirmer, initially placed at $h=2$, as a function of the initial pitching angle $\theta_0$ and the squirmer aspect-ratio $e$. (a) As predicted integrating \eqref{htbeta} numerically, and (b) as predicted using the linearized approximation described in the text. Once again the agreement is reasonable when the initial and final pitching angles are not very large. The final pitching angle reached by wall-impacting swimmers is not monotonic in the body aspect-ratio.}
	\label{Figure10}
	\end{center}
	\end{figure}

Using the far-field theory we may rapidly produce a contour plot of the final pitching angle as a function of both the body aspect-ratio $e$ and initial pitching angle $\theta_0$, where again we initially set the body at a distance $h=2$ from the wall. Figure~\ref{Figure10}a shows the predicted values obtained by integrating \eqref{htbeta} numerically, while Fig.~\ref{Figure10}b shows the analytically predicted values from the simplified approach above. Once again, as in Fig.~\ref{Figure8}, the agreement is {quite good} when the initial and final pitching angles are not very large. The numerical integration of the complete far-field theory indicates that the final pitching angle reached by wall-impacting swimmers is not monotonic in the body aspect-ratio. This effect is not captured in the linearized approximations above. Making predictions about the consequences of wall-impact behavior for non-spherical bodies is complicated by the added repulsion force in the numerical simulations. In particular, the repulsive force breaks a time symmetry in the otherwise time-reversible structure of the Stokes equations, so arguments depending on this symmetry are generally invalid. Nevertheless, when the initial pitching angle is not too large in magnitude the hydrodynamics are often sufficient to keep the swimmer sufficiently far from the wall, and in such situations the repulsive force never plays a role. In these cases we find excellent agreement between predictions and the full simulation results.

Finally, when the body is in contact with the wall the $h$ and $\theta$ dynamics are effectively decoupled. Assuming that during wall contact $h=h_w(\theta)$, and expanding $\dot{\theta}$ [see \eqref{htbeta}] for small pitching angles, we may integrate
\begin{gather}
\dot{\theta}=\frac{3\tilde{\beta}}{8e^4}\left(1+\left[\frac{3}{2}-\frac{2}{e^2}\right]\theta^2\right)+O(\theta^4),
\end{gather}
to find 
\begin{gather}
\theta(t)=\frac{1}{q_e}\tanh\left(\frac{3q_e\tilde{\beta} \,t}{8 e^4}+\tanh^{-1}\left(q_e\theta_w\right)\right)\cdot
\end{gather}
where $q_e=\sqrt{2-3 e^2/2}/e$, and the body is taken to impact the wall at $t=0$ with pitching angle $\theta_w$. The time required for the body to rotate to $\theta=0$ (at which point the body departs from the wall) is then
\begin{gather}
T(\theta_w)=\frac{8 e^4}{3 q_e \tilde{\beta}} \tanh^{-1}\left(-q_e \theta_w\right).
\end{gather}

\section{Bacteria-like polar swimmers: geometric asymmetry}

\subsection{Model swimmers and computational adjustments}

\begin{figure}
\vspace{.15in}
\begin{center}
\includegraphics[width=3.2in]{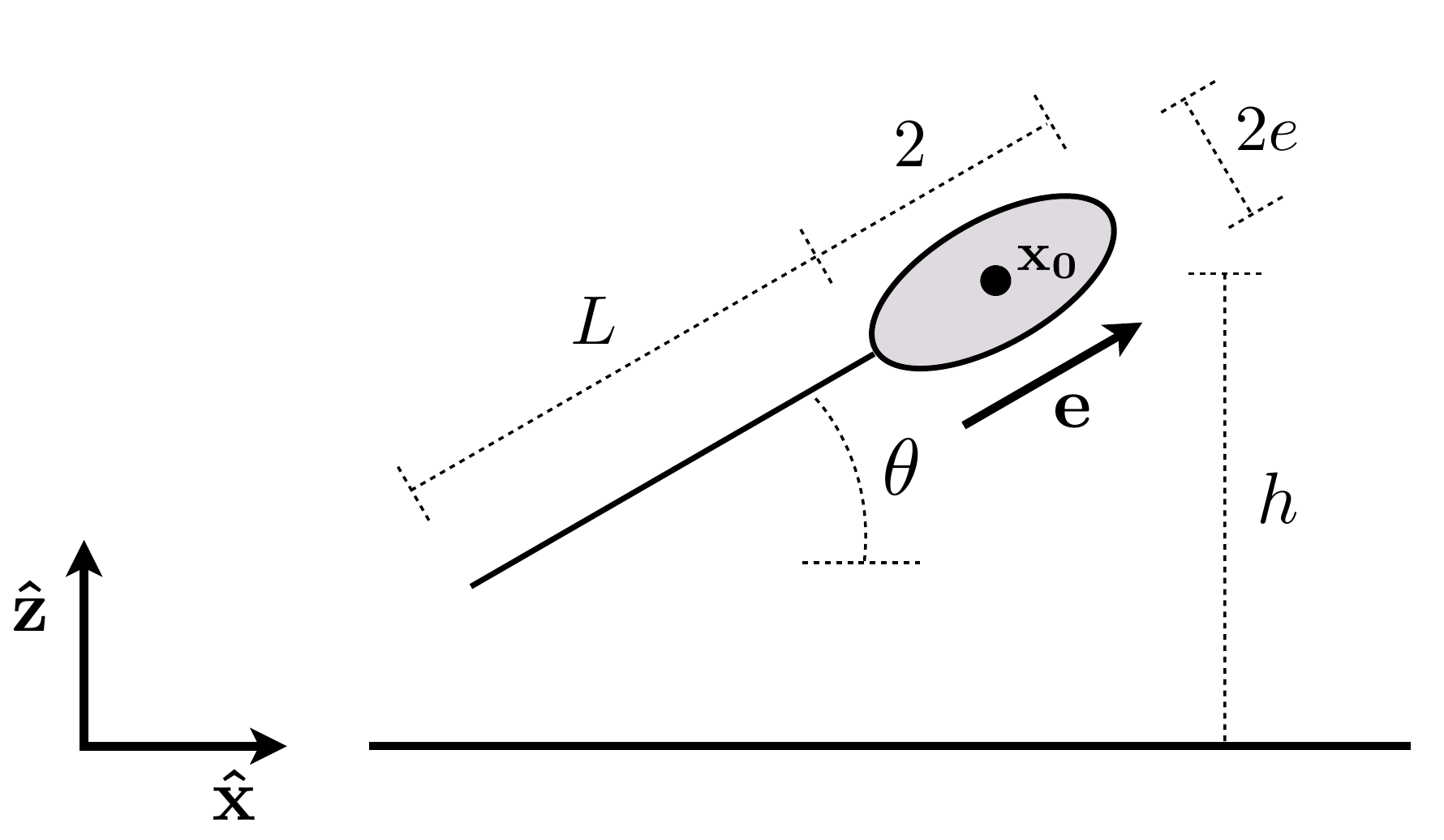}
\caption{A polar bacteria-like model swimmer  {directed along $\v{e}$} composed of a spheroidal head with aspect-ratio $e$ and an actively pushing rod. (The system is still made dimensionless by scaling lengths upon the semi-major axis of the head).}
\label{Figure11}
\end{center}
\end{figure}

In order to probe the role of polar body geometry in hydrodynamic interactions with walls for more biologically relevant swimmers we will enlist the help of a different class of model organisms, as illustrated in Fig. \ref{Figure11}. The body is composed of an inert spheroidal head with centroid $\v{x_0}$ (on which the no-slip condition is applied) and an active propelling rod of dimensionless length $L$. The system is made dimensionless by scaling on the semi-major axis length of the head and on the free-space swimming speed as before. The propulsive mechanism is a prescribed uniform distribution of a tangential force per unit length on the rod, $\v{e}\cdot \v{f}(s)=-\mathcal{F}$, where $s\in[0,L]$ is the arc-length parameter along the rod and $\mathcal{F}$ is determined numerically so that the body swims with unit speed in free-space. The fluid velocity on the rod is thus composed of a rigid body motion in addition to a tangential slip velocity along the long axis, $u_s(s)\v{e}$, which must be determined through force and torque balance.

While a uniform distribution of force along the rod is chosen for the sake of simplicity, variations in force distribution are typical in planar flagellar undulations and can play an important role in setting swimming trajectories. For example, \cite{sb09} have shown that the wavenumber in spermatozoan swimming influences the force distribution along the flagellum, which in turn affects the pitching dynamics of the organism near a surface. Variations in the force distribution can in fact be an order of magnitude larger than their mean values \citep[see][]{jb79,sb09}. To what extent the change in pitching dynamics with wavenumber is due to a change in the time-averaged singularity strengths, or due to the time-variation of those strengths, remains to be seen. It is possible that the assumption of uniform force distribution is more suited to studying bacteria with helical flagella, for instance \citep[see][]{Lighthill96}.

\begin{figure}
\vspace{.15in}
\begin{center}
\includegraphics[width=5.2in]{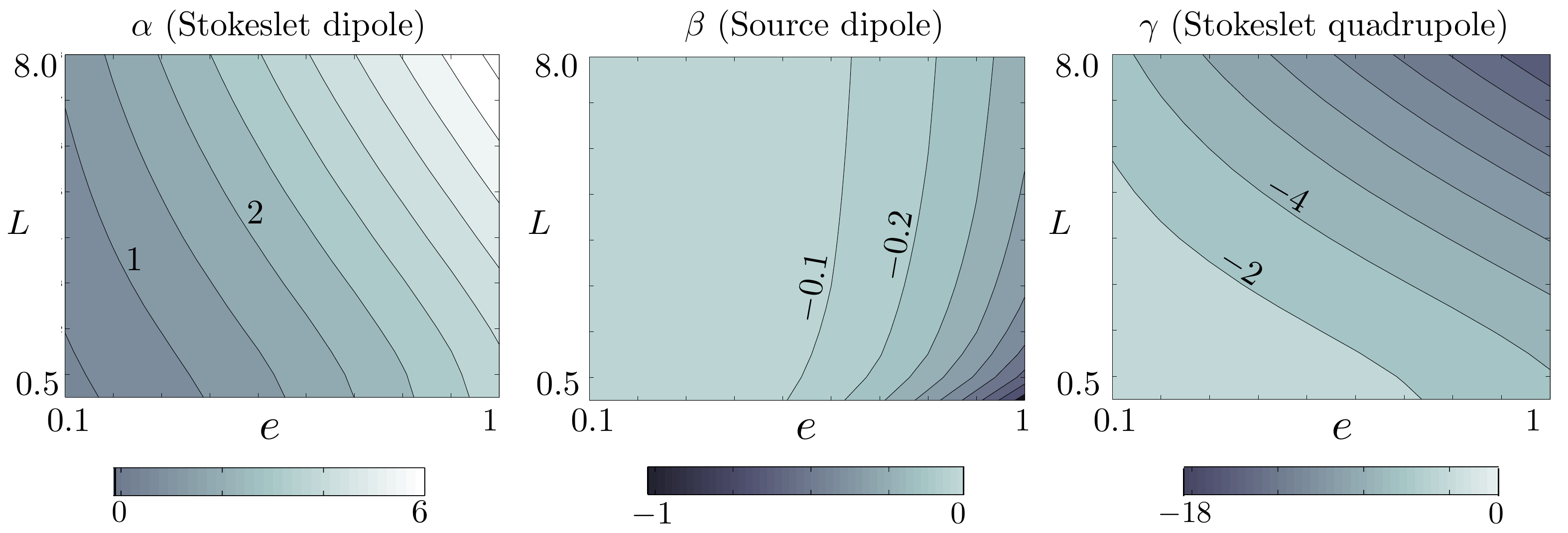}
\caption{Computed singularity strengths (about the sphere centroid $\v{x_0}$) for sphere-rod swimmers of varying rod length $L$ and head aspect-ratio $e$. The quadrupole strength is always negative, due only to the placement of the singularities at $\v{x_0}$.}
\label{Figure12}
\end{center}
\end{figure}

Numerically, the rod is represented simply as a line of Stokeslets, which may be included into \eqref{RegularizedStokeslets} without difficulty. The spacing between the points on the rod is chosen to be the same mean distance as between points on the body, $h_{grid}$. In order to compute the singularity strengths, the integrals \eqref{SingularityStrengths}-\eqref{SingularityStrengths3} are still appropriate for determining the contribution of the head to the singularity strengths (and the $u$ integrals disappear due to the no-slip condition that is applied on the spheroid). But to these values we add contributions from the rod as follows. Again aligning the body along $\v{e=\hat{x}}$ for the sake of presentation, the singularities emanating from the rod center have strengths (from \eqref{SingularityStrengths}-\eqref{SingularityStrengths3}):
\begin{gather}
F'=\frac{1}{8\pi}\int_{s=0}^L f_x(s)\,ds,\\
\alpha'=\frac{1}{8\pi}\int_{s=0}^L (x(s)-\lambda)f_x(s)\,ds,\\
\beta'=0,\\
\gamma'=\frac{1}{16\pi}\int_{s=0}^L (x(s)-\lambda)^2 f_x(s)\,ds,
\end{gather}
where $\lambda=-(1+L/2)$, $x(s)=-(1+s)$, $f_x=\v{\hat{x}\cdot f}$, and we have included $F'$ as the coefficient of the Stokeslet singularity (which balances the Stokeslet singularity strength of the moving sphere due to the zero net force condition). In order to express the singularities as emanating from the spheroid center, $\v{x_0}$, the singularities above may simply be shifted, yielding the natural expressions
\begin{gather}
\alpha_{rod}=\alpha'+\lambda F'=\frac{1}{8\pi}\int_{s=0}^L  x(s) f_x(s),\\
\beta_{rod}=0,\\
\gamma_{rod}=\gamma'+\lambda \alpha'+\frac{\lambda^2}{2}F'=\frac{1}{16\pi}\int_{s=0}^L x(s)^2 f_x(s).
\end{gather}
Adding these last integrals to the expressions in \eqref{SingularityStrengths}-\eqref{SingularityStrengths3} returns the singularity strengths of the entire spheroid-rod swimmer. Similar expressions for the singularity strengths generated by a body composed of multiple ellipsoids may be produced in a similar way, shifting the dipole and quadrupole strengths as required, suggesting a tractable method for constructing the singularity strengths for an arbitrary swimmer geometry. The slenderness of the rod allows us to neglect the slip velocity in the determination of the singularity strengths here. The singularity strengths so computed using the full boundary integral solution to the Stokes equations \eqref{RegularizedStokeslets} are shown in Fig.~\ref{Figure12} for a range of rod lengths $L$ and body aspect-ratios $e$. One simple estimate of the singularity strengths for any body aspect-ratio may be deduced by assuming no hydrodynamic interactions between the spheroid and rod, yielding
\begin{gather}
\alpha=\left(1+\frac{L}{2}\right),\,\,\,\,\,\beta=-\frac{e^2}{3},\,\,\,\,\,\gamma=-\frac{1}{2}\left(1+\frac{L}{2}\right)^2.
\end{gather}

The Stokeslet quadrupole strength is always negative, which seems to be at odds with our exploration of the ellipsoidal Janus swimmers. The discrepancy is due to our choice to write the singularities as emanating from the center of the spheroid, $\v{x_0}$. By shifting the singularities to another point along the body, such as the center of reaction, the Stokeslet quadrupole strength can take on positive or negative values.

\subsection{A modified Faxen Law for spheroid-rod swimmers}

Fax\'en's Law as described by \eqref{FaxenU}-\eqref{FaxenO} is inadequate for the body geometry in Fig.~\ref{Figure11}. While it is common practice to represent microorganisms of arbitrary geometry and propulsion as prolate ellipsoids for the sake of mathematical tractability, this approach in general can lead even to qualitatively incorrect predictions (to be more precise, qualitative departures from the values found in the full simulations). To derive a more appropriate representation of the effect of an ambient flow on the spheroid-rod model swimmer, we apply the usual Fax\'en Law to the sphere, but couple this to a slender body representation of the rod to describe its interaction with the surrounding fluid. For simplicity we neglect hydrodynamic interactions between the two, but require the spheroid-rod system to move as a rigid body while satisfying the zero net force and zero net torque conditions. This approach returns surprisingly good agreement with a ``perfect'' Fax\'en Law for our purposes, in which the full boundary integral formulation is used to solve for the rigid body motion in a prescribed ambient flow. The errors in what follows are due instead primarily to the breakdown of the far-field representation of the wall-induced flow field.

The spheroid is taken to move through the fluid with dimensionless translational velocity $\v{U}$ and rotational velocity $\v{\Omega}$. In this approximation, the motion of the spheroid then applies the following (dimensionless) force and torque  to the surrounding fluid:
\begin{gather}
\v{F_S}=6\pi\left[X^A\v{ee'}+Y^A(\v{I-ee'})\right]\left(\v{U}-\v{u^*}(\v{x_0})\right),
\end{gather}
and
\begin{gather}
\v{T_S}=8\pi\left[X^C\v{ee'}+Y^C(\v{I-ee'})\right]\left(\v{\Omega}-\v{\omega}^*\right)-8\pi Y^H\v{e}\times \t{E^*}\v{(x_0)\cdot e},
\end{gather}
where $X^A=(8/3)\chi^3/(-2\chi+(1+\chi^2)L_\chi)$, $Y^A=(16/3)\chi^3/(2\chi+(3\chi^2-1)L_\chi)$, $X^C=(4/3)\chi^3(1-\chi^2)/(2\chi-(1-\chi^2)L_\chi)$, $Y^C=(4/3)\chi^3(2-\chi^2)/(-2\chi+(1+\chi^2)L_\chi)$, $Y^H=(4/3)\chi^5/(-2\chi+(1+\chi^2)L_\chi)$, $L_\chi=\log[(1+\chi)/(1-\chi)]$, and $\chi=\sqrt{1-e^2}$ \citep{kk91}. Recall that $\v{u^*}$ in our case is the fluid velocity created by the image singularities as described in \S II. 

Meanwhile, a local slender body approximation is used to describe the fluid force imposed by the motion of the rod, in which the force per unit length is written as
\begin{gather}
\v{f}(s)=\frac{8\pi}{c+2}\left(\t{I}-\frac{c-2}{2c}\v{ee'}\right)\left(\v{u}(s)-\v{u}^*(\v{x_0}-s\,\v{e})\right),\label{rft}
\end{gather}
where $c=\log(4/\e^2 e)$, $\e$ is the aspect-ratio of the slender rod, and we estimate the slenderness of the rod by taking $\e=\delta\sqrt{e}/(2L)$ \citep[see][]{gh55,Childress81}. Since the rod moves with rigid body motion, $\v{u}(s)=\v{U}-s\,\v{\Omega}\times \v{e}$. Note that the slip velocity does not enter into the Fax\'en-type approach here. Also, we refer the reader to \cite{bm11}, who explore slender body theory using regularized Stokeslets in much greater detail. 

The motion of the rod applies the following force and torque (about $\v{x_0}$) on the surrounding fluid:
\begin{gather}
\v{F_R}=\int_{s=0}^L \v{f}(s)\,ds,\,\,\,\,\,\v{T_R}=\int_{s=0}^L (\v{x}(s)-\v{x_0})\times \v{f}(s)\,ds.
\end{gather}
The zero net force and zero net torque conditions, $\v{F_S+F_R}=0$, $\v{T_S+T_R}=0$, produce a linear system which may be solved for the resulting translational velocity $\v{U}$ and rotational velocity $\v{\Omega}$. Into these expressions, we simply insert the singularity image systems for the tilted force dipole, source dipole, and force quadrupole as before, as $\v{u^*}$.

\subsection{How accurate is the far-field representation?}

\begin{figure}
\vspace{.15in}
\begin{center}
\includegraphics[width=5.2in]{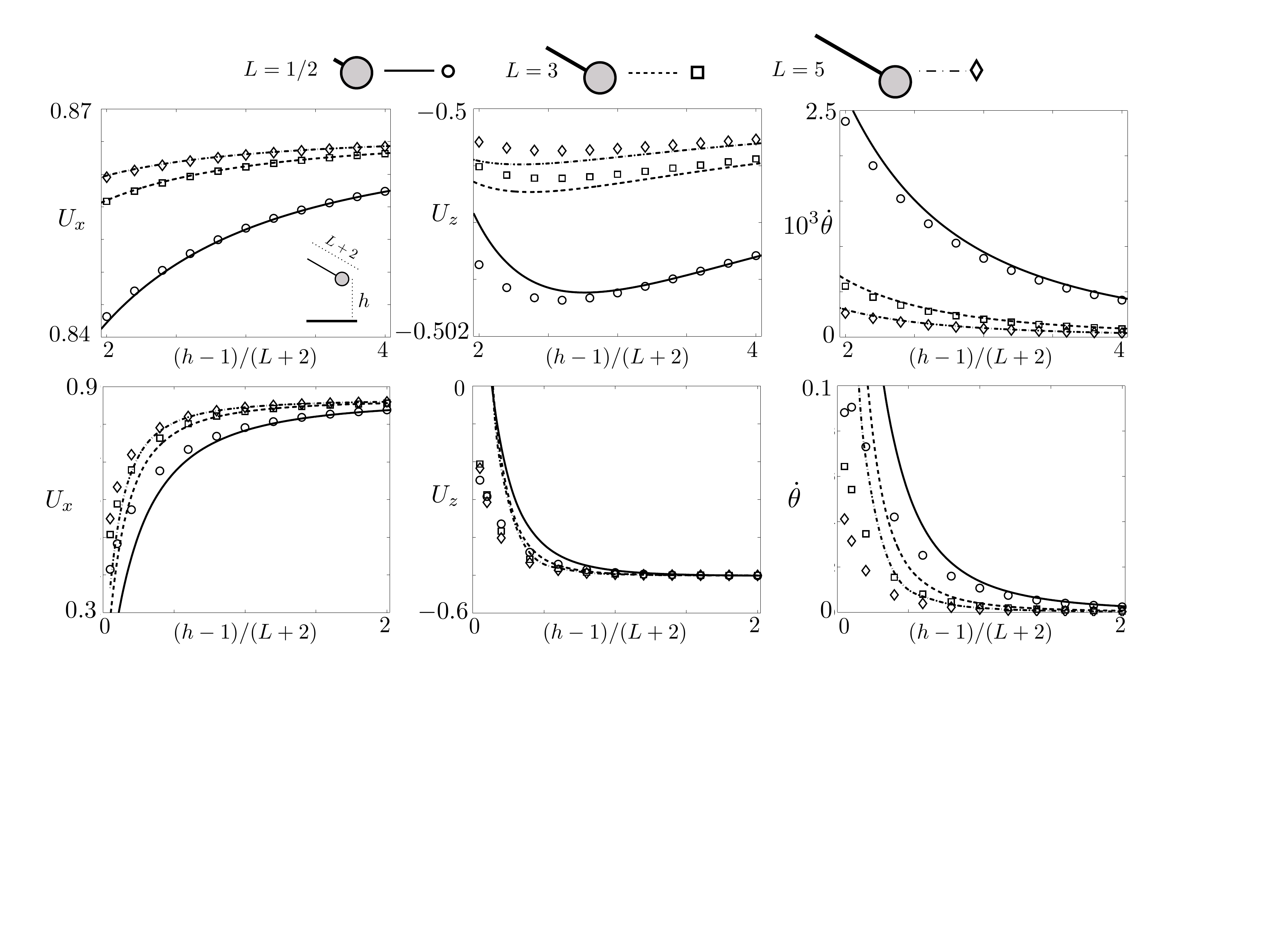}
\caption{Comparison of full simulation and far-field predictions for $\theta=-\pi/6$, with computed values indicated by symbols and far-field predictions indicated by lines. The top row shows results for bodies well separated from the wall, $(h-1)/(L+2)\in[2,4]$, while the bottom row shows results for bodies very close to the wall, $(h-1)/(L+2)\in[0,2]$. The simple Fax\'en Law developed in the text captures well the adjustment to the swimming dynamics when the body is far from the wall, even without the inclusion of hydrodynamic interactions between the sphere and the rod. For bodies very close to the wall the far-field prediction retains qualitative accuracy for a range of distances, but eventually becomes qualitatively incorrect.}
\label{Figure13}
\end{center}
\end{figure}

We are now equipped with the tools to compare the full simulations with the far-field predictions derived above. Fixing the pitch angle to $\theta=-\pi/6$, and the head aspect-ratio to $e=1$, we show in Fig.~\ref{Figure13} (top row) that the simple Fax\'en Law developed above captures very well the adjustment to the swimming dynamics when the body is far from the wall, even without the inclusion of hydrodynamic interactions between the sphere and the rod. Scanning to even smaller distances (Fig.~\ref{Figure13}, bottom row), we find very accurate quantitative similarity between the predictions and the simulation results down to approximately $h=(L+2)$ (the total length of the sphere-rod body). When the body is yet closer to the wall, the far-field prediction retains qualitative accuracy for a range of distances, but eventually becomes qualitatively incorrect. As seen in the comparison of $U_x$ and $U_z$ on the bottom row of Fig.~\ref{Figure13}, the translational velocities predicted by the far-field theory in this particular case take the wrong sign and yield non-physical behavior. For all but these very small distances from the wall, the far-field prediction with the Fax\'en Law of the previous section provides a quick and accurate approach for predicting the wall-induced adjustments to the swimmer trajectories.

Again we have verified by comparison with a ``perfect'' Fax\'en Law (using the full simulations to determine the effect of the image singularities) that this degradation is not due to the breakdown of the modified Fax\'en Law but due to the breakdown in the far-field approximation of the wall effect.

\section{Employment of the reduced model: spheroid-rod swimmers}
\subsection{Wall-induced rotations}

As in \S IV, we now consider the initial effect of the wall on bacteria-like swimmers of varying geometry and propulsive activity. Our previous analysis for ellipsoidal model swimmers suggests that the ratio of head-size to flagellum length might play a critical role in setting the wall-induced rotational behavior.

\begin{figure}
\vspace{.15in}
\begin{center}
\includegraphics[width=5.2in]{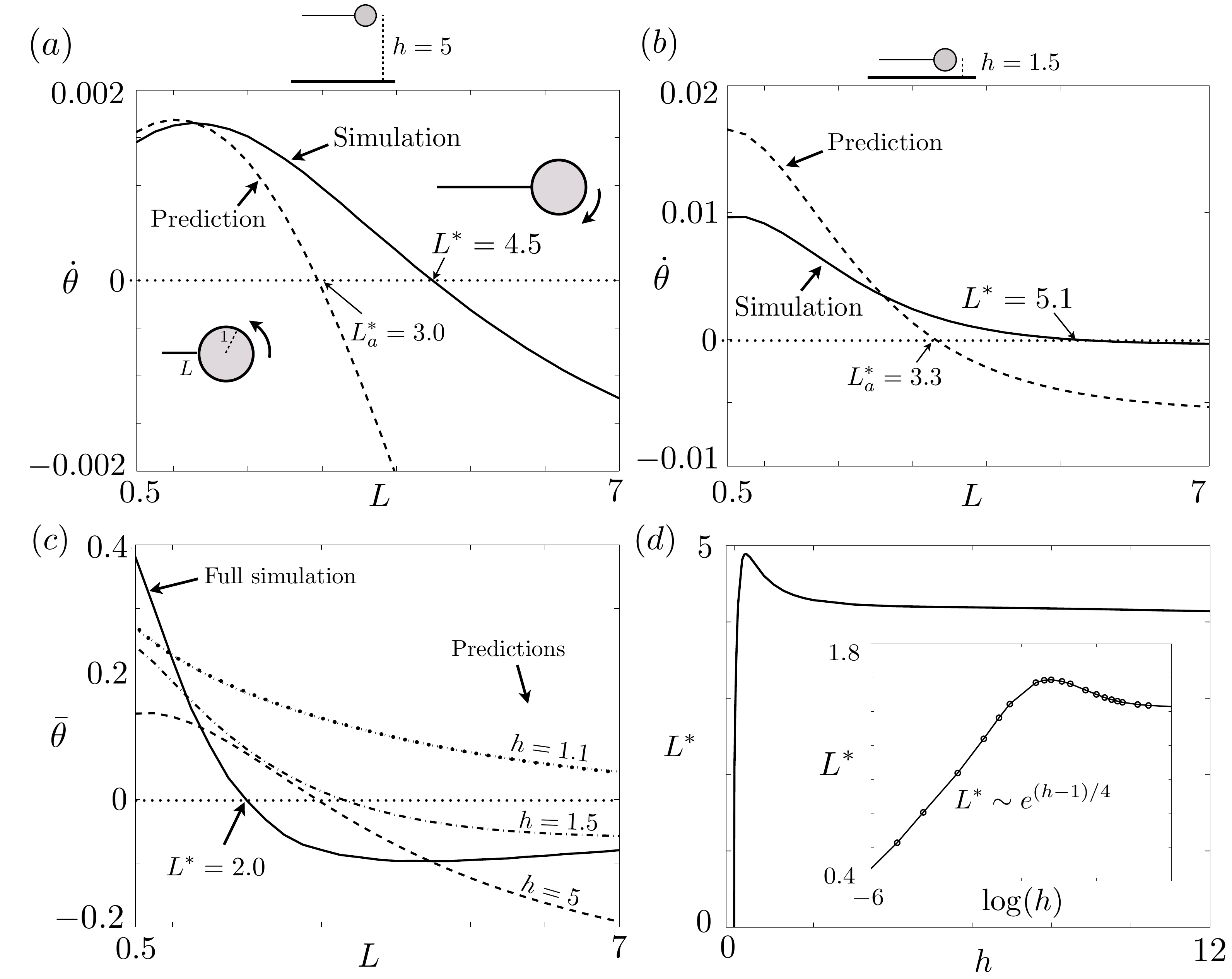}
\caption{(a) Rotation rate for swimmers with spherical heads and varying rod lengths $L$ with $\theta=0$ and $h=5$ fixed. (b) The same, for a fixed distance $h=1.5$. (c) The equilibrium pitching angle determined using fully time-integrated simulations ($h$ and $\theta$ are determined by the dynamics) as a function of $L$ is shown as a dark line. The pitching angle in the steady swimming state is positive for rod lengths $L<L^*\approx 2$, and negative for rod lengths $L>L^*$. The negative equilibrium pitching angle does not vary significantly as the rod length is increased, but takes on a minimum value at $L\approx 4$. The far-field prediction of equilibrium pitching angle for three fixed heights are included for comparison. (d) The critical rod length for which there is rotation direction reversal as a function of the distance to the wall, $h$,  {as determined using full numerical simulations}.}
\label{Figure14}
\end{center}
\end{figure}

We begin by fixing a swimmer with a spherical head to a distance $h=5$ from the wall, and the pitching angle to $\theta=0$, and plot the wall-induced rotation as a function of rod length in Fig.~\ref{Figure14}a. And indeed, we see that there is a critical rod length at which the rotation generated by the proximity to the wall reverses sign. The results suggest that for organisms with small flagella relative to cell-body size, the rotation will act to redirect the swimming away from the wall; similarly, for organisms with longer flagella relative to the cell-body size, the wall-effect will be to redirect the swimming towards the wall, increasing the likelihood of wall impact. This critical rod length was determined by simulations to take the value of approximately $L^*=4.5$, or slightly more than twice the cell-body diameter. The far-field prediction also shows this transition in the direction of rotation. The prediction is in close quantitative agreement with the full simulations for small rod lengths (where the far-field approximation is expected to be more accurate), but slightly underestimates the critical rod length for direction reversal, $L^*_a=3.0$. Nevertheless, the important qualitative features of the wall-effect are captured by this simple means of estimation. 

Now let us  {place the bacteria-like swimmer much closer to the wall}. Fixing $h=1.5$, we plot the rotational velocity as determined by simulation and far-field prediction in Fig.~\ref{Figure14}b, again as a function of rod length. While the quantitative error in the far-field prediction increases, the simple estimation is still able to capture a qualitative difference in pitching dynamics for rod lengths below and above a critical value. The result that there exists a critical rod length for rotational direction reversal appears to be robust, given that the adjustments to this critical rod length from $h=5$ to $h=1.5$ are quite small.

Loosely applying the theoretical predictions for ellipsoidal body geometries described in \S III (for instance the linearized expression shown in \ref{dotthetacompare}), we might expect that a large source dipole strength relative to the Stokeslet quadrupole strength would lead to an induced rotation away from the wall. In drawing this comparison recall that the singularity strengths represented in Fig.~\ref{Figure12} are centered about the centroid of the head, so that the source dipole should be compared instead to an appropriate combination of the Stokeslet quadrupole and a shifted Stokeslet dipole. This rough comparison provides intuition as to why an increase in rod length for a given head shape, which increases the Stokeslet dipole and quadrupole strengths more rapidly than the source dipole strength, can lead to a reversal in the wall-induced rotation. 

In general the rotational effects described above are coupled to the wall-induced attraction or repulsion in setting the complete swimming trajectory of a given organism. We already see, however, that the geometry of a swimming microorganism is very likely to play an important role in such dynamics as wall-entrapment or adhesion and the development of biofilms. While organisms with small flagella relative to cell-body size may still be hydrodynamically (or otherwise) pinned to a solid surface, their orientation in such a state can be directed away from the wall and the entrapment may be unstable to sufficient perturbation. Meanwhile, for microorganisms with longer flagella, the rotational component of the wall-effect is likely to act in tandem with hydrodynamic attraction to greatly increase the likelihood of wall-impact and long-term entrapment. 

\subsection{Equilibrium pitching angles}

We now allow the polar bacteria-like body to swim until it finds an equilibrium pitching angle. From the previous consideration we expect to find a critical rod length below which the equilibrium angle is positive, and above which the equilibrium angle is negative. The bacteria-like swimmers studied here have $\alpha>0$, so that the effect of the wall at leading order (when $\theta\sim 0$) is to attract the swimmer to the surface. For a small enough initial distance $h(t=0)$ we find that these swimmers (with $L\in[0.5,7]$) all become pinned to the wall. For larger initial distances, not all swimmers will impact the wall and find such an equilibrium state. Particularly, for organisms with small flagella relative to the cell body size, there will be a competition between the rotation away from the wall (which redirects the swimming velocity) and the hydrodynamic attraction towards the wall.

Figure~\ref{Figure14}c shows the equilibrium pitching angle determined using fully time-integrated simulations (in which $h$ and $\theta$ evolve dynamically) as a solid line. As in \S V, a repulsion force is included in order to keep the body from making contact with the wall of the form $\v{F}_{rep}=A e^{-B d}/[1-e^{-B d}]\v{\hat{z}}$, where $d$ is the minimum distance between the body surface and the wall, and here we take $A=100$, $B=100$; with this repulsion the body does not come closer than $h\approx1.01$ for any rod length. Starting from various initial distances (close to the wall) and pitching orientations, and evolving in time until a steady equilibrium is reached, the plotted equilibria appear to be the only stable equilibria. The pitching angle in the steady swimming state is seen to be positive for rod lengths $L<L^*\approx 2$, and negative for rod lengths $L>L^*$. The negative equilibrium pitching angle does not vary significantly as the rod length is increased, but takes on a minimum value at $L\approx 4$. 

We include in Fig.~\ref{Figure14}c the equilibrium angles predicted by the far-field theory for fixed distances $h=5$, $h=1.5$, and $h=1.1$. The far-field theory provides a reasonable qualitative prediction of the trapped equilibrium state, but only when the distance to the wall is not taken to be too small. The far-field theory breaks down when the swimmer is close to the wall, and the pitched-down equilibrium states are not predicted in this range of rod lengths. Higher order terms in the multipole expansion not included here (or lubrication effects) are thus found to play a vital role in determining the very-close-contact behavior, just as was generally the case for the model Janus swimmers.

Note that the critical rod length for direction reversal does depend on the distance $h$. At a distance $h=5$ (fixing $\theta=0$), bodies with rods longer than $L^*=4.5$ will rotate downward towards the wall, while only bodies with rod length greater than $L^*=5.1$ rotate towards the wall when $h=1.5$  {(as determined using the full simulations; see Figs.~\ref{Figure14}a-b)}. When the body is in very close contact with the wall, $h=1.1$, then the critical rod length is approximately $L^*=2.0$  {(see Fig.~\ref{Figure14}c)}. The non-monotonicity of the critical rod length for rotation direction reversal is shown in Fig.~\ref{Figure14}d (on a semi-log scale in the inset), and this critical length is found to decrease to zero as the spherical head is moved closer to the wall. The motion of a sphere moving close to a plane wall has received classical attention \citep[see for instance][]{gcb66}, where it has been shown that the effect of lubrication forces act to ``roll''  {a sedimenting} sphere along the surface. Correspondingly in the case studied here, the lubrication forces are expected to contribute to a downward pitching motion, and thus to reduce the critical rod length for direction reversal at very small distances. This is not to suggest that, absent a steric repulsion, all swimmers would eventually turn towards the wall to a negative equilibrium pitching angle. A swimming velocity directed away from the wall can be sufficient to balance hydrodynamic attraction at a distance $h$ such that lubrication forces are not dominant.

\begin{figure}
\vspace{.15in}
\begin{center}
\includegraphics[width=5.2in]{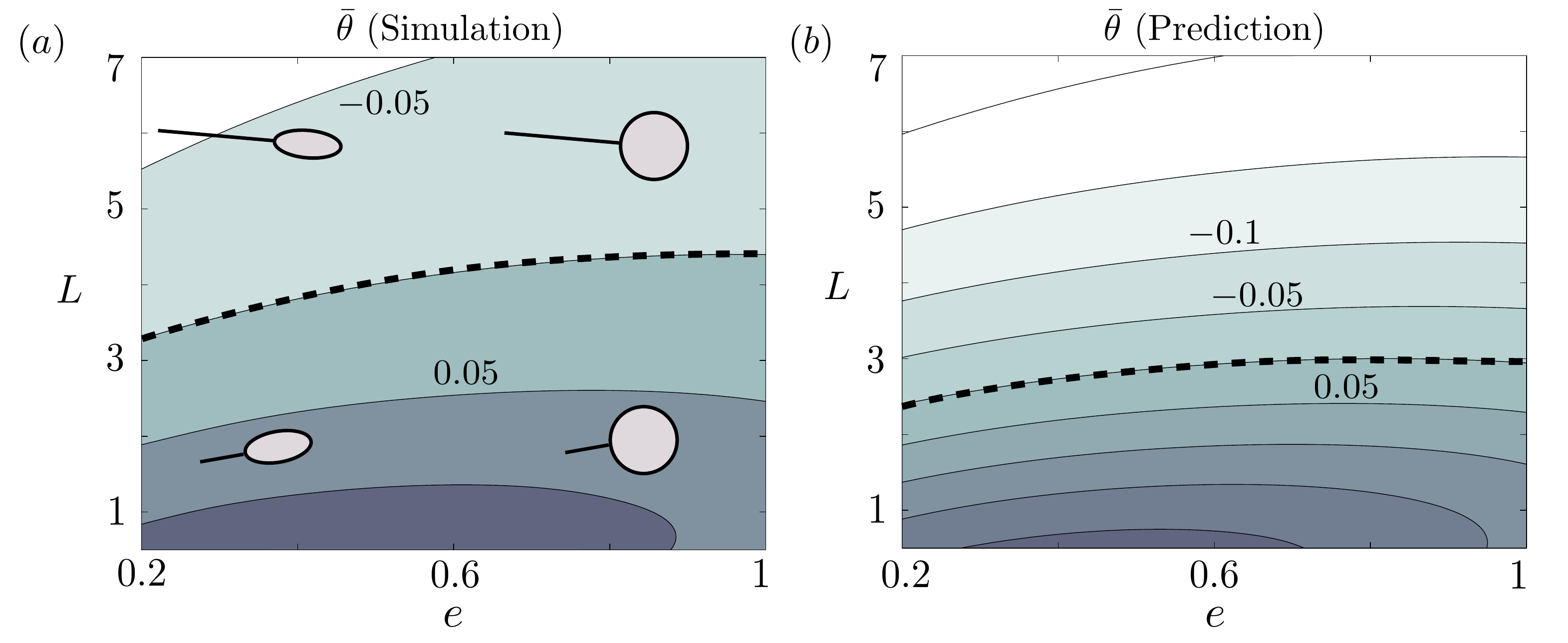}
\caption{(a) Contours of equilibrium pitching angle $\bar{\theta}$ for fixed height $h=5$ computed using the full simulations. (b) The same, but as predicted using the far-field approximation. As in the case of the Janus swimmers, bodies with small no-slip regions (cell bodies) and long active regions (flagella) are pitched downwards passively by the hydrodynamic interaction with the wall, and vice versa for large cell bodies with small flagella.}
\label{Figure15} 
\end{center}
\end{figure}

Finally, another geometrical factor of possible importance in wall-induced swimming behavior is the shape of the cell-body. For our last consideration, let us return to a regime where the far-field theory retains a modest predictive power. Figure~\ref{Figure15}a shows the equilibrium pitching angle for a range of rod lengths $L$ and head aspect-ratios $e$ when the head is fixed at a distance $h=5$ from the wall, as determined through full numerical simulation. The direction of rotation for bodies swimming parallel to the wall at this distance may be inferred by continuity. For a given head shape, we find that there is a critical rod length in the range $L\in(3,5)$ for which the equilibrium pitching angle changes sign, denoted by a dashed line. The equilibrium angles predicted by the far-field theory are shown in Fig.~\ref{Figure15}b. The far-field theory captures the change in direction reversal, though underestimates the critical rod length, and also generally overestimates the magnitude of the equilibrium pitching angle. The far-field theory does capture the decrease in the critical rod length for head shapes of smaller aspect-ratio. 

Of particular note, Figs.~\ref{Figure15}a-b suggest that organisms with sufficiently long flagella may be passively rotated towards the wall regardless of the shape of the cell-body. Similarly, if the flagella is sufficiently short the induced rotation may be oriented away from the wall for any head shape. Once again applying loosely the exact theory for ellipsoidal geometries from \S III, as the cell-body becomes more prolate in shape (and hence decreases in volume given our choice of nondimensionalization), the source dipole strength rapidly decreases in magnitude to zero. This would suggest that the contribution to rotation that is oriented away from the wall diminishes with decreasing aspect-ratio, so that the critical rod length for direction reversal as $e\rightarrow 0$ should also decrease, consistent with our findings. 

\section{Discussion}

In this paper, we have explored a multipole representation of swimming organisms to improve our understanding of swimming behaviors near a surface from a general hydrodynamic perspective. The simplified model has shown itself to be very useful in understanding the behavior of a selection of model swimmers of varying geometry and propulsive activity. Our results present some hope that by determining only a small number of parameters describing a given organism, much of its hydrodynamic interaction with a solid or free surface, or with another similar organism, might be predicted without the need for large scale numerical simulations. The framework described here may be of use in understanding as yet puzzling behaviors in near-wall microorganism behavior, and conversely may be of great use in the design of moving structures to perform tasks near a surface. One of the most general findings in our study has been that swimming bodies with small inert heads and long active propulsive mechanisms are those bodies which rotate nose down towards the wall due to hydrodynamic interactions, while other swimmers rotate nose up away from the wall. Among other reasons, the question of equilibrium geometry (i.e. orientation) is important because it impacts the response of a wall-bound organism to a shear flow, a topic of intended future study along the lines considered herein \citep{hkmk07,kk09}.

Avenues of interest for future study include time-dependent singularity strengths (due perhaps to periodic undulations in propulsion mechanism) and their consequences on wall-induced behaviors and trajectories. Such variations may well change the nature of the equilibrium states near the wall, and this is suggested by the work of \cite{sgs10} and \cite{ekg10}. The matching of far-field predictions with near-field lubrication effects may also prove fruitful in certain cases, which would allow for a more complete swimming trajectory and hydrodynamic entrapment analysis. The framework described above might be pursued further to predict long time trajectories for a number of swimmers.

While not considered here, we have included the stress-free boundary condition velocities in Appendix B, and there are likely many interesting phenomena which can be tuned or explained by varying the boundary conditions at the surface in precisely this way. The qualitative behaviors predicted by the far-field consideration for the Stokeslet dipole, source dipole, and Stokeslet quadrupole do not vary dramatically for such boundary conditions, but the same cannot be stated for the rotlet dipole singularity (as described in \S III).

Among other physical and biological effects, we have ignored here the consequences of Brownian motion in order to take a clearer look at hydrodynamic effects. The interaction of both hydrodynamic and Brownian effects in swimming microorganisms has seen significant recent interest. \cite{ltt08} and \cite{lt09} have shown that wall collisions and rotational Brownian motion alone can give rise to accumulation of microswimmers near a surface. Reintroducing the hydrodynamic effects of the wall, the same authors also show the ways in which Brownian motion is amplified by hydrodynamic interactions with the wall. \cite{am87} have shown that {\it Rhodobacter sphaeroides} is apparently governed significantly by rotational Brownian motion alone. Equipped with a more complete theory for the wall effects on motility, we expect that the interplay between hydrodynamics and stochastic processes may be easier to distinguish in many cases. While Brownian motion and steric effects have been argued to drown out fluid mechanical interactions in swimming bacteria (and bacteria-wall interactions) by \cite{ddcgg11}, this is less likely to be the case for organisms of larger scale.

We acknowledge the support of the NSF through Grant No. CBET-0746285.

\appendix

\section{Vector notation for the fundamental singularities}

The fundamental singularities of the Stokes equations have a long history in the study of fluid dynamics, and their tensorial representations may be found in \cite{poz}. We refer the reader also to the work of \cite{cw75}. Here we include the vector notation of the fundamental singularities for reference. The Stokeslet, and Stokeslet dipole, quadrupole, and octupole may be written as
\begin{align}
&\v{G}(\v{x};\v{e})=\frac{1}{|\v{x}|}\left(\v{e}+\frac{(\v{x\cdot \v{e}})\v{x}}{|\v{x}|^2}\right),\\
&\v{G_D}(\v{x};\v{d},\v{e})=\frac{1}{|\v{x}|^2}\left(\frac{(\v{d\cdot x})\v{e}-(\v{e\cdot x})\v{d}-(\v{d\cdot e})\v{x}}{|\v{x}|}+\frac{3(\v{e\cdot x})(\v{d\cdot x})\v{x}}{|\v{x}|^3}\right),\\
&\v{G_Q}(\v{x};\v{c},\v{d},\v{e})=\frac{1}{|\v{x}|^3}\Big((\v{d\cdot e})\v{c}+(\v{c\cdot e})\v{d}-(\v{c\cdot d})\v{e}-\frac{3g_1}{|\v{x}|^2}+\frac{15(\v{c\cdot x})(\v{d\cdot x})(\v{e\cdot x})\v{x}}{|\v{x}|^4}\Big),\\
&\v{G_O}(\v{x};\v{b},\v{c},\v{d},\v{e})=\frac{3}{|\v{x}|^4}\Big(\frac{g_2}{|\v{x}|}-\frac{5 g_3}{|\v{x}|^3}+\frac{35(\v{b\cdot x})(\v{c\cdot x}) (\v{d\cdot x})(\v{e\cdot x}) \v{x}}{|\v{x}|^5}\Big),
\end{align}
where
\begin{align}
g_{1}=&\left[(\v{d\cdot e})(\v{c\cdot x})+(\v{c\cdot e})(\v{d\cdot x})+(\v{c\cdot d})(\v{e\cdot x})\right]\v{x}\nonumber\\
&+(\v{d\cdot x})(\v{e\cdot x})\v{c}+(\v{c\cdot x})(\v{e\cdot x})\v{d}-(\v{c\cdot x})(\v{d\cdot x})\v{e},\\
g_2=&[(\v{c\cdot d})(\v{e\cdot x})+(\v{c\cdot e})(\v{d\cdot x})+(\v{d\cdot e})(\v{c\cdot x})]\v{b} \nonumber\\
&+[(\v{d\cdot e})(\v{b\cdot x})+(\v{b\cdot e}) (\v{d\cdot x})+(\v{b\cdot d}) (\v{e\cdot x})] \v{c}\nonumber\\
&+ [(\v{c\cdot e})(\v{b\cdot x})+(\v{b\cdot e}) (\v{c\cdot x})+(\v{b\cdot c})(\v{e\cdot x}) ]\v{d}\nonumber\\
&-[(\v{c\cdot d})(\v{b\cdot x})+(\v{b\cdot d}) (\v{c\cdot x})+(\v{b\cdot c})(\v{d\cdot x})]\v{e}\nonumber\\
&+ [(\v{c\cdot d})(\v{b\cdot e})+(\v{c\cdot e})(\v{b\cdot d})+(\v{b\cdot c})(\v{d\cdot e}) ]\v{x},\\
g_3=&(\v{c\cdot x})(\v{d\cdot x}) (\v{e\cdot x})\v{b}+(\v{b\cdot x})(\v{d\cdot x}) (\v{e\cdot x})\v{c}\nonumber\\
&+(\v{b\cdot x})(\v{c\cdot x}) (\v{e\cdot x})\v{d}- (\v{b\cdot x})(\v{c\cdot x}) (\v{d\cdot x})\v{e}\nonumber\\
 & +\Big[(\v{b\cdot e})(\v{c\cdot x}) (\v{d\cdot x})+(\v{b\cdot d})(\v{c\cdot x}) (\v{e\cdot x})+(\v{b\cdot c})(\v{d\cdot x})(\v{e\cdot x})\nonumber\\
 &+(\v{c\cdot d})(\v{b\cdot x}) (\v{e\cdot x})+(\v{c\cdot e})(\v{b\cdot x})(\v{d\cdot x})+(\v{d\cdot e})(\v{b\cdot x}) (\v{c\cdot x})\Big]\v{x}.
\end{align}
Meanwhile, the source doublet and higher order source singularities in vector form are
\begin{align}
&\v{D}(\v{x;e})=\frac{1}{|\v{x}|^3}\left(-\v{e}+\frac{3(\v{x \cdot e})\v{x}}{|\v{x}|^2}\right),\\
&\v{Q}(\v{x};\v{d},\v{e})=\frac{-3}{|\v{x}|^4}\left(\frac{(\v{d\cdot x})\v{e}+(\v{e\cdot x})\v{d}+(\v{d\cdot e})\v{x}}{|\v{x}|}-\frac{5(\v{e\cdot x})(\v{d\cdot x})\v{x}}{|\v{x}|^3}\right),\\
&\v{O}(\v{x};\v{c},\v{d},\v{e})=\frac{3}{|\v{x}|^5}\Big((\v{c\cdot d})\v{e}+(\v{c\cdot e})\v{d}+(\v{d\cdot e})\v{c}-\frac{5g_4}{|\v{x}|^2}+\frac{35(\v{c\cdot x})(\v{d\cdot x})(\v{e\cdot x})\v{x}}{|\v{x}|^4}\Big),
\end{align}
where
\begin{gather}
g_4=(\v{d\cdot x})(\v{e\cdot x})\v{c}+(\v{c\cdot x})(\v{e\cdot x})\v{d}+(\v{c\cdot x})(\v{d\cdot x})\v{e}\nonumber\\
+\left[(\v{d\cdot e})(\v{c\cdot x})\v{x}+(\v{c\cdot d})(\v{e\cdot x})+(\v{c\cdot e})(\v{d\cdot x})\right]\v{x}.
\end{gather}
Finally, the rotlet and rotlet dipole may be written as
\begin{align}
&\v{R}(\v{x;e})=\frac{\v{e\times x}}{|\v{x}|^3},\\
&\v{R_D}(\v{x;d,e})=\frac{\v{d\times e}}{|\v{x}|^3}+\frac{3(\v{d\cdot x})(\v{e\times x})}{|\v{x}|^5}\cdot
\end{align}

\section{``Tilted'' image systems of the fundamental singularities}

\begin{figure}
\vspace{.15in}
\begin{center}
\includegraphics[width=2.5in]{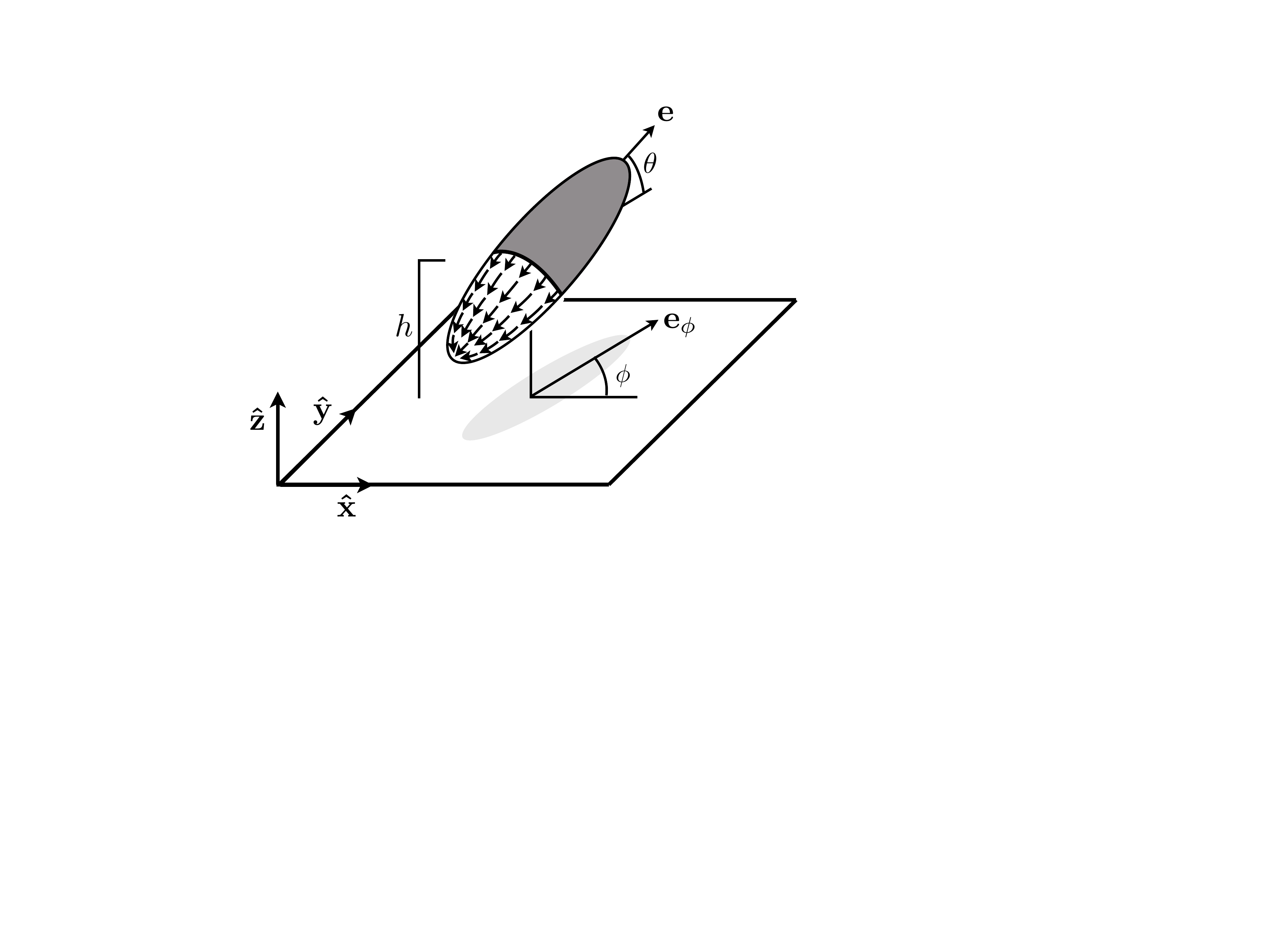}
\caption{A model swimmer near a wall. The body swims in a direction $\v{e}$, which may be decomposed into a horizontal part along $\v{e}_\phi$ and a vertical part along $\v{\hat{z}}$. The pitching angle $\theta=\cos^{-1}(\v{e\cdot e}_\phi)$ is indicated, as is the distance from the wall $h$ of the body centroid. $\phi$ denotes the angle of the horizontal swimming component relative to the $\v{\hat{x}}$ axis, and for the present study any variations in $\phi$ are decoupled from the pitching and height dynamics of $\theta$ and $h$. Ignoring the effect of the rotlet dipole, we may set $\phi=0$ and $\v{e}_\phi=\v{\hat{x}}$ without loss of generality.}
\label{Figure16}
\end{center}
\end{figure}

The image systems for the fundamental singularities (those axisymmetric about the swimming direction $\v{e}$) are dependent upon the distance of the singularity to the wall, $h$, and the orientation, $\v{e}$. The orientation $\v{e}$ is written in polar and azimuthal angles describing the swimming orientation via $\v{e}=\cos(\theta)\cos(\phi)\v{\hat{x}}+\cos(\theta)\sin(\phi)\v{\hat{y}}+\sin(\theta)\v{\hat{z}}$, as shown in Fig.~\ref{Figure16}. We proceed to show the effect of a no-slip wall and a stress-free surface on a body of aspect-ratio $e$ for each of the fundamental singularities considered in \eqref{Farfieldu}.

\subsection{Tilted Stokeslet image system}

The image system for a Stokeslet near a wall is well known \citep{bc74}. The image system for a Stokeslet directed along $\v{e}=\cos(\theta)\cos(\phi)\v{\hat{x}}+\cos(\theta)\sin(\phi)\v{\hat{y}}+\sin(\theta)\v{\hat{z}}$ in the fluid is simply a linear combination of these known image systems, and we write
\begin{align}
\v{G}^*(\v{x-x_0^*;e})=&\cos(\theta)\Big(-\v{G}(\v{e}_\phi)+2h \v{G_D}(\v{e}_\phi,\v{\hat{z}})-2 h^2\v{D}(\v{e}_\phi)\Big)\nonumber\\
&+\sin(\theta)\Big(-\v{G}(\v{\hat{z}})-2h\v{G_D}(\v{\hat{z},\hat{z}})+2 h^2 \v{D}(\v{\hat{z}})\Big),
\end{align}
where $\v{e}_\phi=\cos(\phi)\v{\hat{x}}+\sin(\phi)\v{\hat{y}}$. For the image systems described here and below, we use the shorthand notation $\v{G}(\v{e})=\v{G}(\v{x-x_0^*};\v{e})$ (the singularities are placed at the image point inside the wall).

The addition of the Stokeslet singularity and the image set above placed at the image location $\v{x_0^*}$ renders the fluid velocity zero on the wall exactly for arbitrary distances h from the wall,
\begin{gather}
\v{u}(\v{x})|_{z=0}=\frac{f}{8\pi\mu}\Big(\v{G}(\v{x-x_0};\v{e})+\v{G}^*(\v{x-x_0^*};\v{e})\Big)\Big|_{z=0}=0,
\end{gather}
where $\v{x_0}^*=\v{x_0}-2h\v{\hat{z}}$ is the position of the image singularity (inside the wall). Fax\'en's Law \eqref{FaxenU}-\eqref{FaxenO} is now used to determine the wall-induced translational ($\v{\tilde{u}}$) and rotational ($\v{\tilde{\Omega}}$) velocities of a (prolate ellipsoidal) body with a Stokeslet placed at its centroid $\v{\hat{z}}\cdot \v{x_0}=h$:
\begin{gather}
\tilde{u}_\phi=-\frac{3\cos(\theta)}{4h},\,\,\,\,\tilde{u}_z=-\frac{3\sin(\theta)}{2h},\,\,\,\,\tilde{\Omega}_{\phi^\perp}=-\frac{3 \Gamma  \cos(\theta)(1+\sin^2(\theta))}{8 h^2}\cdot
\end{gather}
Here we have written $\v{\tilde{u}}=u_\phi \v{e}_\phi+u_z \v{\hat{z}}$ and $\v{\tilde{\Omega}}=\tilde{\Omega}_{\phi^\perp}(\v{\hat{z}}\times\v{e}_\phi)$ (so that $\dot{\theta}=-\tilde{\Omega}_{\phi^\perp}$). There is no rotation outside of the $\v{\hat{z}}\times\v{e}_\phi$ component, which is assured by symmetry.

It is simpler to determine the velocities induced by the presence of another swimmer symmetrically placed opposite an imaginary plane, or equivalently by the presence of a stress-free boundary, at $z=0$; we need only apply an image system which consists solely of the mirror image of the primary singularity, or $\v{G}(\v{x-x_0^*};\v{e^*})$, where $\v{e^*}=\v{e}-2(\v{e\cdot \hat{z}})\v{\hat{z}}$. These induced velocities are
\begin{gather}
\tilde{u}_\phi=\frac{\cos(\theta)}{2h},\,\,\,\,\tilde{u}_z=-\frac{\sin(\theta)}{h},\,\,\,\,\tilde{\Omega}_{\phi^\perp}=-\frac{\cos(\theta) \left(1+3 \Gamma  \sin^2(\theta)\right)}{4 h^2}\cdot
\end{gather}
A Stokeslet aligned parallel to a stress-free boundary therefore increases the wall-parallel velocity component, while the same Stokeslet induces a decrease in the wall-parallel velocity near a no-slip wall. 

\subsection{Tilted Stokeslet dipole image system}

The image system for a tilted symmetric Stokeslet dipole near a wall is not simply a linear combination of wall-parallel and wall-perpendicular Stokeslet dipole image systems, as noted by \cite{bc74}. Instead, we may determine the image system for a tilted Stokeslet dipole by taking derivatives of the Stokeslet image system, where derivatives along $\v{\hat{z}}$ incorporate the h-dependence of the singularity coefficients. The resulting image system is
\begin{align}
\v{G^*_D}&(\v{x-x_0^*;e,e})=\cos^2(\theta)\left(-\v{G_D}(\v{e}_\phi,\v{e}_\phi)+2 h \v{G_Q}(\v{e}_\phi,\v{e}_\phi,\v{\hat{z}})-2 h^2\v{Q}(\v{e}_\phi,\v{e}_\phi)\right) \nonumber \\
&+\sin^2(\theta)\left(-\v{G_D}(\v{\hat{z}},\v{\hat{z}})+4 h \v{D}(\v{\hat{z}})+2 h \v{G_Q}(\v{\hat{z}},\v{\hat{z}},\v{\hat{z}})-2 h^2\v{Q}(\v{\hat{z}},\v{\hat{z}})\right)\nonumber \\
&+\sin(\theta)\cos(\theta)\left(\v{G_D}(\v{e}_\phi,\v{\hat{z}})+\v{G_D}(\v{\hat{z}},\v{e}_\phi)-4 h \v{D}(\v{e}_\phi)-4 h \v{S_Q}(\v{e}_\phi,\v{\hat{z}},\v{\hat{z}})+4 h^2\v{Q}(\v{e}_\phi,\v{\hat{z}})\right).
\end{align}

Fax\'en's Law returns the wall-induced translational and rotational velocities now of 
\begin{gather}
\tilde{u}_\phi=\frac{3\sin(2\theta)}{8h^2},\,\,\,\,\tilde{u}_z=-\frac{3(1-3\sin^2(\theta))}{8h^2},\,\,\,\,\tilde{\Omega}_{\phi^\perp}=\frac{3\sin(2\theta)}{16h^3}\left(1+\frac{\Gamma}{2}\left(1+\sin^2(\theta)\right)\right).
\end{gather}
Once again, as for all the symmetrized singularities about $\v{e}$, there is no rotation outside of the $\v{\hat{z}}\times\v{e}_\phi$ component, and we have written $\v{\Omega}=\Omega_{\phi^\perp}(\v{\hat{z}}\times\v{e}_\phi)$. For a swimming body with a positive Stokeslet dipole component (a pusher), for small angles $\theta$, the wall induces an attraction and a rotation acting to align the body with its long axis parallel to the wall.

For induced velocities near a mirror-image swimmer or near a stress-free surface we instead find  {the qualitatively similar effects:} 
\begin{gather}
\tilde{u}_\phi=0,\,\,\,\,\tilde{u}_z=-\frac{1}{4h^2}\left(1-3\sin^2(\theta)\right),\,\,\,\,\tilde{\Omega}_{\phi^\perp}=\frac{3\sin(2\theta)}{16h^3}\left(1+\Gamma \sin^2(\theta)\right).
\end{gather}

\subsection{Tilted Stokeslet quadrupole image system}

Similarly, directional derivatives of the tilted Stokeslet dipole system yield the image of the axisymmetric Stokeslet quadrupole:
\begin{align}
\v{G^*_Q}&(\v{x-x_0^*;e,e})=\cos^3(\theta)\left(-\v{G_Q}(\v{e}_\phi,\v{e}_\phi,\v{e}_\phi)+2 h \v{G_O}(\v{e}_\phi,\v{e}_\phi,\v{e}_\phi,\v{\hat{z}})-2 h^2\v{O}(\v{e}_\phi,\v{e}_\phi,\v{e}_\phi)\right) \nonumber \\
&+\sin^3(\theta)\left(3\v{G_Q}(\v{\hat{z}},\v{\hat{z}},\v{\hat{z}})+4 \v{D}(\v{\hat{z}})-2 h \v{G_O}(\v{\hat{z}},\v{\hat{z}},\v{\hat{z}},\v{\hat{z}})-8 h \v{Q}(\v{\hat{z}},\v{\hat{z}})+2 h^2\v{O}(\v{\hat{z}},\v{\hat{z}},\v{\hat{z}})\right)\nonumber \\
&+\cos^2(\theta)\sin(\theta)\Big[3\v{G_Q}(\v{e}_\phi,\v{e}_\phi,\v{\hat{z}}) +2\v{G_Q}(\v{\hat{z}},\v{e}_\phi,\v{e}_\phi)-8h \v{Q}(\v{e}_\phi,\v{e}_\phi)\nonumber\\
&-6h \v{G_O}(\v{e}_\phi,\v{e}_\phi,\v{\hat{z}},\v{\hat{z}})+6 h^2 \v{O}(\v{e}_\phi,\v{e}_\phi,\v{\hat{z}})\Big]\nonumber \\
&+\sin^2(\theta)\cos(\theta)\Big[-\v{G_Q}(\v{\hat{z}},\v{\hat{z}},\v{e}_\phi)-6\v{G_Q}(\v{e}_\phi,\v{\hat{z}},\v{\hat{z}})-4 \v{D}(\v{e}_\phi)+6h \v{G_O}(\v{e}_\phi,\v{\hat{z}},\v{\hat{z}},\v{\hat{z}})\nonumber\\
&+16h \v{Q}(\v{e}_\phi,\v{\hat{z}})-6 h^2\v{O}(\v{e}_\phi,\v{\hat{z}},\v{\hat{z}})\Big].
\end{align}

The wall-induced translational and rotational velocities associated with the tilted Stokeslet quadrupole are surprisingly simple:
\begin{gather}
\tilde{u}_\phi=\frac{\cos(\theta)}{16h^3}\left(7-27\sin^2(\theta)\right),\,\,\,\,\tilde{u}_z=\frac{\sin(\theta)}{4h^3}\left(7-9\sin^2(\theta)\right),\\
\tilde{\Omega}_{\phi^\perp}=\frac{3 \cos(\theta)}{8 h^4} \left(1-3 \sin^2(\theta)+\frac{\Gamma}{4} \left(11-3\sin^4(\theta)\right)\right).
\end{gather}

The induced velocities near a mirror-image swimmer or near a stress-free surface are
\begin{gather}
\tilde{u}_\phi=\frac{\cos(\theta)}{8h^3}\left(1-3\sin^2(\theta)\right),\,\,\,\,\tilde{u}_z=\frac{\sin(\theta)}{h^3}\left(1-\frac{3}{2}\sin^2(\theta)\right),\\
\tilde{\Omega}_{\phi^\perp}=\frac{3\cos(\theta)}{16h^4}\left(1-5\sin^2(\theta)+\Gamma\left(2+\sin^2(\theta)-3\sin^4(\theta)\right)\right).
\end{gather}

\subsection{Tilted source dipole image system}

As in the case of the tilted Stokeslet singularity, the tilted dipole image system is simply a linear combination of wall-parallel and wall-perpendicular image systems (see \cite{bc74}),
\begin{align}
\v{D}^*(\v{x-x_0^*;e})=&\cos(\theta)\Big(\v{D}(\v{e}_\phi)+2 \v{G_Q}(\v{e}_\phi,\v{\hat{z}},\v{\hat{z}})-2 h\v{Q}(\v{e}_\phi,\v{\hat{z}})\Big)\nonumber\\
&+\sin(\theta)\Big(-3\v{D}(\v{\hat{z}})-2\v{G_Q}(\v{\hat{z},\hat{z},\hat{z}})+2 h \v{Q}(\v{\hat{z}},\v{\hat{z}})\Big).
\end{align}

The wall-induced translational and rotational velocities are
\begin{gather}
\tilde{u}_\phi=-\frac{\cos(\theta)}{4h^3},\,\,\,\,\tilde{u}_z=-\frac{\sin(\theta)}{h^3},\,\,\,\,\tilde{\Omega}_{\phi^\perp}=-\frac{3\cos(\theta)}{8h^4}\left(1+\frac{3\Gamma}{2}\left(1+\sin^2(\theta)\right)\right).
\end{gather}

The induced velocities near a mirror-image swimmer or near a stress-free surface are
\begin{gather}
\tilde{u}_\phi=-\frac{\cos(\theta)}{8h^3},\,\,\,\,\tilde{u}_z=-\frac{\sin(\theta)}{4h^3},\,\,\,\,\tilde{\Omega}_{\phi^\perp}=-\frac{3\Gamma\cos(\theta)}{16h^4}(1+\sin^2(\theta)).
\end{gather}

\subsection{Tilted rotlet image system}

The image system for a tilted rotlet is given by
\begin{align}
\v{R^*}(\v{x-x_0^*;e})=&\cos(\theta)\left(-\v{R}(\v{e}_\phi)+2h \v{D}(\v{e}_\phi^\perp)-\v{G_D}(\v{e}_\phi^\perp,\v{\hat{z}})-\v{G_D}(\v{\hat{z}},\v{e}_\phi^\perp) \right)-\sin(\theta)\v{R}(\v{\hat{z}}),
\end{align}
where $\v{e}^\perp_\phi=-\sin(\phi)\v{\hat{x}}+\cos(\phi)\v{\hat{y}}$. The wall-induced velocities on the body are
\begin{gather}
\v{\tilde{u}}=0,\,\,\,\,\v{\tilde{\Omega}}=-\frac{\cos(\theta)}{16h^3}\left(5-3\Gamma\sin^2(\theta)\right)\v{\hat{\phi}}-\frac{\sin(\theta)}{16h^3}\left(2+3\Gamma\cos^2(\theta)\right)\v{\hat{z}}.
\end{gather}

The image singularity for motion near a mirror-image swimmer or near a stress-free surface are qualitatively different than those required near the no-slip wall. The wall-parallel component of the image system is simply
\begin{gather}
(\t{I}-\v{\hat{z}}\v{\hat{z}}')\v{R^*}(\v{x-x_0^*,e})=(\t{I}-\v{\hat{z}}\v{\hat{z}}')\v{R}(-\v{e}+2(\v{e\cdot z})\v{\hat{z}}).
\end{gather}
Meanwhile, the mirror-image of the singularity that is perpendicular to the boundary is exactly the same singularity, but shifted to the image point,
\begin{gather}
\v{\hat{z}}\cdot\v{R^*}(\v{x-x_0^*;e})=\v{\hat{z}}\cdot\v{R}(\v{e}).
\end{gather}
Combining the above, the induced velocities are in this case
\begin{gather}
\v{\tilde{u}}=\frac{\cos(\theta)}{4h^2}\v{\hat{\phi}^\perp},\,\,\,\,\v{\tilde{\Omega}}=\frac{\cos(\theta)}{16h^3}\left(1+3\Gamma\sin^2(\theta)\right)\v{\hat{\phi}}+\frac{\sin(\theta)}{16h^3}\left(2-3\Gamma\cos^2(\theta)\right)\v{\hat{z}}.
\end{gather}

Analagous to the qualitative change for a wall-parallel Stokeslet near a free-slip or no-slip surface, here there is an increase in the rotation rate for a rotating body near a stress-free surface, and a decrease in the rotation rate for a rotating body near a no-slip wall. Interestingly, there are angles $\theta$ for which the direction of rotational precession about the $\v{\hat{z}}$ axis depends upon the body aspect-ratio.

\subsection{Tilted rotlet doublet image system}

The image system for a tilted (axisymmetric) rotlet is given by
\begin{align}
\v{R_D^*}(\v{x-x_0^*};\v{e},\v{e})&=\cos^2(\theta)\left(-\v{R_D}(\v{e}_\phi,\v{e}_\phi)+2h \v{Q}(\v{e}_\phi,\v{e}_\phi^\perp)-\v{G_Q}(\v{e}_\phi,\v{e}_\phi^\perp,\v{\hat{z}})-\v{G_Q}(\v{e}_\phi,\v{\hat{z}},\v{e}_\phi^\perp) \right)\nonumber\\
&+\sin^2(\theta)\v{R_D}(\v{\hat{z}},\v{\hat{z}}),\nonumber\\
&+\cos(\theta)\sin(\theta)\Big[\v{R_D}(\v{\hat{z}},\v{e}_\phi) -\v{R_D}(\v{e}_\phi,\v{\hat{z}})\nonumber\\
&+2\v{D}(\v{e}_\phi^\perp)+\v{G_Q}(\v{e}^\perp_\phi,\v{\hat{z}},\v{\hat{z}})+\v{G_Q}(\v{e}^\perp_\phi,\v{e}_\phi^\perp,\v{\hat{z}})-2h\v{Q}(\v{e}_\phi^\perp,\v{\hat{z}}) \Big].
\end{align}

The induced velocities on the body are
\begin{subequations}
\begin{gather}
\v{\tilde{u}}=0,
\end{gather}
\begin{gather}
\v{\tilde{\Omega}}=\frac{3\sin(2 \theta)}{64h^4} \left(6-\Gamma(1+3\sin^2(\theta)\right)\v{\hat{\phi}}-\frac{3}{32 h^4}\left(1-3\sin^2(\theta)-\Gamma\cos^2(\theta)(1+3\sin^2(\theta))\right)\v{\hat{z}}.
\end{gather}
\end{subequations}

The image singularity for motion near a mirror-image swimmer or near a stress-free surface is given by 
\begin{gather}
\v{R_D^*}(\v{x-x_0^*};\v{e,e})=\v{R_D}(\v{e}-2(\v{e\cdot z})\v{\hat{z}},-\v{e}+2(\v{e\cdot z})\v{\hat{z}}).
\end{gather}
The induced velocities are in this case
\begin{gather}
\v{\tilde{u}}=-\frac{3\sin(2\theta)}{16h^3}\v{\hat{\phi}^\perp},
\end{gather}
\begin{align}
\v{\tilde{\Omega}}=&-\frac{3\sin(2 \theta)}{64h^4} \left(2+\Gamma(1+3\sin^2(\theta)\right)\v{\hat{\phi}}+\frac{3}{32 h^4}\left(1-3\sin^2(\theta)+\Gamma\cos^2(\theta)(1+3\sin^2(\theta))\right)\v{\hat{z}}.
\end{align}

\subsection{Stresslet image system}

The image system of the Stresslet singularity which appears in the double layer integral in \eqref{BoundaryIntegral}, may be deduced by decomposing it into the fundamental singularities considered above. Consider the $\v{d,e}$-Stresslet, as defined in \eqref{Stresslet},
\begin{gather}
\v{T}(\v{x};\v{d,e})=\frac{-6(\v{d\cdot x})(\v{e\cdot x})\v{x}}{|\v{x}|^5}\cdot
\end{gather}
Decomposing the image singularities into operators acting on the components of $\v{e}$, we write
\begin{gather}
\v{T^*}(\v{x;d,e})=\v{T}^*\Big|_{e_x} e_x+\v{T}^*\Big|_{e_y} e_y+\v{T}^*\Big|_{e_z} e_z,
\end{gather}
where
\begin{align}
\v{T}^*\Big|_{e_x}&=-2d_x \v{U}+4h \,d_z\v{D(\hat{x})}+4h\,d_x\v{D(\hat{z})}\nonumber\\
&+2d_{x}\v{G_D(\hat{x},\hat{x})}-\,4d_x \v{G_D(\hat{z},\hat{z})}+d_y\Big( \v{G_D(\hat{x},\hat{y})}+\v{G_D(\hat{y},\hat{x})}\Big)\nonumber\\
&-d_{z}\Big(\v{G_D(\hat{x},\hat{z})}+\v{G_D(\hat{z},\hat{x})}\Big)+4 h^2 \,d_x\v{Q(\hat{x},\hat{x})}+4 h^2\,d_y \v{Q(\hat{x},\hat{y})}\nonumber\\
&-4 h^2\,d_z \v{Q(\hat{x},\hat{z})}+4h\,d_z\v{G_Q(\hat{x},\hat{z},\hat{z})}-4h\, d_x\v{G_Q(\hat{x},\hat{x},\hat{z})}-4h\,d_y\v{G_Q(\hat{x},\hat{y},\hat{z})},
\end{align}

\begin{align}
\v{T}^*\Big|_{e_y}&=-2d_y \v{U}+4h \,d_z\v{D(\hat{y})}+4h\,d_y\v{D(\hat{z})}\nonumber\\
&+2d_{y}\v{G_D(\hat{y},\hat{y})}-\,4d_y \v{G_D(\hat{z},\hat{z})}+d_x\Big( \v{G_D(\hat{x},\hat{y})}+\v{G_D(\hat{y},\hat{x})}\Big)\nonumber\\
&-d_{z}\Big(\v{G_D(\hat{y},\hat{z})}+\v{G_D(\hat{z},\hat{y})}\Big)+4 h^2 \,d_y\v{Q(\hat{y},\hat{y})}+4 h^2\,d_x \v{Q(\hat{x},\hat{y})}\nonumber\\
&-4 h^2\,d_z \v{Q(\hat{y},\hat{z})}+4h\,d_z\v{G_Q(\hat{y},\hat{z},\hat{z})}-4h\, d_y\v{G_Q(\hat{y},\hat{y},\hat{z})}-4h\,d_x\v{G_Q(\hat{x},\hat{y},\hat{z})},
\end{align}

\begin{align}
\v{T}^*\Big|_{e_z}&=-2d_z \v{U}+4h \,d_x\v{D(\hat{x})}+4h \,d_y\v{D(\hat{y})}-4h\,d_z\v{D(\hat{z})}\nonumber\\
&-2d_{z}\v{G_D(\hat{z},\hat{z})}-d_x\Big( \v{G_D(\hat{x},\hat{z})}+\v{G_D(\hat{z},\hat{x})}\Big)-d_y\Big( \v{G_D(\hat{y},\hat{z})}+\v{G_D(\hat{z},\hat{y})}\Big)\nonumber\\
&+4 h^2 \,d_z\v{Q(\hat{z},\hat{z})}-4 h^2\,d_x \v{Q(\hat{x},\hat{z})}-4 h^2\,d_y \v{Q(\hat{y},\hat{z})}\nonumber\\
&+4h\,d_x\v{G_Q(\hat{x},\hat{z},\hat{z})}+4h\,d_y\v{G_Q(\hat{y},\hat{z},\hat{z})}-4h\,d_z\v{G_Q(\hat{z},\hat{z},\hat{z})}.
\end{align}
Recall that all the singularities above are placed at the image point $\v{x=x_0^*}$, as in Appendix B. 

When the Stresslet appears in the boundary integral representation in \eqref{BoundaryIntegral}, we have $\v{d}=\v{n}(\v{y})$ and $\v{e}=\v{u}(\v{y})-\v{u}(\v{x})$. Note that the single layer integral in \eqref{BoundaryIntegral} may be eliminated with the introduction of an auxiliary Stresslet singularity density $\v{q}(\v{x})$ \citep{pm87,poz}. In this case, $\v{d}=\v{n(y)}$ is still determined by the geometry but $\v{e}=\v{q}(\v{y})-\v{q}(\v{x})$ is unknown. The framework above may then be used to construct the linear operator acting on $\v{q}$, an approach amenable to highly accurate numerical schemes which may be made spatially adaptive at no theoretical cost, unlike approaches based on the single layer integral which are Fredholm equations of the first kind and thus lack the foundational theory underlying the suggested method.


\begin{thebibliography}{}

\bibitem[Ainley {\it et al.}(2008)]{adebc08} \textsc{Ainley, J.,  Durkin, S., Embid, R., Biondala, P., \& Cortez, R.} 2008 {The method of images for regularized Stokeslets}. \textit{J. Comp. Phys.} \textbf{227}, 4600--4616.

\bibitem[Armitage \& Macnab(1987)]{am87} \textsc{Armitage, J. P. \& Macnab, R. M.} 1987 {Unidirectional, intermittent rotation of the flagellum of rhodobacter sphaeroides}. \textit{J. Bacteriol.} \textbf{169}, 514--518.

\bibitem[Bees \& Croze(2010)]{bc10} \textsc{Bees, M. A.\& Croze, O.A.} 2010 {Dispersion of biased swimming micro-organisms in a fluid flowing through a tube}. \textit{Proc. R. Soc. A} \textbf{466}, 2057--2077.

\bibitem[Berg \& Turner(1990)]{bt90} \textsc{Berg, H.C. \& Turner, L.} 1990 {Chemotaxis of bacteria in glass capillary arrays}. \textit{Biophys. J.} \textbf{58}, 919--930

\bibitem[Berke {\it et al.}(2008)]{btbl08} \textsc{Berke, A. P.,  Turner, L. , Berg, H. C., \& Lauga, E.} 2008 {Hydrodynamic attraction of swimming microorganisms by surfaces}. \textit{Phys. Rev. Lett.} \textbf{101}, 038102.

\bibitem[Blake(1971)]{Blake71} \textsc{Blake, J.R.} 1971 {A spherical envelope approach to ciliary propulsion}. \textit{J.~Fluid Mech.} \textbf{46}, 199--208.

\bibitem[Blake \& Chwang(1974)]{bc74} \textsc{Blake, J.R. \& Chwang, A.T.} 1974 {Fundamental singularities of viscous flow}. \textit{J. Eng. Math.} \textbf{8}, 23--29.

\bibitem[Bouzarth \& Minion(2011)]{bm11} \textsc{Bouzarth, E.L. \& Minion, M.L.} 2011 {Modeling slender bodies with the method of regularized Stokeslets}. \textit{J. Comp. Phys.} \textbf{230}, 3929--3947.

\bibitem[Brady \& Bossis(1985)]{bb85} \textsc{Brady, J.F. \& Bossis, G.} 1985 {The rheology of concentrated suspensions of spheres in simple shear flow by numerical simulation}. \textit{J. Fluid Mech.} \textbf{155}, 105--129.

\bibitem[Brenner(1961)]{Brenner61} \textsc{Brenner, H.} 1961 {Effect of finite boundaries on the Stokes resistance of an arbitrary particle}. \textit{J.~Fluid Mech.} \textbf{8}, 35--48.

\bibitem[Childress(1981)]{Childress81}
\textsc{Childress, S.} 1981 \emph{Mechanics of Swimming and Flying}. Cambridge University Press, Cambridge, U.K..

\bibitem[Chwang  \& Wu(1975)]{cw75} \textsc{Chwang, A.T. \& Wu, T.Y.T.} 1975 {Hydromechanics of low-Reynolds-number flow. Part 2. Singularity method for Stokes flows}. \textit{J. Fluid Mech.} \textbf{67}, 787--815.

\bibitem[Cisneros {\it et al.}(2007)]{cgdk07} \textsc{Cisneros, L., Dombrowski, C., Goldstein, R.E., \& Kessler, J. O.} 2007 {Reversal of bacteria at an obstacle}. \textit{Phys. Rev. E} \textbf{73}, 030901(R).

\bibitem[Cortez(2002)]{Cortez02} \textsc{Cortez, R.} 2002 {The Method of Regularized Stokeslets}. \textit{SIAM J. Sci. Comput.} \textbf{23}, 1204--1225.

\bibitem[Crowdy \& Or(2010)]{co10} \textsc{Crowdy, D.G. \& Or, Y.} 2010 {Two-dimensional point singularity model of a low-Reynolds-number swimmer near a wall}. \textit{Phys. Rev. E} \textbf{81}, 036313.

\bibitem[Crowdy(2011)]{Crowdy11} \textsc{Crowdy, D.G.} 2011 {Treadmilling swimmers near a no-slip wall at low Reynolds number}. \textit{Int. J. Non-Linear Mech.} \textbf{46}, 577--585.

\bibitem[Di Leonardo \textit{et al.}(2011)]{dldaai11} \textsc{Di Leonardo, Dell'Arciprete, D., Angelani, L. \& Iebba, V.} 2011 {Swimming with an Image}. \textit{Phys. Rev. Lett.} \textbf{106}, 038101.

\bibitem[Di Leonardo \textit{et al.}(2010)]{dladariscmdadf10} \textsc{Di Leonardo, Angelani, L.,  Dell'Arciprete, D., Ruocco, G., Iebba, V., Schippa, S., Conte, M.P., Mecarini, F., De Angelis, F., \& Di Fabrizio, E.} 2010 {Bacterial ratchet motors}. \textit{Proc. Natl. Acad. Sci. USA} \textbf{107}, 9541--9545.


\bibitem[Drescher {\it et al.}(2009)]{dltipg09} \textsc{Drescher, K., Leptos, K. C., Tuval, I., Ishikawa, T., Pedley, T.J., \& Goldstein, R.E.} 2009 {Dancing Volvox: Hydrodynamic Bound States of Swimming Algae}. \textit{Phys. Rev. Lett.} \textbf{102}, 168101.

\bibitem[Drescher {\it et al.}(2010)]{dgmpt10} \textsc{Drescher, K., Goldstein, R.E., Michel, N., Polin, M., \& Tuval, I.} 2010 {Direct Measurement of the Flow Field around Swimming Microorganisms}. \textit{Phys. Rev. Lett.} \textbf{105}, 168101.

\bibitem[Drescher {\it et al.}(2011)]{ddcgg11} \textsc{Drescher, K., Dunkel, J., Cisneros, L.H., Ganguly, S., \& Goldstein, R.E.} 2011 {Fluid dynamics and noise in bacterial cell-cell and cell-surface scattering}. \textit{Proc. Nat. Acad. Sci. USA} \textbf{108}, 10940--10945.

\bibitem[Dreyfus {\it et al.}(2005)]{dbrfsb05} \textsc{Dreyfus R., Baudry J., Roper M. L., Fermigier M., Stone H. A. \& Bibette J.} 2005 {Microscopic artificial swimmers}. \textit{Nature} \textbf{437}, 862--865.

\bibitem[Elgeti \& Gompper(2009)]{eg09} \textsc{Elgeti, J. \& Gompper, G.} 2009 {Self-propelled rods near surfaces}. \textit{Euro. Phys. Lett.} \textbf{85}, 38002.

\bibitem[Elgeti, Kaupp \& Gompper(2010)]{ekg10} \textsc{Elgeti, J., Kaupp, U. B. \& Gompper, G.} 2010 {Hydrodynamics of sperm cells near surfaces}. \textit{Biophys. J.} \textbf{99}, 1018--1026.

\bibitem[Evans \& Lauga(2010)]{el10} \textsc{Evans, A. \& Lauga, E.} 2010 {Propulsion by passive filaments and active flagella near boundaries}. \textit{Phys. Rev. E} \textbf{82}, 041915.

\bibitem[Fauci \& McDonald(1995)]{fm95} \textsc{Fauci, L.J. \& McDonald, A.} 1995 {Sperm motility in the presence of boundaries}. \textit{Bull. Math. Biol.} \textbf{57}, 679--699.


\bibitem[Fournier-Bidoz {\it et al.}(2005)]{fbamo05} \textsc{Fournier-Bidoz S., Arsenault A. C., Manners I. \& Ozin G. A.} 2005 {Synthetic self-propelled nanorotors}. \textit{Chem. Commun.} \textbf{4}, 441--443.

\bibitem[Galajda {\it et al.}(2007)]{gkca07} \textsc{Galajda P., Keymer J., Chaikin P. and Austin R.} 2007 {A Wall of Funnels Concentrates Swimming Bacteria}. \textit{J. Bacteriol.} \textbf{189}, 8704--8707.


\bibitem[Ghosh \& Fischer(2009)]{gf09} \textsc{Ghosh, A. \& Fischer, P.} 2009 {Controlled Propulsion of Artificial Magnetic Nanostructured Propellers}. \textit{Nanoletters} \textbf{9}, 2243--2245.

\bibitem[Giacch\'e, Ishikawa \& Yamaguchi(2010)]{giy10} \textsc{Giacch\'e, D., Ishikawa, T., \& Yamaguchi, T.} 2010 {Hydrodynamic entrapment of bacteria swimming near a solid surface}. \textit{Phys. Rev. E} \textbf{82}, 056309.

\bibitem[Golestanian, Liverpool, \& Ajdari(2007)]{gla07} \textsc{Golestanian, R., Liverpool, T. B. \& Adjari, A.} 2007 {Designing phoretic micro- and nano-swimmers}. \textit{New J. Phys.} \textbf{9}, 126.

\bibitem[Goldman, Cox, \& Brenner(1966)]{gcb66} \textsc{Goldman, A. J., Cox, R. G. \& Brenner, H.} 1966 {Slow viscous motion of a sphere parallel to a plane wall -- I Motion through a quiescent fluid}. \textit{Chem. Eng. Sci.} \textbf{22}, 637--651.

\bibitem[Goto \textit{et al.}(2005)]{gnbnm05} \textsc{Goto, T., Nakata, K., Baba, K., Nishimura, M., \& Magariyama, Y.} 2005 {A fluid-dynamic interpretation of the asymmetric motion of singly flagellated bacteria swimming close to a boundary}. \textit{Biophys. J.} \textbf{89}, 3771--3779.

\bibitem[Gray \& Hancock(1955)]{gh55} \textsc{Gray, J. \& Hancock, G. J.} 1955 {The Propulsion of Sea-Urchin Spermatozoa}. \textit{J. Exp. Biol.} \textbf{32}, 802--814. 

\bibitem[Guasto {\it et al.}(2010)]{gjg10} \textsc{Guasto, J. S., Johnson, K. A., \& Gollub, J. P.} 2010 {Oscillatory Flows Induced by Microorganisms Swimming in Two Dimensions}. \textit{Phys. Rev. Lett.} \textbf{105}, 168102.

\bibitem[Happel \& Brenner(1965)]{hb65}
\textsc{Happel, J. and Brenner, H.} 1965 \emph{Low Reynolds Number Hydrodynamics}. Prentice Hall, Inc., Englewood Cliffs, N.J..

\bibitem[Harkes, Dankert \& Feijen(1992)]{hdf92} \textsc{Harkes, G., Dankert, J. \& Feijen, J.} 1992 {Bacterial migration along solid surfaces}. \textit{Appl. Environ. Microbiol. } \textbf{58}, 1500--1505.


\bibitem[Harshey(2003)]{Harshey03} \textsc{Harshey, R. M.} 2003 {Bacterial Motility on a Surface: Many Ways to a Common Goal}. \textit{Ann. Rev. Microbiol.} \textbf{57}, 249--273.

\bibitem[Harvey \& Young(1980)]{hy80} \textsc{Harvey, R. W., and L. Y. Young} 1980 {Enumeration of particle-bound and unattached respiring bacteria in the salt marsh environment}. \textit{Appl. Environ. Microbiol.} \textbf{40}, 156--160.

\bibitem[Hernandez-Ortiz {\it et al.}(2009)]{houg09} \textsc{Hernandez-Ortiz, J.P., Underhill, P. T., \& Graham, M.D.} 2009 {Dynamics of confined suspensions of swimming particles}. \textit{J. Phys.: Condens. Matter} \textbf{21}, 204107.

\bibitem[Hernandez-Ortiz {\it et al.}(2005)]{hosg05} \textsc{Hernandez-Ortiz, J.P., Stoltz, C.G. \& Graham, M.D.} 2005 {Transport and Collective Dynamics in Suspensions of Confined Swimming Particles}. \textit{Phys. Rev. Lett.} \textbf{95}, 204501.

\bibitem[Hill \textit{et al.}(2007)]{hkmk07} \textsc{Hill, J., Kalkanci, O., McMurry, J. L., \& Koser, H.} 2007 {Hydrodynamic surface interactions enable Escherichia Coli to seek efficient routes to swim upstream}. \textit{Phys. Rev. Lett.} \textbf{98}, 068101.

\bibitem[Hohenegger \& Shelley (2010)]{hs10} \textsc{Hohenegger, C. \& Shelley, M. J.} 2010 {Stability of active suspensions}. \textit{Phys. Rev. E} \textbf{81}, 046311.

\bibitem[Ishikawa \& Pedley(2007)]{ip07} \textsc{Ishikawa, T. \& Pedley, T. J.} 2007 {Diffusion of swimming model micro-organisms in a semi-dilute suspension}. \textit{J.~Fluid Mech.} \textbf{588}, 437--462.

\bibitem[Ishikawa, Simmonds, \& Pedley(2006)]{isp06} \textsc{Ishikawa, T., Simmonds, M. P., \& Pedley, T. J.} 2006 {Hydrodynamic interaction of two swimming model micro-organisms}. \textit{J.~Fluid Mech.} \textbf{568}, 119--160.

\bibitem[Jiang, Yoshinaga, \& Sano(2010)]{jys10} \textsc{Jiang, H.-R., Yoshinaga, N., \& Sano, M.} 2010 {Active motion of a Janus particle by self-thermophoresis in a defocused laser beam}. \textit{Phys. Rev. Lett.} \textbf{105}, 268302.

\bibitem[Johnson \& Brokaw(1979)]{jb79} \textsc{Johnson, R. E., \& Brokaw, C. J.} 1979 {Flagellar hydrodynamics. A comparison between resistive-force theory and slender-body theory}. \textit{Biophys. J.} \textbf{25}, 113--127.

\bibitem[Kanevsky, Shelley, \& Tornberg(2010)]{kst10} \textsc{Kanevsky, A., Shelley, M. J., \& Tornberg, A.-K.} 2010 {Modeling simple locomotors in Stokes flow}. \textit{J. Comput. Phys.} \textbf{229}, 958--977.

\bibitem[Kaya \& Koser(2009)]{kk09} \textsc{Kaya, T. \& Koser, H.} 2009 {Characterization of hydrodynamic surface interactions of Escherichia coli cell bodies in shear flow}. \textit{Phys. Rev. Lett.} \textbf{103}, 138103.

\bibitem[Keller  \& Wu(1975)]{kw77} \textsc{Keller, S.R. \& Wu, T.Y.T.} 1977 {A porous prolate-spheroidal model for ciliated micro-organisms}. \textit{J. Fluid Mech.} \textbf{80}, 259--278.

\bibitem[Kim \& Karrila(1991)]{kk91}
\textsc{Kim, S. and Karrila, S.J.} 1991 \emph{Microhydrodynamics: Principles and Selected Applications}. Dover Publications, Inc., Mineola, N.Y.

\bibitem[Klapper \& Dockery(2010)]{kd10} \textsc{Klapper, I. \& Dockery, J.} 2010 {Mathematical Description of Microbial Biofilms}. \textit{SIAM Review} \textbf{52}, 221--265.

\bibitem[Kolter \& Greenberg(2006)]{kg06} \textsc{Kolter, R. \& Greenberg, E. P.} 2006 {The superficial life of microbes}. \textit{Nature} \textbf{441}, 300--302.

\bibitem[Lauga {\it et al.}(2006)]{ldlws06} \textsc{Lauga, E., DiLuzio, W. R., Whitesides, G. M. \& Stone, H. A.} 2006 {Swimming in circles: Motion of bacteria near solid boundaries}. \textit{Biophys. J.} \textbf{90}, 400--412.

\bibitem[Lauga \& Powers(2009)]{lp09} \textsc{Lauga, E. \& Powers, T.} 2009 {The hydrodynamics of swimming microorganisms}. \textit{Rep. Prog. Phys.} \textbf{72}, 096601.

\bibitem[Li \& Tang(2009)]{lt09} \textsc{Li, G. \& Tang, J. X.} 2009 {Amplified effect of Brownian motion in bacterial near-surface swimming}. \textit{PNAS} \textbf{105}, 18355--18359.

\bibitem[Li, Tam \& Tang(2008)]{ltt08} \textsc{Li, G., Tam, L.-K. \& Tang, J. X.} 2008 {Accumulation of microswimmers near a surface mediated by collision and rotational Brownian motion}. \textit{Phys. Rev. Lett.} \textbf{103}, 078101.

\bibitem[Liao {\it et al.}(2011)]{lsdkw07} \textsc{Liao, Q., Subramanian, G., DeLisa, M. P., Koch, D. L., \& Wu, M.} 2007 {Pair velocity correlations among swimming Escherichia coli bacteria are determined by force-quadrupole hydrodynamic interactions}. \textit{Phys. Fluids} \textbf{19}, 061701.


\bibitem[Lighthill(1952)]{Lighthill52} \textsc{Lighthill, M. J.} 1952 {On the squirming motion of nearly spherical deformable bodies through liquids at very small Reynolds numbers}. \textit{Comm. Pure Appl. Math.} \textbf{5}, 109--118.

\bibitem[Lighthill(1996)]{Lighthill96} \textsc{Lighthill, M. J.} 1996 {Helical distributions of Stokeslets}. \textit{J. Eng. Math.} \textbf{30}, 35--78.


\bibitem[Lin, Thiffeault, \& Childress(2011)]{ltc11} \textsc{Lin, Z., Thiffeault, J-L., \& Childress, S.} 2011 {Stirring by squirmers}. \textit{J. Fluid Mech.} \textbf{669}, 167--177.

\bibitem[Llopis \& Pagonabarraga(2010)]{lp10} \textsc{Llopis, I. \& Pagonabarraga, I.} 2011 {Hydrodynamic interactions in squirmer motion: Swimming with a neighbour and close to a wall}. \textit{J. Non-Newt. Fluid Mech.} \textbf{165}, 946--952.

\bibitem[Lynch, Lappin-Scott \& Costerton(2003)]{llsc03}
\textsc{Lynch, J. F., Lappin-Scott, H. M., \& Costerton, J. W.} 2003 \emph{Microbial Biofilms}. Cambridge University Press, Cambridge, U.K..

\bibitem[Magnaudet {\it et al.}(2003)]{Magnaudet03} \textsc{Magnaudet, J., Takagi, S., \& Legendre, D.} 2003 \emph{Drag, deformation and lateral migration of a buoyant drop moving near a wall}. \textit{J. Fluid Mech.} \textbf{476}, 115--157.

\bibitem[Michelin \& Lauga(2010)]{ml10} \textsc{Michelin, S. \& Lauga, E.} 2010 {Efficiency optimization and symmetry-breaking in a model of ciliary locomotion}. \textit{Phys. Fluids} \textbf{22}, 111901.

\bibitem[O'Toole(2000)]{otkk00} \textsc{OÕToole, G., Kaplan, H.B. \& Kolter, R.} 2000 {Biofilm Formation as Microbial Development}. \textit{Annu. Rev. Microbiol.} \textbf{54}, 49--79.

\bibitem[Pak {\it et al.}(2011)]{pgwl11} \textsc{Pak, O.S., Gao, W., Wang, J. \& Lauga, E.} 2011 {High-Speed Propulsion of Flexible Nanowire Motors: Theory and Experiments}. \textit{Soft Matter} \textbf{7}, 8169--8181.

\bibitem[Paxton {\it et al.}(2004)]{pkossacmlc04} \textsc{Paxton, W. F., Kistler, K. C., Olmeda, C. C., Sen, A., St. Angelo, S. K., Cao, Y., Mallouk, T. E., Lammert, P. E. \& Crespi, V. H. } 2004 {Catalytic Nanomotors: Autonomous Movement of Striped Nanorods}. \textit{J. Am. Chem. Soc.} \textbf{126}, 13424.

\bibitem[Poortinga {\it et al.}(2002)]{pbnb02} \textsc{Poortinga, A. T., Bos, R., Norde, W. \& Busscher, H. J.} 2002 {Electric double layer interactions in bacterial adhesion to surfaces}. \textit{Surf. Sci. Rep.} \textbf{47}, 1--32.

\bibitem[Power \& Miranda(1987)]{pm87} \textsc{Power, H. \& Miranda, G.} 1987 {Second kind integral equation formulation of Stokes' flows past a particle of arbitrary shape}. \textit{SIAM J. Appl. Math.} \textbf{47}, 689--698.

\bibitem[Pozrikidis(1992)]{poz}
\textsc{Pozrikidis C.} 1992 \emph{Boundary Integral and Singularity Methods for Linearized Viscous Flow}. Cambridge University Press, Cambridge, U.K..

\bibitem[Purcell(1977)]{Purcell77} \textsc{Purcell, E. M.} 1977 {Life at Low Reynolds Number}. \textit{Am. J. Phys.} \textbf{45}, 3--11.

\bibitem[Rothschild(1963)]{Rothschild63} \textsc{Rothschild, L. J.} 1963 {Non-random distribution of bull spermatazoa in a drop of sperm suspension}. \textit{Nature (London)} \textbf{198}, 122--1222.

\bibitem[R\"uckner \& Kapral(2007)]{rk07} \textsc{R\"uckner G. \& Kapral, R.} 2007 {Chemically Powered Nanodimers}. \textit{Phys. Rev. Lett.} \textbf{98}, 150603.

\bibitem[Saintillan \& Shelley(2008)]{ss08} \textsc{Saintillan, D. \& Shelley, M.J.} 2008 {Instabilities and Pattern Formation in Active Particle Suspensions: Kinetic Theory and Continuum Simulation}. \textit{Phys. Rev. Lett.} \textbf{100}, 178103.

\bibitem[Shum, Gaffney \& Smith(2010)]{sgs10} \textsc{Shum, H., Gaffney, E.A., \& Smith, D.J.} 2010 {Modelling bacterial behaviour close to a no-slip plane boundary: the influence of bacterial geometry}. \textit{Proc. R. Soc. A} \textbf{466}, 1725--1748.

\bibitem[Smith \& Blake(2009)]{sb09} \textsc{Smith, D.J. \& Blake, J.R.} 2009 {Surface accumulation of spermatozoa: a fluid dynamic phenomenon}. \textit{The Mathematical Scientist} \textbf{34} 74--87.

\bibitem[Smith \textit{et al.}(2009)]{sgbk09} \textsc{Smith, D.J., Gaffney, E.A., Blake, J.R., \& Kirkman-Brown, J.C.} 2009 {Human sperm accumulation near surfaces: a simulation study}. \textit{J. Fluid Mech.} \textbf{621}, 289--320.

\bibitem[Spagnolie \& Lauga(2010)]{sl10} \textsc{Spagnolie, S. E. \& Lauga, E.} 2010 {Jet propulsion without inertia}. \textit{Phys. Fluids}, \textbf{22}, 081902.

\bibitem[Swan \& Khair(2008)]{sk08} \textsc{Swan, J. W. \& Khair, A. S.} 2008 {On the hydrodynamics of `slip--stick' spheres}. \textit{J. Fluid Mech.}, \textbf{606}, 115--132.

\bibitem[Taylor(1951)]{Taylor51} \textsc{Taylor, G.I.}  1951 {Analysis of the swimming of microscopic organisms}. \textit{Proc. Roy. Soc. Lond.} \textbf{A209}, 447--461.

\bibitem[Tuval {\it et al.}(2005)]{tcdwkg05} \textsc{Tuval, I., Cisneros, L., Dombrowski, C., Wolgemuth, C. W., Kessler, J.O., \& Goldstein, R. E.} 2005 {Bacterial swimming and oxygen transport near contact lines}. \textit{Proc. Natl. Acad. Sci. USA} \textbf{102}, 2277--2282.

\bibitem[Van Loosdrecht \textit{et al.}(1990)]{vllnz90} \textsc{Van Loosdrecht, M.C.M., Lyklema, J., Norde, W., \& Zehnder, A.J.B.} 2003 {Influence of interfaces on microbial activity}. \textit{Microbiol. Rev.} \textbf{54}, 75--87.

\bibitem[Wang(2009)]{Wang09} \textsc{Wang, J.} 2009 {Can Man-Made Nanomachines Compete with Nature Biomotors?} \textit{ACS Nano} \textbf{3}, 4--9.

\bibitem[Zargar, Najafi, \& Miri(2009)]{znm09} \textsc{Zargar, R., Najafi, A., \& Miri, M.} 2009 {Three-sphere low-Reynolds-number swimmer near a wall}. \textit{Phys. Rev. E} \textbf{80}, 026308.

\bibitem[Zhu et al.(2011)]{zdqlb11} \textsc{Zhu, L., Do-Quang, M., Lauga, E. and Brandt, L.} 2011 {Locomotion by tangential deformation in a polymeric fluid}. \textit{Phys. Rev. E} \textbf{83}, 011901.


\end{thebibliography}
\end{document}